\documentclass[11pt]{cmuthesis}

\newcommand{\eg}{{\em e.g.,}}
\newcommand{\ie}{{\em i.e.,}}
\newcommand{\etc}{{\em etc.}}
\newcommand{\etal}{{\em et~al.}}

\usepackage{epsfig} 
\usepackage{graphicx}
\usepackage{multirow}
\usepackage{tabu}
\usepackage{amsmath}
\usepackage{setspace}
\usepackage{mathtools}

\usepackage[ruled]{algorithm2e} 

\SetAlFnt{\small}
\SetAlCapFnt{\small}
\SetAlCapNameFnt{\small}
\SetAlCapHSkip{0pt}
\IncMargin{-\parindent}

\newcommand*\mean[1]{\bar{#1}}

\usepackage{rotating}

\newcommand{\red}[1]{\textcolor{black}{#1}}

\makeatletter
\newcommand{\algorithmfootnote}[2][\footnotesize]{%
  \let\old@algocf@finish\@algocf@finish
  \def\@algocf@finish{\old@algocf@finish
    \leavevmode\rlap{\begin{minipage}{\linewidth}
    #1#2
    \end{minipage}}%
  }%
}
\makeatother

\begin{document}
\frontmatter
\pagestyle{plain}

\title{Enhancing the Structural Performance of Additively Manufactured Objects}

\author{Erva Ulu}

\bsdegree{Mechanical Engineering, Middle East Technical University}
\msdegree{Mechanical Engineering, Bilkent University}

\Month{May}

\Year{2018}

\permission{\textit{All Rights Reserved}}

\maketitle


\begin{abstract}
The ability to accurately quantify the performance an additively manufactured (AM) product is important for a widespread industry adoption of AM as the design is required to: (1) satisfy geometrical constraints, (2) satisfy structural constraints dictated by its intended function, and (3) be cost effective compared to traditional manufacturing methods. Optimization techniques offer design aids in creating cost-effective structures that meet the prescribed structural objectives. The fundamental problem in existing approaches lies in the difficulty to quantify the structural performance as each unique design leads to a new set of analyses to determine the structural robustness and such analyses can be very costly due to the complexity of in-use forces experienced by the structure. This work develops computationally tractable methods tailored to maximize the structural performance of AM products. A geometry preserving build orientation optimization method as well as data-driven shape optimization approaches to structural design are presented. Proposed methods greatly enhance the value of AM technology by taking advantage of the design space enabled by it for a broad class of problems involving complex in-use loads. 
\end{abstract}


\begin{acknowledgments}

I would like to thank my advisor Prof. Levent Burak Kara for giving me this opportunity. Prof. Kara, your guidance, support and encouragement have been crucial for me to successfully complete all the work in this thesis. 

Special thanks go to the committee members, Prof. Kate Whitefoot, Prof. James McCann, and Prof. O. Burak Ozdoganlar. I had a chance to work with you closely on several projects. Your input has been very helpful in shaping up those projects along the way as well as this thesis. It is an honor for me to work with such experts as you during my PhD. I also would like to thank all the collaborators and the Visual Design and Engineering Lab members who I had the chance to work with, discuss and get advice from. It has been a great pleasure to work with you.

I have made many friends through this journey. It would not be possible for me to get through all the challenges without them. 

I would like to thank my family for their endless love, support and encouragement. I am grateful to have such a family who I felt by my side at all times even though they were miles away.

And my wife, Nurcan Gecer Ulu. I am more than grateful to share this experience with you. Without your love, support and patience, it would not be possible. You made my life enjoyable here, the entire time.

This work was supported by America Makes Project \#4058, NSF CMMI 1235427, Siemens Corporate Research and Carnegie Mellon Manufacturing Futures Initiative.

\end{acknowledgments}
\tableofcontents
\listoffigures
\listoftables
\mainmatter


\chapter{Introduction}
\label{chp:intro}

\section{Motivation}

\begin{figure*}
\centering
\includegraphics[width=\textwidth]{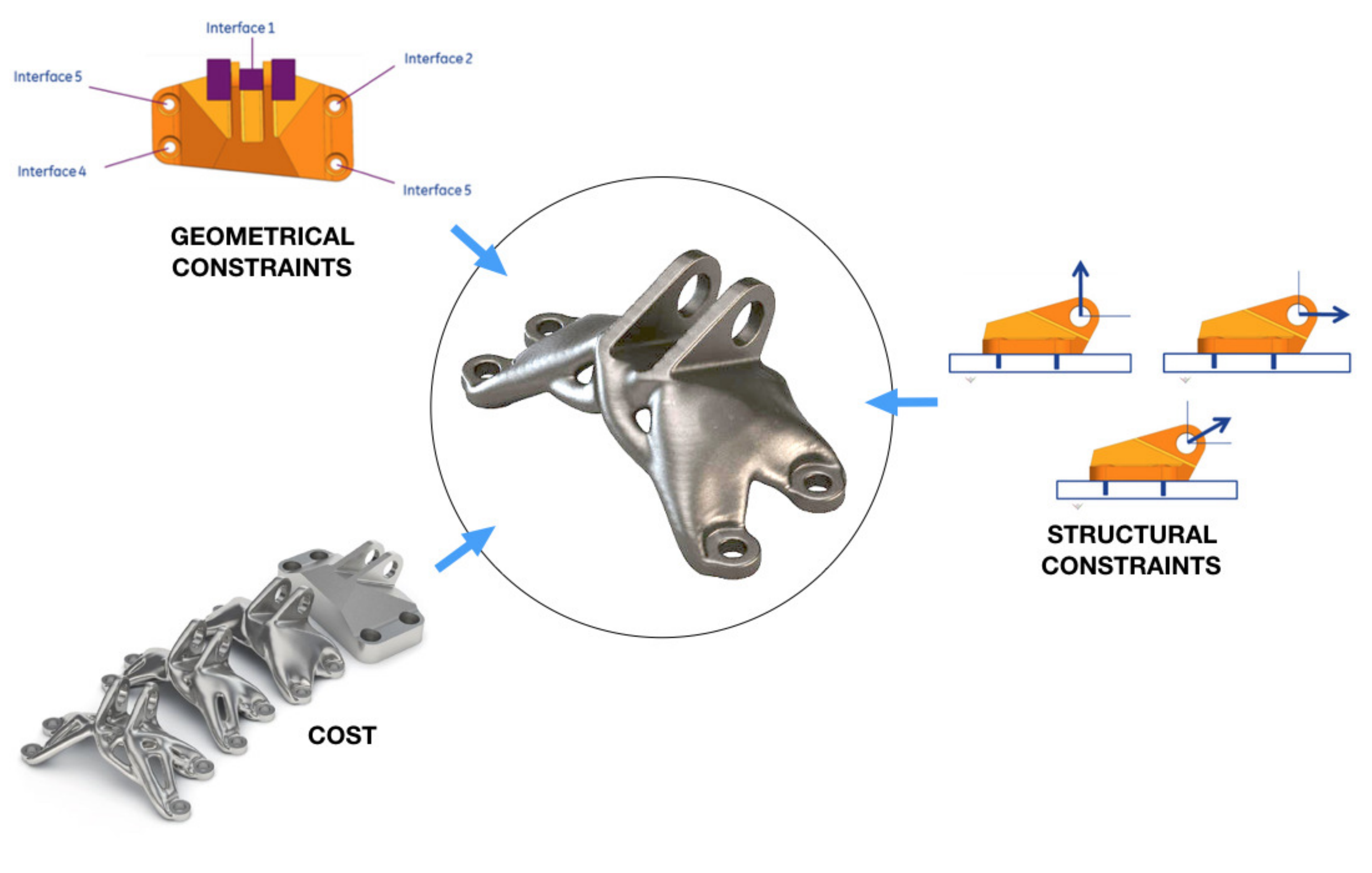} 
\caption{Geometrical constraints, structural constraints and cost are the major factors leading the design process for additively manufactured products. Image courtesy of GrabCAD, 3D Systems and Generate.}
\label{fig:Introduction:Motivation}
\end{figure*}

Additive manufacturing (AM) has been emerging as a powerful technique to manufacture three-dimensional (3D) objects \cite{scott2012additive,vaezi2013areview}. There is a growing interest in AM due to its applicability to complex geometries, rapid design-to-fabrication turnaround, and its widening spectrum of material choice, making it suitable in a myriad of engineering applications \cite{scott2012additive, vaezi2013areview, gibson2010additive}. In such applications, however, design of an additively manufactured product is particularly important for widespread industry adoption as it is required to: (1) satisfy geometrical constraints, (2) satisfy structural constraints dictated by its intended functional purpose,  and (3) be cost effective compared to traditional manufacturing methods (Figure \ref{fig:Introduction:Motivation}). 

Common approach in designing a structure meeting these requirements usually start by defining a design envelope so that the geometrical constraints are always satisfied for any design that fits into this design space. Then, the design is altered to satisfy the structural constraints so that the resulting product is robust under the in-use forces it experiences while keeping the cost minimum by reducing the material usage. 

The fundamental problem, here, lies in the difficulty to quantify the structural performance. Each design requires a new structural analysis to determine the structural robustness and such analyses can be very costly due to the complexity of in-use forces experienced by the structure. 
Existing approaches address this by making overly conservative simplifications, resulting in over-engineered solutions. Hence, they cannot take advantage of the design space enabled by additive manufacturing.
As a result, a new suite of practical design and simulation technologies is required to enhance the industrial value of AM . Therein, the overarching goal of this thesis is \textit{to develop and evaluate computationally practical methods that improve the structural performance of additively manufactured products}. We investigate geometry-preserving build-orientation selection, data-driven shape optimization, and reduced order topology optimization approaches.

\section{Scientific Challenges}

Optimization techniques offer design aids in creating structures that meet the prescribed structural needs. Recent studies highlight the flexibility of 3D printing in structural optimization \cite{lee2012stress,stava2012stress,wang2013cost}. Broadly defined, the principle scientific challenges in structural optimization are related to computational cost of repeated structural analysis operations, complexities in the load configurations and high-dimensionality of the shape parametrization.




\paragraph{Repeated costly FEA operations} Main impediment in structural optimization approaches is the challenge in computing the structural robustness. Each unique set of design parameters leads to a new structural analysis to compute the stress, strain and displacement fields and determine whether the current design satisfies the structural constraints under the loads it experiences. Combined with the geometric complexities and anisotropic material behavior, large number of iterations involving repeated FEA operations often introduce expensive computational bottlenecks.

Moreover, even with very small alterations in the design problem, a structurally optimum result can not be predicted directly by a human from physical principles due to the complex nature of the problem. Hence, the entire optimization process needs to be performed from scratch.


\paragraph{Complicated in-use loads} A common assumption in structural optimization approaches is that the applied loads are purely static. In many real world applications, however, the structural loads are much more complicated; load application points, load magnitudes or both might vary during the use of the object. A naive approach to optimizing a structure under such uncertainties in the force configurations would be to compute an optimal structure for every possible force configuration and select the best one. However, finding the best structure among all candidates requires an expensive verification step to ensure that the structure is safe for all other loading configurations, hence making the problem combinatorial. Moreover, there is no guarantee that any optimum solution obtained for a certain load configuration will be structurally sound for the remaining cases \cite{banichuk2013introduction}. Given the high computational cost of 3D optimization for a single load configuration, the combinatorial nature of the standard approach makes the solution computationally intractable for practical use.

\paragraph{Shape parametrization} Although there is a large body of work investigating various parametrization methods for shape optimization \cite{stava2012stress, lu2014build, musialski2015reduced},  they commonly suffer from the fact that the geometry needs to be discretized repeatedly as the structure evolves along the optimization path. This process tends to introduce non-linearities into the optimization problem as well as it creates an extra computational burden. Topology optimization approaches \cite{christiansen2015combined, bendsoe2003topology} addresses this problem by using a fixed volumetric mesh as the parameterization. However, dimensionality in such parametrization method is usually very large, thereby making the optimization process prohibitively expensive.

\section{Methodology}

\begin{figure*}
\centering
\includegraphics[width=\textwidth]{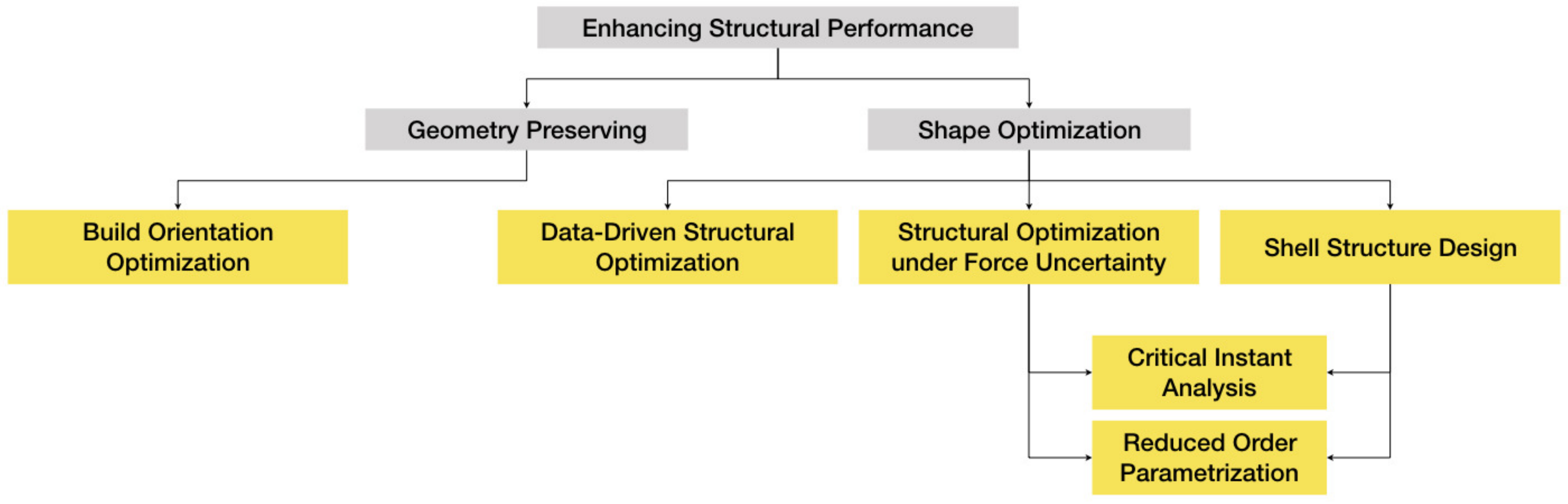} 
\caption{We investigate geometry preserving as well as shape optimization methods. Our contributions in this domain are highlighted in yellow.}
\label{fig:Introduction:Contributions}
\end{figure*}

This thesis presents computational methods for enhancing structural performance of additively manufactured objects. Structural performance of an additively manufactured product can be best described by strength-to-weight ratio as the two principal design constraints, structural robustness and cost, are conflicting in material usage. As more material is used, creating stronger designs far from failure, manufacturing as well as life-cycle costs get worse \cite{zhang2015efficient,huang2017cost}. However, studies have demonstrated that optimizing the geometry of AM parts can help decrease material usage, while maintaining part strength. It has been also shown experimentally that better structural performance can be obtained by exploiting the anisotropy in material properties while keeping the geometry, therefore the material usage, the same. In this thesis, we build upon these prior work (discussed in Chapter~\ref{chp:relatedwork}) and alleviate some of the fundamental limitations associated with current computational tools and practices. Figure~\ref{fig:Introduction:Contributions} illustrates the methods we investigate in this thesis.

First, in Chapter~\ref{chp:BuildOrientation}, we introduce a build orientation optimization algorithm where we seek to exploit the anisotropy in material properties of additively manufactured products. Our surrogate modeling based optimization approach provides a practical solution to this highly non-linear and complex problem by avoiding large number of costly FEA operations. 

In Chapter~\ref{chp:DataDrivenTopOpt}, we investigate a data-driven approach to conventional topology optimization problems. Here, we address the computational inefficiencies in structural optimization algorithms by learning from known solutions to topology optimization problems. This allows quick estimation of a new design for a novel set of design constraints. We also demonstrate that these estimations can serve as effective initial conditions that facilitate faster convergence in conventional topology optimization problems. 

Aforementioned methods we investigated in this thesis are formulated based on the assumption that the external forces can be modeled as known and fixed quantities. However, a critical observation that arise here is that the external forces' contact locations and magnitudes may exhibit significant variations during the use of the object in many real world applications. Hence, in Chapter~\ref{chp:ForceLocationUncertainty}, we address the computational complexities of structural optimization problems in which there is uncertainty in force locations. We present an efficient critical instant analysis
approach which determines the most critical force contact location responsible for creating the highest stress within the current shape hypothesis. Combined with our low dimensional shape parametrization, our method provides a practical solution to the structural optimization problem.

Finally, in Chapter~\ref{chp:ProposedWork}, we develop a method for designing shell structures. Building upon the lightweighting methods introduced in Chapter~\ref{chp:DataDrivenTopOpt} and Chapter~\ref{chp:ForceLocationUncertainty}, our aim is to control how the structure evolves through the optimization by introducing a new shape parametrization that enforces shell-like structures. Combined with our critical instant analysis, the resulting structures are aimed to withstand worst-case loading scenarios. 

\section{Contributions}

Aligned with our motivation and the aforementioned challenges, the set of contributions demonstrated in this thesis includes:
\begin{itemize}
\item A novel build orientation selection algorithm for AM that maximizes the minimum factor of safety under prescribed loading and boundary conditions.
\item A surrogate-based optimization approach that minimizes the number of FE simulations in build orientation optimization.
\item A framework to experimentally determine the process dependent anisotropic material properties of additively manufactured products.
\item A novel data-driven approach for structural topology optimization problems.
\item A comparison of mapping methods between the loading configurations and the optimal topologies.
\item A practical method for estimation of initial topologies to conventional topology optimization approaches.
\item A novel formulation for  structural optimization problems under force location uncertainty.
\item A method we call \textit{critical instant analysis} that identifies the critical load instant quickly.
\item A practical reduced order lightweighting method using the above two ideas.
\item A novel formulation for shell structure design involving structural mechanics.
\item A gradient-free shape optimization approach to arbitrary 3D problems with uncertainties in force configurations.
\item A heat based shape parametrization method that allows large variations in thickness while guaranteeing self-intersection free boundaries in the resulting structure.
\end{itemize}

\subsection{Publication List}

\begin{enumerate}

 \item \textbf{Erva Ulu}, Runze Huang, Levent Burak Kara and Kate S. Whitefoot, 
"Concurrent Structure and Process Optimization for Minimum Cost Metal Additive Manufacturing",
 \textit{Journal of Mechanical Design}, 2018. (Under review) \cite{ulu2018concurrent}

\item Yining Wang, \textbf{Erva Ulu}, Aarti Singh and Levent Burak Kara,
"Efficient Load Sampling for Worst-Case Structural Analysis Under Force Location Uncertainty",
 \textit{ASME International Design Engineering Technical Conferences and Computers and Information in Engineering Conference}, 2018. \cite{wang2018efficient}

\item \textbf{Erva Ulu}, James McCann and Levent Burak Kara,
"Lightweight Structure Design Under Force Location Uncertainty",
 \textit{ACM Transactions on Graphics}, 2017.  \cite{ulu2017lightweight}

\item Runze Huang, \textbf{Erva Ulu}, Levent Burak Kara and Kate S. Whitefoot,
"Cost Minimization in Metal Additive Manufacturing Using Concurrent Structure and Process Optimization",
 \textit{ASME International Design Engineering Technical Conferences and Computers and Information in Engineering Conference}, 2017. \cite{huang2017cost}
 
\item Erhan Batuhan Arisoy, Suraj Musuvathy, \textbf{Erva Ulu} and Nurcan Gecer Ulu, 
"Methods and System to Predict Hand Positions for Multi-hand Grasps of Industrial Objects",
\textit{Patent Publication Number:WO2017132134 A1}, 2017. \cite{arisoy2017methods}

  \item Erhan Batuhan Arisoy, Guannan Ren, \textbf{Erva Ulu}, Nurcan Gecer Ulu and Suraj Musuvathy,
"A Data-Driven Approach to Predict Hand Positions for Two-Hand Grasps of Industrial Objects",
 \textit{ASME International Design Engineering Technical Conferences and Computers and Information in Engineering Conference}, 2016. \cite{arisoy2016data}
 
  \item \textbf{Erva Ulu}, Rusheng Zhang and Levent Burak Kara,
"A Data-driven Investigation and Estimation of Optimal Topologies Under Variable Loading Configurations",
 \textit{Computer Methods in Biomechanics and Biomedical Engineering: Imaging \& Visualization}, 2016. \cite{ulu2015adata}
 
 \item \textbf{Erva Ulu}, Emrullah Korkmaz, Kubilay Yay, O. Burak Ozdoganlar and Levent Burak Kara,
"Enhancing the Structural Performance of Additively Manufactured Objects Through Build Orientation Optimization",
 \textit{Journal of Mechanical Design}, 2015. \cite{ulu2015enhancing}

  \item \textbf{Erva Ulu}, Rusheng Zhang, Mehmet Ersin Yumer and Levent Burak Kara,
"A Data-driven Investigation and Estimation of Optimal Topologies Under Variable Loading Configurations",
 \textit{Computational Modeling of Objects Presented in Images. Fundamentals, Methods, and Applications}, 2014. \cite{ulu2014adata}

\end{enumerate}

\chapter{Background}
\label{chp:relatedwork}

\section{Build Orientation in AM}
There is a growing interest in AM technologies in the fields of computational design, process design and material science. Here, we focus on the studies that highlight the \emph{directional dependencies} in AM, computational design with \emph{structural concerns} and \emph{build orientation selection} for AM.

\subsection{Directional Dependencies}

The most commonly studied effects of print-induced anisotropy  include dimensional accuracy and surface roughness~\cite{xu1999considerations,thrimurthulu2004optimum}, build time and cost~\cite{ahn2007fabrication, canellidis2009genetic}, the amount of support material~\cite{alexander1998part, vanek2014clever} and the mechanical properties (\eg strength, elastic modulus)~\cite{ahn2002anisotropic, bagsik2011mechanical}. Most relevant to our work, we focus on the studies examining the anisotropy in the structural properties of AM parts.

Various studies have experimentally shown that 3D printed parts exhibit directional dependencies in their mechanical properties.  Ahn \etal~\cite{ahn2002anisotropic} characterize the anisotropic mechanical properties of ABS parts manufactured using fused deposition modeling (FDM). Similarly,  El-Gizawy \etal~\cite{elgizawy2011process} and Hill and Haghi~\cite{hill2014deposition} investigate the mechanical properties of polyetherimide and polycarbonate when used in FDM. Barclift and Williams~\cite{barclift2012examining} and Kesy and Kotlinski~\cite{kesy2010Mechanical} experimentally study the effects of process parameters on material properties in polyjet printing. Similarly, Galeta \etal~\cite{galeta2013influence} study powder based AM. These experimental studies demonstrate that AM induces a significant structural anisotropy for many process and material combinations. Moreover, these works have shown that the resulting anisotropy can be represented very well using an orthotropic material model.     

Several computational methods have also been proposed to address this problem. Hildebrand \etal~\cite{hildebrand2013orthogonal} minimize the directional bias by partitioning the model into parts and selecting a build direction individually for each part. However, they investigate the geometric accuracy only. Zhou \etal~\cite{zhou2013worst} take a worst-case analysis approach to identify the structurally weak parts of a design where a constrained optimization problem is solved to obtain the worst loading configuration with the orthotropic material assumption. Umetani and Schmidt~\cite{umetani2013cross} address the structural anisotropy in FDM with the assumption that the vertical bonds between the layers are much weaker than the in-layer bonds. Based on this assumption, a cross sectional heuristic analysis is formulated to find an orientation that maximizes mechanical strength. Our approach builds upon these prior works by enabling a orthotropic material model with unique properties in each of the three principal directions. Additionally, in our approach, we maximize the factor of safety by considering the prescribed external loads and boundary conditions without making simplifying assumptions about the analysis.

\subsection{Structural Concerns}

Several studies have recently focused on the computational design for AM addressing structural concerns. In these studies, a common approach is to deform or modify the initial design to overcome its structural problems. Luo \etal~\cite{Luo2012Chopper} partition large objects into 3D printable smaller parts where each partition's impact on the overall structural robustness is evaluated using FE analysis, which informs the strategy for subsequent partitions. Similarly, Stava \etal~\cite{stava2012stress} evaluate hand-held objects' structural weakness using FE analysis to determine parts of the designs that require thickening, hollowing, or strut placement. Analysis is restricted to boundary conditions representing gravity and gripping using two fingers at heuristically predicted locations. Based on this analysis, several automatic shape modifications are proposed.  

Recent works have also focused on cost-effective 3D printing strategies while still addressing structural concerns. Wang \etal~\cite{wang2013cost} replace the solid interior of the object with a truss structure to reduce the amount of material used in the printing process.  Lu \etal~\cite{lu2014build} use a hollowing approach based on Voronoi diagrams to obtain lightweighted structures that can sustain prescribed stresses.

In our approach, we preserve the input design and do not perform shape modification. Instead, for an input design with prescribed boundary conditions, we optimize the build orientation to maximize the stress-based factor of safety in the resulting fabricated object.  However if needed, the above cost-effective methods can be used as a pre-processing step to reduce the amount of material used in AM.

\subsection{Build Orientation Selection}

Although build orientation selection with respect to geometrical features is very well studied for AM applications, there is only a limited amount of work that directly addresses structural concerns. Suh and Wozny~\cite{suh1995integration} account for the critical features (\eg thin walls and slender protrusions) that need to be appropriately oriented due to potential failure problems. They use a purely geometric approach to ensure that the critical features lie in the layer accumulation direction and do not consider the loading conditions on the designed object. In an inspiring work, Thompson and Crawford~\cite{thompson1995optimizing} introduce a build orientation selection algorithm that considers the load and boundary conditions together with the material properties. To this end, they use the Tsai-Wu failure criterion to determine whether the object is safe or unsafe for a candidate build orientation. However, this binary objective only ensures safe orientations and does not maximize the factor of safety. Umetani and Schmidt~\cite{umetani2013cross} suggest the best build orientation for a given geometry by analyzing the structural weakness at different cross-sections assuming that the primary mode of loading is bending. The part is oriented such that the weakest cross-sections are as perpendicular as possible to the layer accumulation direction. This approach assumes that the material behaves isotropically within a single layer, hence the in-layer orientation does not affect mechanical strength. 

Our approach is inspired by the studies presented in~\cite{thompson1995optimizing} and~\cite{umetani2013cross} in that the actual loading conditions determine the build orientation if the structural robustness is the main concern. However, unlike~\cite{thompson1995optimizing}, we maximize the mechanical strength over the entire geometry instead of incorporating the failure criterion as a constraint when assessing candidate orientations. Moreover, our work differs from~\cite{umetani2013cross} in that we do not assume in-layer isotropy and allow all modes of loading configurations (bending, torsion, compression \etc) to be jointly considered.

\section{Data Driven Structure Design}
In this section, we review the relevant literature on structural topology optimization techniques, use of data analysis and dimensionality reduction approaches as well as data mapping methods.  

\subsection{Topology Optimization} 
Topology optimization is one of the most powerful technologies in structural design \cite{bendsoe2004topology,dijk2013levelset}. It optimizes the shape and material connectivity of a domain through the use of finite element methods together with various optimization techniques \cite{norato2007atopological}.

Density-based topology optimization approaches including homogenization methods \cite{bendsoe1988generating,suzuki1991ahomogenization} and solid isotropic microstructure with penalty (SIMP) methods \cite{bendsoe1989optimal,rozvany1992generalized} are one of the most popular methods in the literature. These methods approach  topology optimization  in  a way that defines geometry by optimizing  material distribution in the domain. A detailed review on density based topology optimization methods can be found in \cite{hassani1998areview,rozvany2001aims,rozvany2009acritical}. Another approach for structural topology optimization is based on topological derivatives and level-sets \cite{norato2007atopological,sethian2000structural,wang2003alevel}. The optimization process utilizes the implicit description of the boundary to numerically represent the geometry. A recent work \cite{dijk2013levelset} discusses the level-set based topology optimization methods more deeply. In \cite{rozvany2009acritical}, topological derivative and level-set based methods in the literature are claimed to be very promising although they are not widely embraced by industries. Aside from the above methods, evolutionary approaches are also used for topology optimization, e.g. \cite{chapman1994genetic,jakiela2000continuum}. However, the use of genetic algorithms are computationally expensive, thus they are suitable for only small scale problems \cite{rozvany2009acritical}.

Since topology optimization is an iterative and computationally demanding process, an efficient implementation of the above mentioned methods in various programming languages is also important for designers. In \cite{andreassen2011efficient,sigmund2001a99}, authors present two different versions of an efficient {MATLAB} code for structural topology optimization of classical Messerschmitt-B\"{o}lkow-Blohm (MBB) beam problem. As an optimization technique, they implemented an available {SIMP} approach with slight modifications involving filters. In our approach, we utilize the available code in \cite{andreassen2011efficient} to generate the initial optimized topologies for different loading conditions as a way to generate the pool of training data.

\subsection{Data Analysis and Dimensionality Reduction} 
In data-driven methods, a pre-analysis of available data to extract informative characteristics is essential, especially for large multivariate data sets. Such methods are commonly used in engineering design and computer science, e.g. \cite{sirovich1987lowdimensional,kirby1990application,turk1991eigenfaces,allen2003thespace}. Although the design approach is not data driven, dimensionality reduction idea is also used in structural topology optimization in \cite{guest2010reducing} to reduce the computational cost by decreasing the number of independent design variables. 

In data driven design context, the most commonly used dimensionality reduction methods include principal component analysis ({PCA}) \cite{jolliffe2005principal}, multidimensional scaling ({MDS}) \cite{cox1994multidimensional}, isomaps \cite{tenenbaum1998advances} and locally linear embedding ({LLE}) \cite{roweis2000nonlinear}. PCA is an eigenvector based approach that uses an orthogonal transformation to convert the original data into linearly independent components. Dimensionality reduction is accomplished by representing data in terms of the linearly independent components that best explain the variance in the data. In MDS, high dimensional data is embedded into low dimensional space in such a way that pairwise distances between data points are preserved. Isomaps aim to preserve the geodesic distances in the manifold formed by the data. LLE is a neighborhood-preserving dimensionality reduction method. It projects high-dimensional data into lower dimensional global coordinates by utilizing different linear embeddings for each data point locally. In the proposed work, we use PCA to analyze the dominant characteristics of our data set and to reduce the dimensionality accordingly. However, the aforementioned dimensionality reduction methods could be adopted into the workflow of the proposed techniques without loss of generality.

One important aspect of the proposed work is the mapping between an input configuration (in our case the loading configuration) and the resulting optimal topology. Note that a PCA-based learning and topology reconstruction is readily implementable with the available training images. However, the key need is to be able to specify a novel loading configuration, from which  the optimal topology can be estimated. In previous work, most methods employ a linear mapping between the input feature vectors and the resulting PCA reconstructions \cite{allen2003thespace,blanz1999amorphable}. However, the relationship between the input loading configurations and the resulting topology reconstructions in our domain is highly non-linear as demonstrated in the following sections. To address this challenge, we present a mapping technique that uses feed-forward neural networks. This generative method provides a significant improvement over linear regression models by covering non-linearities automatically without requiring any explicit information about the design space complexity.

\section{Structure Design Under Uncertainties}
Our review focuses on  studies that highlight  fabrication oriented design, lightweight structure synthesis, and structural analysis, with an emphasis on approaches involving additive fabrication.

\subsection{Fabrication Oriented Design} 
A large body of work has investigated automatic techniques for 3D shape design and additive fabrication subject to a variety of  functional requirements.
Recent examples include designing for prescribed deformation behaviors~\cite{bickel2010design,skouras2013computational,schumacher2015microstructures,panetta2015elastic,ulu2018designing}, balancing models~\cite{prevost2013make}, spinnable objects~\cite{bacher2014spinit} and broader methods that can handle multiple requirements~\cite{chen2013spec2fab,musialski2015reduced,christiansen2015combined}.
Our problem falls under the general category of weight-optimal structure design subject to external forces \cite{bendsoe1989optimal, wang2013cost, lu2014build, christiansen2015combined}. However, our approach addresses a more general class of problems in which the precise force locations cannot be prescribed apriori, or the structure experiences forces that can contact its surface at a multitude of locations.

\subsection{Lightweight Structure Synthesis}
Cellular structure \cite{medeiros2015adaptive}, honeycomb-like structure \cite{lu2014build}, truss element based skin-frame structure \cite{wang2013cost}, beam element based tree-like structure \cite{zhang2015medial} generation methods and topology optimization methods \cite{christiansen2015combined,bendsoe2003topology} are among the recent lightweight internal structure synthesis techniques that consider durability as one of the primary constraints.
However, these methods assume a prescribed static force configuration for structural design.
Although driven by similar motivations, our work addresses a more general problem of structural design under force location uncertainty. On the other hand, our formulation is also complementary in that it may facilitate the extension of these previous methods to problems involving force uncertainties. 

Langlois~\etal~\cite{langlois2016stochastic} performs structural optimization by predicting the failure modes of objects in real world use.
Their stochastic finite element model uses contact force samples generated by rigid body simulations to predict  failure probabilities.
They perform weight minimization while limiting the failure probability below a prescribed threshold.
While their method is applicable to scenarios where loading is stochastic in nature (such as dropping and collisions),
it is not streamlined for deterministic scenarios where the set of possible force configurations are known and no failure is tolerated for any of them. However, their method is extremely well-suited to automatically generating our contact regions, thereby allowing stochastic scenarios they consider to be addressed using structural guarantees our approach enables.

Model reduction has been used for material \cite{xu2015interactive} design, with a primary emphasis on controlling deformation behavior. Our approach is similar to traditional topology optimization methods \cite{bendsoe2003topology,lee2012stress} in that we optimize the material distribution using a fixed volumetric mesh as the parameterization. However, structural optimization under force location uncertainties introduces computational challenges that make a full dimensional analysis using the original shape parameterization to be prohibitively expensive. We are thus inspired by the above reduction method for shape synthesis, and use this in our implementation in conjunction with  our new critical instant analysis.

Musialski~\etal~\cite{musialski2015reduced} introduce the idea of offset surfaces for hollowing out a solid object.
This method serves as another shape parameterization for functional optimization. In our work, we use this method to form a fixed, ingrown boundary shell, and use the remaining internal volume for shape optimization. 

\subsection{Structural Analysis} 
In structural optimization, stress and deformation analysis using Finite Element Analysis (FEA) often introduce expensive computational  bottlenecks. Simple elemental structures  such as trusses \cite{smith2002creating,rosen2007design,wang2013cost} and beams \cite{zhang2015medial} have been used to alleviate this issue. For  cases where the structure cannot be represented by these simple elements, Umetani and Schmidt~\cite{umetani2013cross} simplify the problem into 2D cross-sections and extend the Euler-Bernoulli model into free-from 3D objects to facilitate analysis. 

Zhou~\etal~\cite{zhou2013worst} extend modal analysis used in dynamic systems (such as vibrations) to static problems to identify the potential regions of a structure that may fail under arbitrary force configurations. Our critical instant analysis builds upon this approach; we use modal analysis to determine the weak regions in a similar manner.
It allows our method to determine possible failure points based purely on geometry, \ie~independent of the loading.
We incorporate the weak region analysis into our structural optimization to focus on only a small region in the object to monitor the stress, thereby helping the convergence. 

In bridge (traffic load) and building (wind load) design, an equivalent uniformly distributed static load can be used to perform simple approximate analysis to account for force location uncertainty \cite{choi2002structural}. However, this approach is limited to simple geometries, making it unsuitable for our purposes.

\chapter[Build Orientation Optimization]{Build Orientation Optimization}
\label{chp:BuildOrientation}
\blindfootnote{This chapter is based on Ulu \etal, 2015 \cite{ulu2015enhancing}.}

Additively manufactured objects often exhibit directional dependencies in their structure due to the layered nature of the printing process. While this dependency has a significant impact the object's functional performance, the problem of finding the best build orientation to maximize structural robustness remains largely unsolved. We introduce an optimization algorithm that addresses this issue by identifying the build orientation that maximizes the factor of safety of an input object under prescribed loading and boundary configurations. First, we conduct a minimal number of physical experiments to characterize the anisotropic material properties. Next, we use a surrogate-based optimization method to determine the build orientation that maximizes the minimum factor safety. The surrogate-based optimization starts with a small number of finite element solutions corresponding to different build orientations. The initial solutions are progressively improved with the addition of new solutions until the optimum orientation is computed. We demonstrate our method with physical experiments on various test models from different categories. We evaluate the advantages and limitations of our method by comparing the failure characteristics of parts printed in different orientations.

\section{Introduction} 
\label{sec:BuildOrientation:Introduction}

There is a growing interest in additive manufacturing (AM) due to its applicability to complex geometries, rapid design-to-fabrication turnaround, and its widening spectrum of material choice, making it suitable in a myriad of engineering applications \cite{campbell2011could, vaezi2013areview, scott2012additive, gibson2010additive, seepersad2014challenges, lipson2013fabricated}. In the context of structural and geometric design, recent works have investigated automatic techniques to achieve prescribed functions such as designing for desired deformations \cite{bickel2010design, chen2013spec2fab}, designing for prescribed appearances \cite{hasan2010physical, dong2010fabricating}, balancing models \cite{prevost2013make} and generating spinnable objects \cite{bacher2014spinit}.

The \emph{layered} nature of AM has major implications on the resulting objects. To date, there have been many studies highlighting the impact of build orientation (\ie how the part is oriented in the print workspace) on aspects such as surface quality, the amount of required support material, geometric accuracy, build time, and overall fabrication cost \cite{alexander1998part, xu1999considerations, ahn2007fabrication, canellidis2009genetic}. However, the build orientation has a major impact on the structural properties of additively manufactured parts. This is commonly manifested in the form of anisotropically printed objects, making structural performance highly dependent on the build orientation. While this intricacy has been observed and experimentally demonstrated in a limited fashion, to date, no attempts have been made to  engineer its impact to improve structural robustness.

In this work, we introduce a new build orientation selection algorithm for polymer-based AM processes that aims to maximize an input object's resistance to failure under prescribed external loads. We define an increased resistance to failure as one that increases the  direction-dependent material yield strength relative to the stresses generated within the object. We thus formulate a new build orientation optimization problem where the optimal orientation is achieved by maximizing the minimum factor of safety observed in the object. The problem, however, is difficult to solve using conventional gradient-based methods. This is because the build orientation impacts several structural parameters including the elastic moduli, the yield strengths, and the material's Poisson's ratios. Additionally, unless the domain is particularly simple, where appropriate analytical functions can be utilized, the relationship between the build orientation and the resulting stress tensor field is difficult to establish in closed form. This difficulty makes the gradient and the Hessian of the objective function very difficult to precompute for arbitrary geometries and loading configurations. 

\begin{figure*}
\centering
\includegraphics[width=\textwidth]{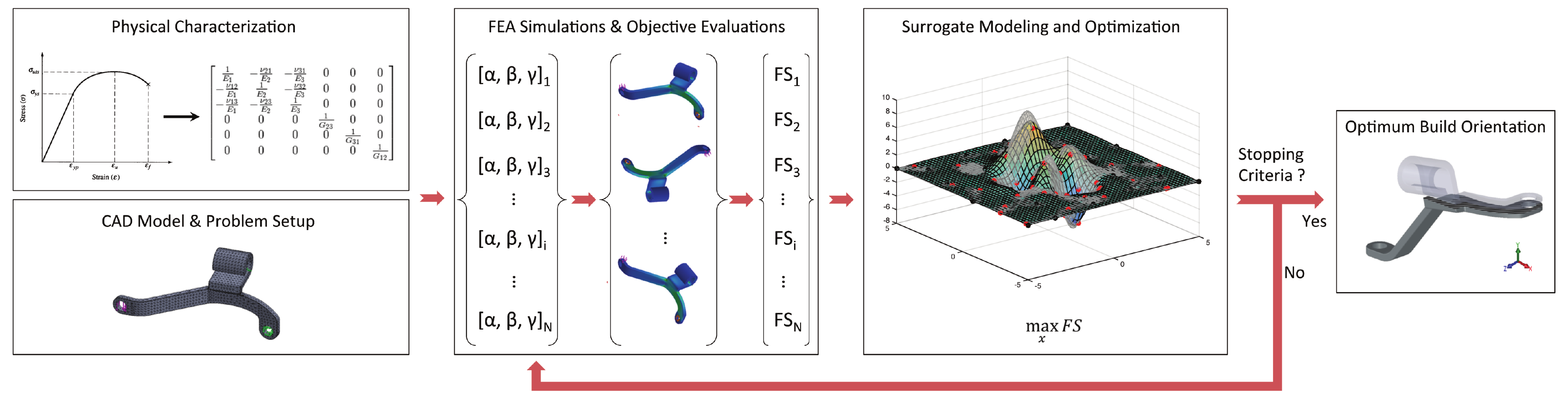} 
\caption{Our approach takes as input a 3D model of an object with the corresponding loading/boundary configurations and anisotropic (orthotropic) material properties, then calculates an optimum build orientation that maximizes the factor of safety (FS). The build orientation is defined by three Euler angles $[\alpha, \beta, \gamma]$. A surrogate model between the candidate orientations and the objective function is constructed. The surrogate model is progressively improved with the addition of new candidate orientations until the optimal orientation is found.}
\label{fig:BuildOrientation:Fig1}
\end{figure*} 

On the other hand, a brute force approach (\eg uniform parameter sweep) will typically require a large number of finite element (FE) simulations to appropriately cover the design space, which can be computationally prohibitive. To address this challenge, we use a surrogate-based optimization method that starts with a small number of FE simulations for various build orientations to model the design space. Then, the initial surrogate model is iteratively improved with the addition of new evaluation points until the optimum orientation is found. In each iteration, the functional form of the surrogate model enables a gradient-based search on this proxy model, thereby accelerating the optimization process. 

Our optimization algorithm uses an orthotropic material model to establish the compliance matrix. To identify the parameters of this matrix, we perform a set of physical experiments on a small set of test specimens that are printed using the target object's material and print settings (Fig.~\ref{fig:BuildOrientation:Fig1}). This choice enables the process and environment dependent properties to be accounted for during our solutions.

Our primary contributions are:

\begin{itemize}
\item a novel build orientation selection algorithm for AM that maximizes the minimum factor of safety under prescribed loading and boundary conditions,
\item a surrogate-based optimization approach that minimizes the number of FE simulations,
\item a framework to experimentally determine the process dependent anisotropic material properties.
\end{itemize}

\begin{figure}
\centering
\includegraphics[width=5.0in]{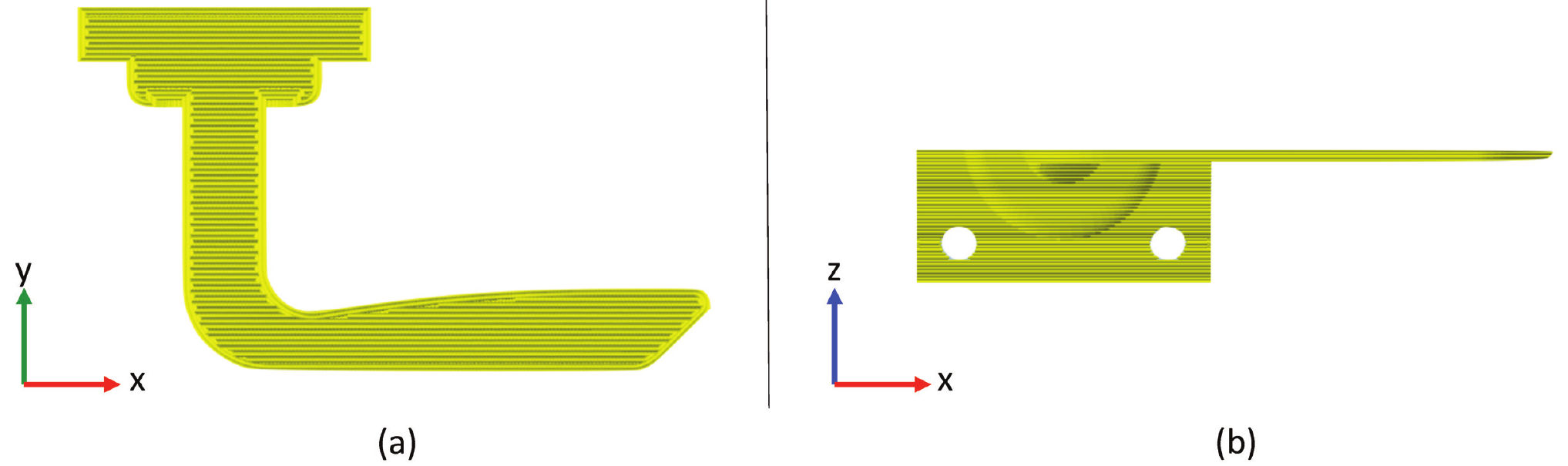} 
\caption{Principal directions in the orthotropic material model (a) A single layer in AM process where $x$ and $y$ are the in-plane principal directions. (b) The layer accumulation (build) direction, $z$.}
\label{fig:BuildOrientation:Fig2}
\end{figure} 

\section{Preliminaries}
\label{sec:BuildOrientation:Preliminaries}

We begin by introducing our material model, our FE simulation infrastructure and the techniques  to physically characterize the anisotropy in 3D printed parts.

\subsection{Material Model and Analysis}

We base our approach on an orthotropic material model which is commonly used in AM due to the 3-orthogonal nature of the print process. Figure~\ref{fig:BuildOrientation:Fig2} illustrates the three principal directions with the coordinate frame $x \perp y \perp z$. Here, $x$ and $y$ correspond to the orthogonal in-layer directions and $z$ corresponds to the layer accumulation direction.

In the orthotropic material model, a total of nine parameters need to be determined experimentally. These parameters are the Young's moduli, shear moduli and Poisson's ratios for the three principal directions. Additionally, to compute the factor of safety, the tensile yield strength, compressive yield strength and shear strength need to be determined experimentally for these principal directions. The orthotropic material model enables all such parameters to be determined with a minimal number of tests using well established metrology techniques including tensile, compressive and shear tests. Further details of the material characterization experiments are explained in Section~\ref{sec:BuildOrientation:Sec3.3}.  

\begin{figure}
\centering
\includegraphics[width=3.5in]{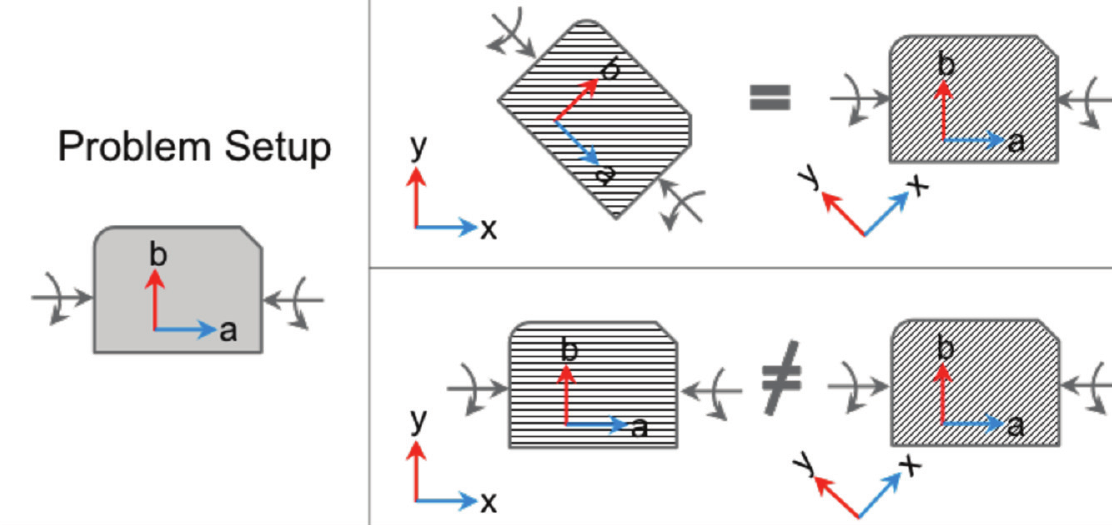}
\caption{Geometry $a,b$ and material $x,y$ coordinate frames. The top row shows an example for the equivalent representations of the same physical problem. The bottom row illustrates our case where there are two distinct build orientations and separate FE simulations are required to obtain the stress information for each configuration. Hence, while stress transformation formulas seem to be applicable here, they are indeed not applicable in our problem.}
\label{fig:BuildOrientation:Fig3}
\end{figure}

\subsection{Finite Element Analysis}
We use FE simulations to calculate  the stress tensor field for a given geometry and boundary conditions. With the orthotropic material assumption, a new FE simulation is required for each candidate build orientation. Figure~\ref{fig:BuildOrientation:Fig3} illustrates this issue on a simple two dimensional example. For a given geometry and boundary conditions, we assign a local coordinate frame, $a \perp b$ ($\perp c$ for 3D), which is attached to the \emph{geometry}. We also establish a global coordinate frame, $x \perp y$ ($\perp z$ for 3D) that represents the \emph{material} orientation. 

In this work, we use a script based ANSYS Mechanical Parametric Design Language  (APDL) to run FE simulations required in our optimization scheme. During optimization, a new material coordinate frame is established that operates on a fixed geometry, mesh, and boundary conditions. The different build orientations are thus evaluated by adjusting the material coordinate frame. After each FE simulation, the computed stress tensor information is encoded in the geometry coordinate frame, thus a stress transformation is required to evaluate the stress values in the material coordinate frame where the structural properties are known. This transformation facilitates the factor of safety calculation at each element in the domain as will be shown later.

\subsection{Material Characterization}
\label{sec:BuildOrientation:Sec3.3}

To demonstrate the integration of physical anisotropy characterization into our optimization scheme, we use a high-resolution ($30 \mu m$ vertical and $42 \mu m$ lateral) Objet Connex 350 multi-material 3D printing system. Thin layers of photosensitive resins ($30 \mu m$) are deposited onto a build tray (350 mm x 350 mm x 200 mm) by inkjet printing. The deposited layer is then immediately cured using a UV light source for photo-polymerization, which is coupled to the print head and solidifies each liquid material layer. During the curing process, a roller levels the liquid polymers making the material immediately ready to be built upon with successive layers. The building process uses two kinds of material: object (two different materials can be used and different digital materials can be obtained through a mixture of these materials) and support. It is possible to build the final product with and without the support material around the features. 

\begin{figure}
\centering
\includegraphics[width=4.0in]{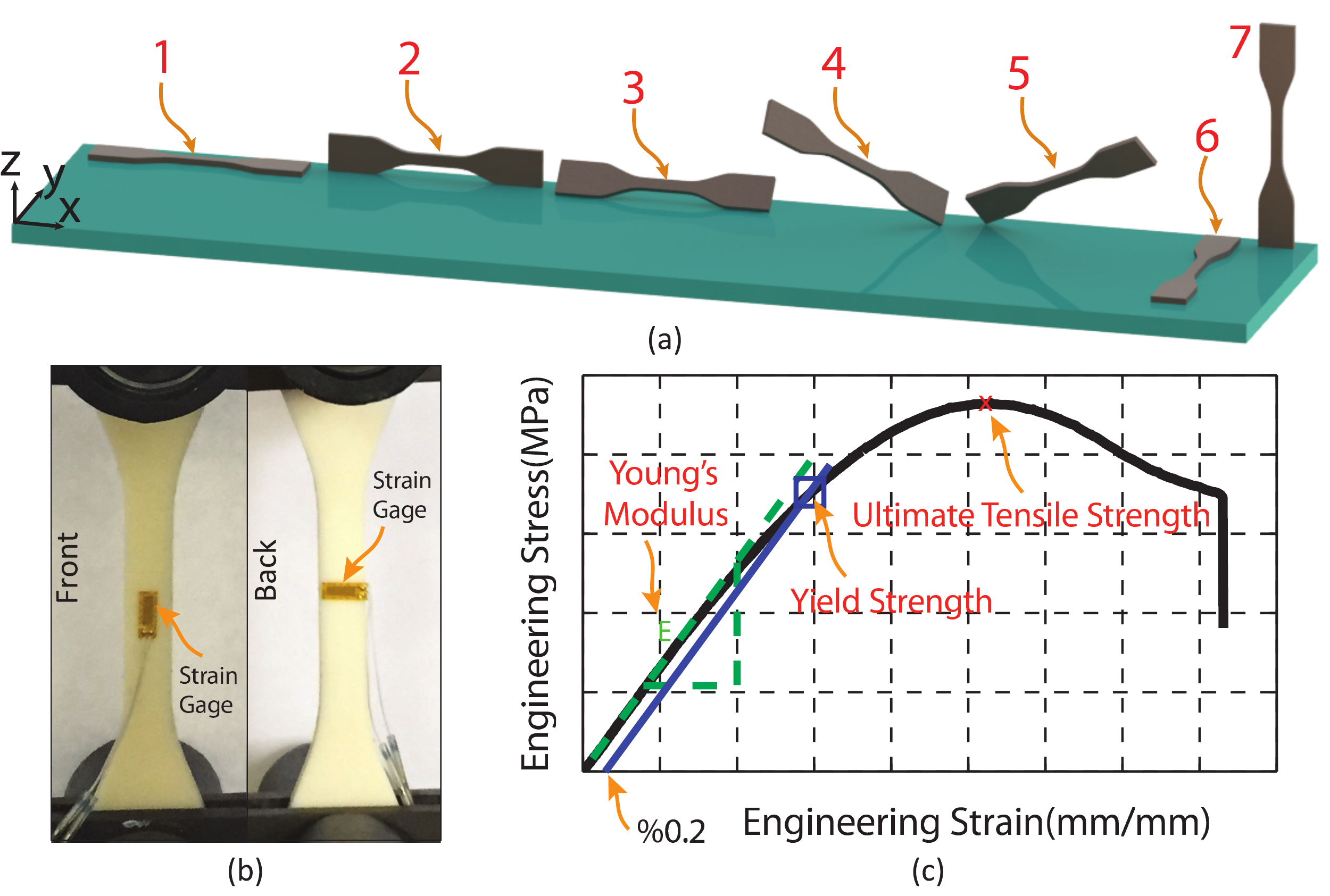} 
\caption{Material characterization. (a) Print directions considered, (b) experimental setup for tensile tests and (c) a typical stress-strain curve showing a subset of the material properties to be extracted.}
\label{fig:BuildOrientation:Fig4}
\end{figure}  

\begin{table*}[t]\small
\caption{Results of physical characterization tests for the three principal directions.}
  \centering
  \begin{tabular}{>{\centering\arraybackslash}m{1.5cm} || >{\centering\arraybackslash}m{1.5cm} | >{\centering\arraybackslash}m{3cm} | >{\centering\arraybackslash}m{2.0cm} | >{\centering\arraybackslash}m{1.5cm} | >{\centering\arraybackslash}m{1.5cm} }
      \hline
    \bf Principal Dir. & \bf Young's Modulus [GPa] & \bf Yield Strength [MPa] (Tensile/Compressive) & \bf Shear Modulus* [GPa] & \bf Shear Strength [MPa]& \bf Poisson's Ratio \\
    \hline \hline 
    x  & 1.16 & 35.86/52.46 & 0.51  & 4.38 & 0.09 \\
    y  & 1.05 & 25.52/37.63 & 0.28  & 4.38 & 0.37\\
    z  & 0.52 & 8.77/13.58  & 0.30  & 4.38 & 0.31\\
    \hline
    \multicolumn{6}{r}{*~Determined analytically using~\cite{zhou2013worst}.}    
  \end{tabular}
  \label{tab:BuildOrientation:Tab1}
\end{table*}

\begin{figure*}
\centering
\includegraphics[width=\textwidth]{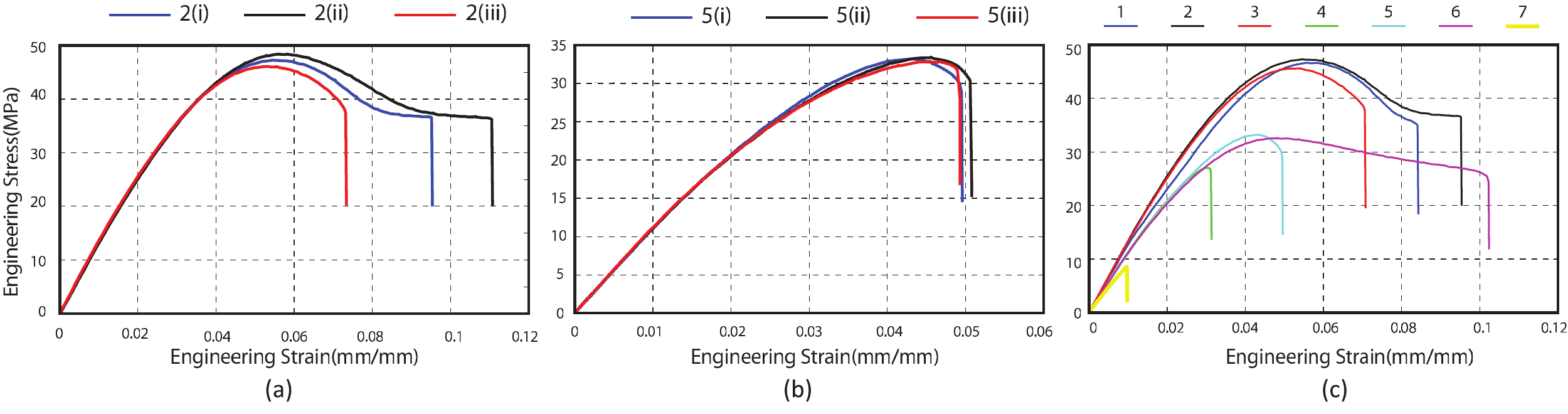} 
\caption{Repeatability tests for (a) orientation 2, (b) orientation 5 and (c) the stress-strain curves for different build orientations shown in Fig.~\ref{fig:BuildOrientation:Fig4}.}
\label{fig:BuildOrientation:Fig5}
\end{figure*}

Using this setup, we print a test specimen in seven different orientations and perform a tensile test for each orientation to reveal the directional dependency of the material properties (Fig.~\ref{fig:BuildOrientation:Fig4}). For each direction, we print three copies of ASTM D638 standard tensile test specimen using VeroWhite\texttrademark photosensitive resin. For each specimen, we conduct a tensile test using an INSTRON 4467 instrument and two strain gauges mounted to the front and back faces of the specimen to obtain the engineering stress-strain curves for each specimen.

Figure~\ref{fig:BuildOrientation:Fig5}(a) and (b) show the stress-strain curves for the three identical specimens printed along directions 2 and 5 (see Fig.~\ref{fig:BuildOrientation:Fig4}(a)), respectively. The results show that the stress-strain curves of similarly oriented specimens behave similarly up to their ultimate tensile strength. Indeed, the deviation in the elastic moduli and yield strength values are less than 3$\%$. Thus, the material properties are consistent within a given orientation. On the other hand, when the parts are printed in different directions, significant differences are observed in the material properties. Fig.~\ref{fig:BuildOrientation:Fig5}(c) shows the stress-strain curves for the specimens printed in the seven different directions revealing  the directional dependency of material properties. 

We extract the material properties required for our optimization algorithm from the stress-strain curves. To this end, the Young's moduli and tensile yield strengths (0.2$\%$ strain offset) as well as the Poisson's ratios (the ratio between the slopes of the stress-axial strain and stress-transverse strain) are obtained for the directions of 1, 6, and 7 shown in Fig.~\ref{fig:BuildOrientation:Fig4}(a). These directions correspond to our standardized principal directions and are listed in Tab.~\ref{tab:BuildOrientation:Tab1}. Furthermore, we perform compression tests on the standard test specimens (ASTM D395)  using an INSTRON 4469 compression instrument to obtain the compressive yield strengths shown in Tab.~\ref{tab:BuildOrientation:Tab1}. For the orthotropic material model, it is also necessary to determine the shear related material properties. For this, we calculate the  required shear moduli using the approach in~\cite{zhou2013worst}. The corresponding shear strengths are assumed to be $50\%$ of the lowest yield strength value according to the maximum shear theory~\cite{shigley2004mechanical} for conservative estimates. However, shear strengths can also be experimentally determined to enhance the precision of our approach without loss of generality.

\section{Build Orientation Selection}

We quantify the structural robustness of an object using the factor of safety (FS) criterion. The overall goal is to choose a build orientation that maximizes the FS over the entire geometry. 

For each element $i$, each FE simulation computes a stress tensor $\pmb{\sigma}_i$ containing six unique components: $\sigma_{X}^i,~\sigma_{Y}^i,~\sigma_{Z}^i,~\tau_{YZ}^i,~\tau_{XZ}^i$ and $~\tau_{XY}^i$. Here,  $\sigma_m^i$ and $\tau_{mn}^i$ terms are the normal and shear stresses, respectively. Based on the maximum stress theory, a conservative approach to assign a single FS  to an element is to compute six independent FS values for each stress component and choose the minimum one as the FS for that element. In our approach, we use this principle to formulate our optimization problem as follows:

\begin{equation}
\label{eq:BuildOrientation:Eq1}
\begin{aligned}
& \underset{{\bf x}}{\text{minimize}}
& & f({\bf x}) = \sum_{i=1}^{n} \left[\sum_{k=1}^{6} \left(\frac{1}{{\bf FS}_i^k({\bf x})}\right)^\kappa \right] \\
& \text{subject to}
& & \alpha , \gamma \in \left[-\pi, \; \pi \right] \; \text{and} \;\beta \in \left[0, \; \pi\right],\\
& \text{where}
& & {\bf x} = [\alpha, \beta, \gamma]^{T}.
\end{aligned}
\end{equation}

\noindent Here, ${\bf FS}_i$ is the 6$\times$1 vector of safety factor values (${\bf FS}_i^k$) for the $i$'th element, and ${\bf x}$ is the vector of design variables where $\alpha$, $\beta$ and $\gamma$ are the intrinsic Euler angles representing a sequential rotation about the global z, x and z axes, respectively. $\kappa$ is a large positive number and $n$ is the number of elements in the FE analysis. The goal is to find ${\bf x}$ that minimizes our objective $f({\bf x})$. In our approach, we calculate ${\bf FS}_i({\bf x})$ for an element as follows:  

\begin{equation}
\label{eq:BuildOrientation:Eq2}
{\bf FS}_i(\bf x) = \pmb{\sigma^Y} / \pmb{\sigma}_i'({\bf x}) \; \;
\text{where} \; \; \pmb{\sigma}_i'(\bf x) = {\bf R}(\bf x)\pmb{\sigma}_i{\bf R^T}(\bf x) 
\end{equation} 

\noindent where $\pmb{\sigma^Y}$ is the 6$\times$1 vector of yield strengths for the anisotropic material obtained for the principal directions (material coordinates). $\pmb{\sigma}_i$ is the stress tensor in the geometry coordinates and $\pmb{\sigma}_i'(\bf x)$ is its transformation to the material coordinates. ${\bf R}(\bf x)$ is the transformation matrix from geometry to material coordinates. 

\begin{figure*}
\centering
\includegraphics[width=\textwidth]{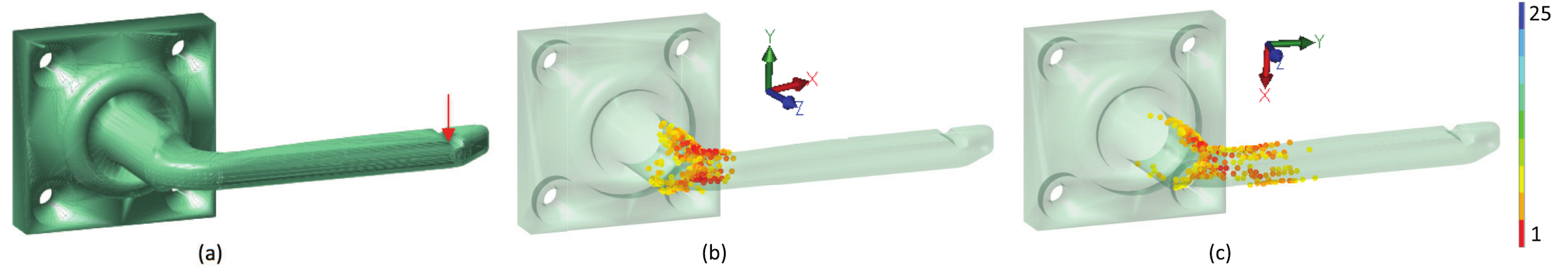} 
\caption{Performance of our objective function. (a) Problem configuration. Elements with the lowest 300 safety factor values are highlighted for the initial (b), and the optimized (c), build orientations.}
\label{fig:BuildOrientation:Fig6}
\end{figure*} 

\begin{figure}
\centering
\includegraphics[width=4.0in]{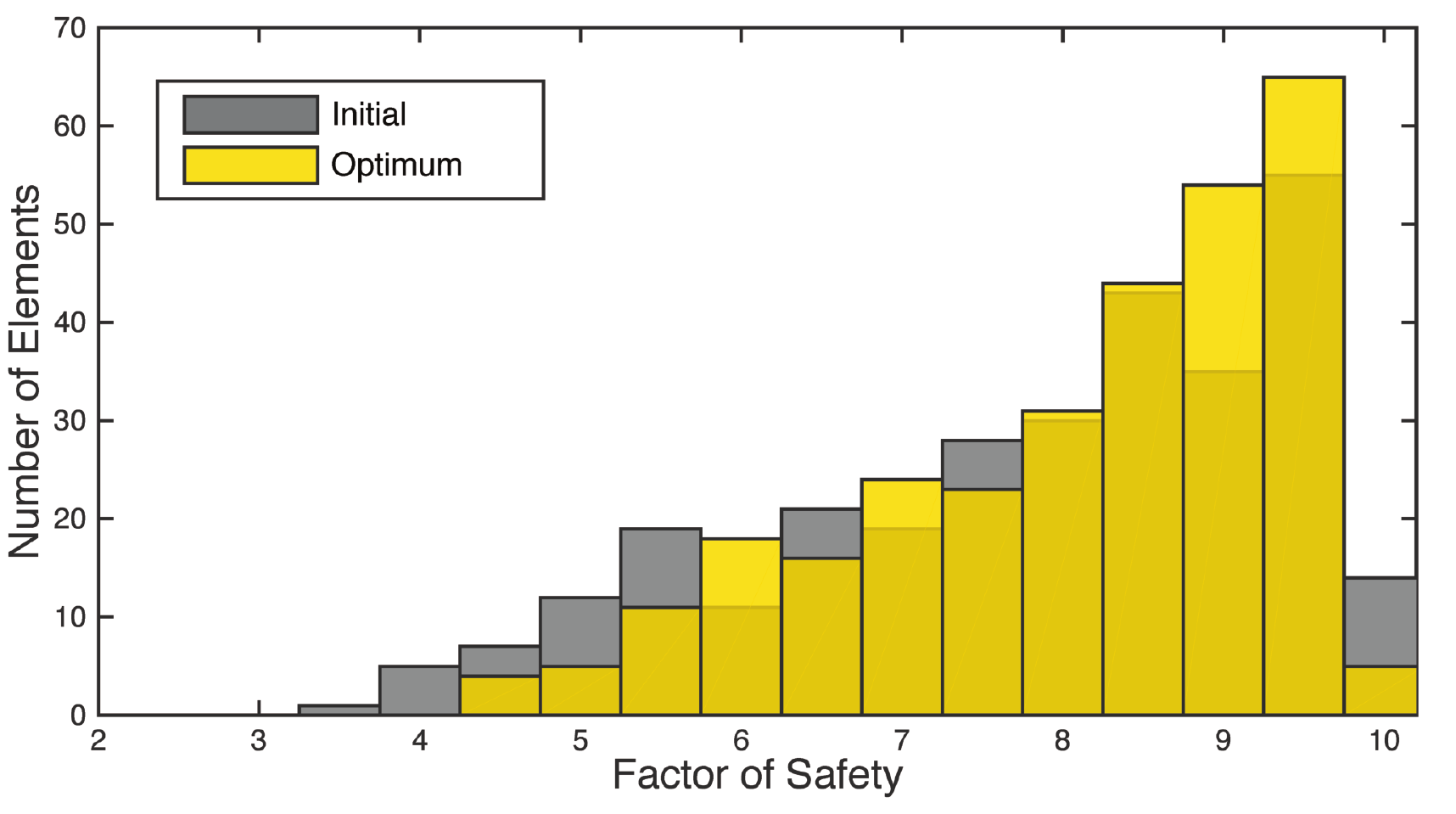} 
\caption{Histograms of the lowest 300 FS values for the initial and optimized problem configurations of Fig.~\ref{fig:BuildOrientation:Fig6}. Note the improvement in the minimum FS, as well as the general shift towards the right.}
\label{fig:BuildOrientation:Fig7}
\end{figure}

One advantage of our formulation given in \eqref{eq:BuildOrientation:Eq1} is that the elements with lower FS values contribute more heavily to the objective function compared to those with high FS values. Hence, in each iteration, the optimization desirably focuses more on increasing the FS of the most critical elements. Figure~\ref{fig:BuildOrientation:Fig6} illustrates the performance of our objective function. For the given door handle, the FS for the most critical element is increased from 3.5 to 4.6 using our approach.  Figure~\ref{fig:BuildOrientation:Fig7} shows the histograms of the lowest 300 FS values for the problem configuration shown in Fig.~\ref{fig:BuildOrientation:Fig6}. As the orientation is optimized, the number of elements with low FS values decreases and the distribution shifts to the right.

\section{Surrogate-based Optimization}
\label{sec:BuildOrientation:Sec5}

Because our objective function is based on the stress values obtained from an FE analysis applied for each candidate orientation, finding the optimum orientation can be very expensive using conventional methods due to the large number simulations. This effort can be even more prohibitive for complex geometries with a large number of elements. Hence, it is critical to determine useful evaluation points to restrict the number of FE runs as much as possible. To address this challenge, we employ a surrogate modeling approach that approximates the design space with a proxy response surface. 

Surrogate models (metamodels) are commonly used in engineering and design optimization when each function evaluation involves costly simulations~\cite{simpson2001kriging, queipo2005surrogate, wang2007review}. In design optimization, these expensive objective functions or constraints are replaced with surrogate models that serve as approximations to the original functions. 

We use surrogate modeling to approximate the design space represented by the build orientation $\bf x$ and the corresponding objective function $f({\bf x})$ in \eqref{eq:BuildOrientation:Eq1}. This objective function is highly dependent on the geometry, loading configuration and orthotropic material properties, with no access to an analytical relationship between the design variables and the objective function. In all the examples presented in this work, we have applied a brute force parametric sweeping as a benchmark and have found the resulting objective functions to be non-convex. We thus formulate our problem as a black-box global approximation problem and employ a surrogate-based optimization method. Specifically, we use MATLAB's Surrogate Model Toolbox (MATSuMoTo) presented in~\cite{mueller2014matsumoto}.
 
The main steps are as follows:
 \begin{enumerate}
 \item {\it Design of Experiments}: Select the number of initial orientations ($\bf x$'s) and evaluate the corresponding objective functions ($f({\bf x})$'s) using FE simulations.
 \item {\it Surrogate Modeling}: Construct the surrogate model mapping $\bf x$'s to $f({\bf x})$'s.
 \item {\it New Samples}: Select new samples using the surrogate model and perform new FE simulations.
 \item {\it Iterate}: Iterate until the maximum number of function evaluations is reached or the improvement $f({\bf x})$ ceases. 
\end{enumerate}  

\begin{figure*}
\centering
\includegraphics[width=\textwidth]{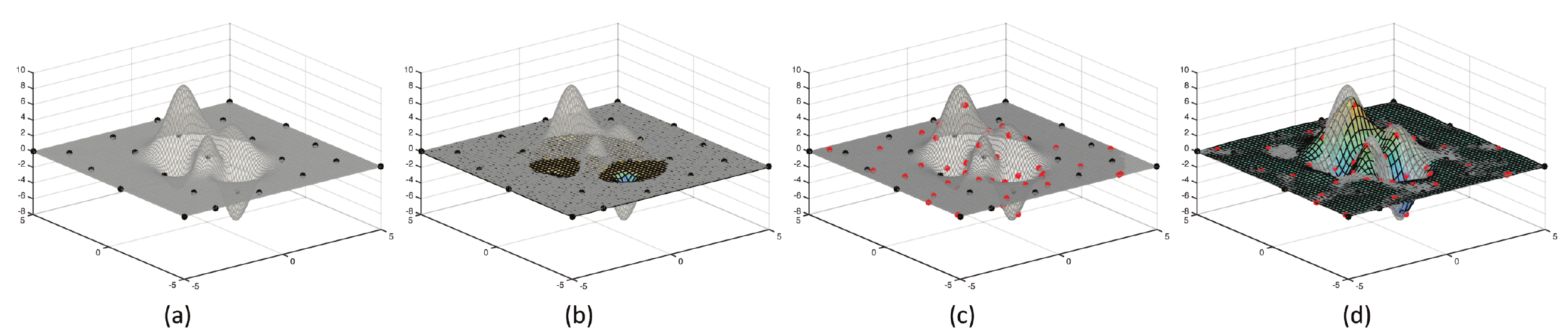} 
\caption{(a)~Initial design of experiments, (b)~constructing a surrogate model, (c)~selecting new samples and (d)~enhancing the surrogate model. (b) and (c) are repeated until a certain stopping criteria is satisfied. Black and red dots represent initial and new samples, respectively. Transparent surface is the exact objective function and colored surface is the surrogate model constructed with the selected samples.}
\label{fig:BuildOrientation:Fig8}
\end{figure*} 
 
Figure~\ref{fig:BuildOrientation:Fig8} shows an example for a two dimensional problem. The third dimension shows the objective values. Here, the transparent surface represents the exact values of the objective function in the specified design space. As the number of iterations (\ie the number of samples evaluated using the expensive objective function) increases, the surrogate model converges to the exact values of the objective function.
 
In Step~1, in order to determine the initial orientations, we use the Latin hypercube sampling (LHS) method with 'maximin' criterion that allows a wider and more uniform coverage of the design space by maximizing the pairwise distances between the sample points. This is a statistical method commonly used for design of experiments. Because there is no a priori information about the design space, LHS is a suitable method that spreads the sample points evenly across the design space. In order to evaluate the objective function at the selected orientations, we use ANSYS Mechanical APDL, and obtain the stress tensor field and calculate the FS values using \eqref{eq:BuildOrientation:Eq2}. 

For surrogate modeling, there are several methods available in the literature including polynomial regression models~\cite{myers1995response}, radial basis functions (RBF)~\cite{simpson2001metamodels, powell1990thetheory, mullur2005extended}, neural networks~\cite{haykin1999neural}, kriging~\cite{simpson2001kriging, martin2004ontheuse} and support vector machines~\cite{girosi1998anequivalence}. The best choice for the surrogate modeling method is usually problem dependent. In this work, we use a cubic RBF (with leave-one-out cross validation) to construct the surrogate model because of its simplicity, robustness to different problem settings and high performance for small sample sizes~\cite{jin2001comparative}. However, it is possible to use other methods or combinations based on the problem setup and the characteristics of the design space, if known apriori. Comparative studies addressing this challenge can be found in the literature~\cite{jin2001comparative, muller2014influence, mullur2006metamodeling}. In Step~3, the constructed surrogate model is used to approximate the objective function in the remainder of the design space without performing costly FE simulations. For the next iteration, we use the randomized global candidate point search as our sampling strategy. Here, in addition to the set of candidate points around the minimum  of the surrogate model, a number of uniformly distributed samples are selected across the entire domain. \red{We choose this strategy to avoid possible local minima by allowing the optimization algorithm continue the search globally even after a local optimum has been detected. With this sampling strategy, it has been shown that the surrogate modeling approach is asymptotically complete, \ie~the algorithm will find the global optimum with probability one for an indefinitely long run-time and exact calculations~\cite{mueller2014matsumoto}. However, for practicality, number of function evaluations are restricted and thus the resulting orientation may not be the exact global optimum.}

\begin{figure*}
\centering
\includegraphics[width=\textwidth]{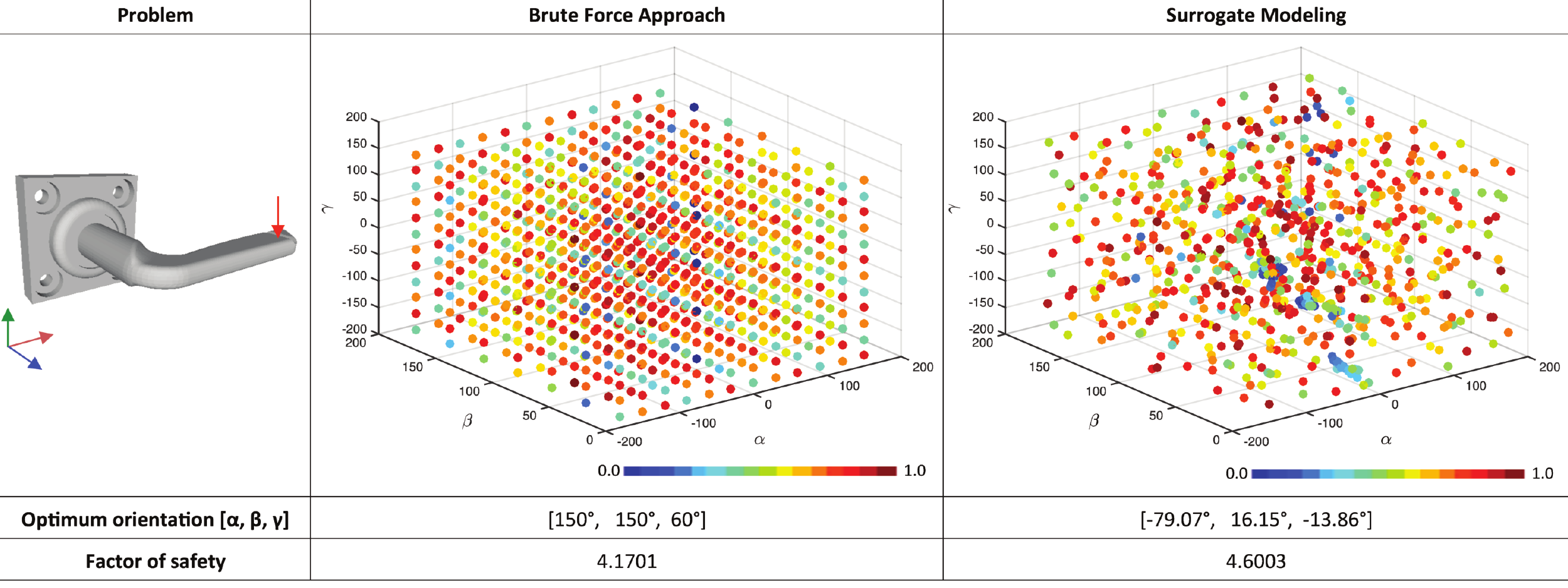} 
\caption{Comparison of surrogate-based optimization with brute force approach. Samples representing different build orientations are shown with dots. The color of each dot represents the normalized objective function value for the corresponding orientation. The best orientations obtained with the two methods and  the corresponding minimum FS values are also shown.}
\label{fig:BuildOrientation:Fig9}
\end{figure*}  

Figure~\ref{fig:BuildOrientation:Fig9} compares our surrogate based optimization with the brute force approach of uniformly sampling the design space. Here, 1008 objective function evaluations are performed using both the brute force and the surrogate-based approaches hence making the computational effort identical in both cases. This number represents a uniform grid of 30 degree increments in each of the design variables in the brute force method. With the surrogate-based optimization, the minimum FS obtained for the optimum orientation after 1008 function evaluations is  approximately 10\% better than that computed by the brute force approach. It can also be observed that the surrogate modeling enables a more efficient sampling strategy where regions of local minima  are more rigorously explored (dense regions with many blue points in Fig.~\ref{fig:BuildOrientation:Fig9}).

\begin{figure}
\centering
\includegraphics[width=5.0in]{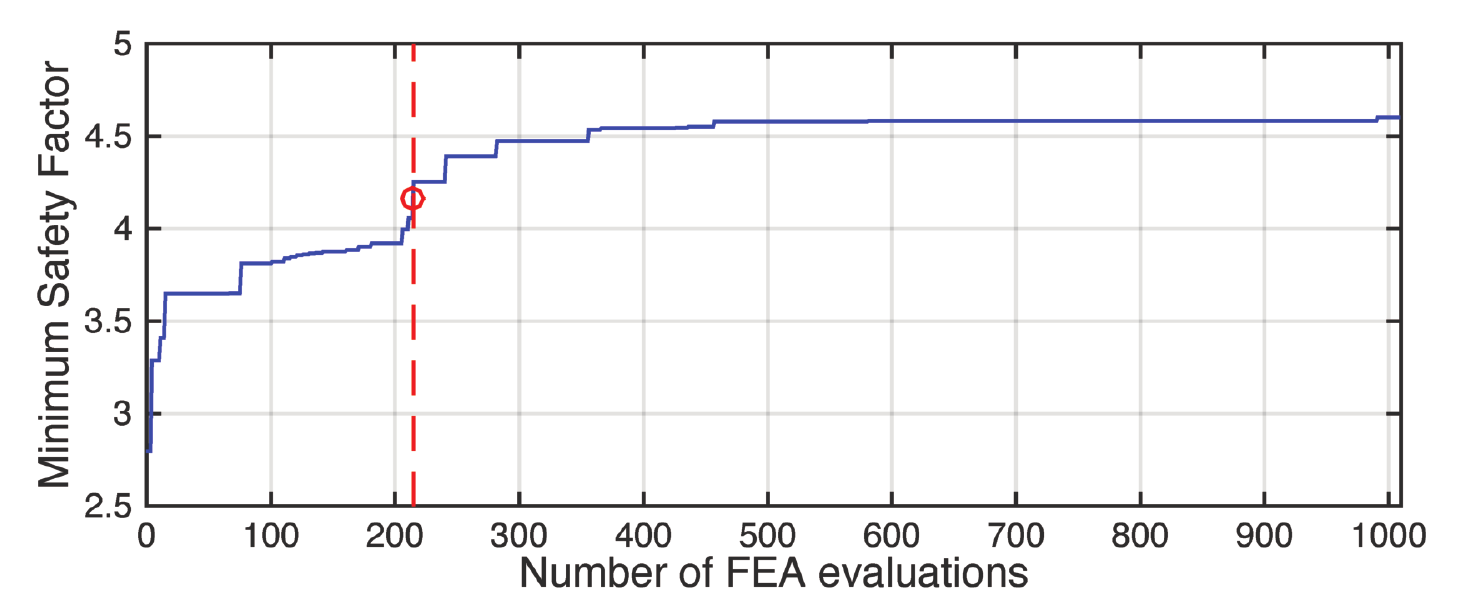} 
\caption{Performance of surrogate-based optimization for problem configuration in Fig.~\ref{fig:BuildOrientation:Fig9} with respect to the sample size. Red circle shows the best objective value that can be obtained using the brute force approach. With surrogate-based optimization, the same performance level of brute force approach can be obtained with only 215 samples.}
\label{fig:BuildOrientation:Fig10}
\end{figure} 

Figure~\ref{fig:BuildOrientation:Fig10} illustrates the performance of the surrogate modeling approach as a function of evaluation points for the problem shown in Fig.~\ref{fig:BuildOrientation:Fig9}. It can be observed that using only 215 function evaluations, the surrogate model can attain the best FS computed by the brute force approach which uses 1008 samples. Although the numerical values here might be problem dependent and may vary, similar gains are expected to be observed for various problem settings.

\section{Results and Discussion}
\label{sec:BuildOrientation:Results}

\begin{figure}
\centering
\includegraphics[width=4.0in]{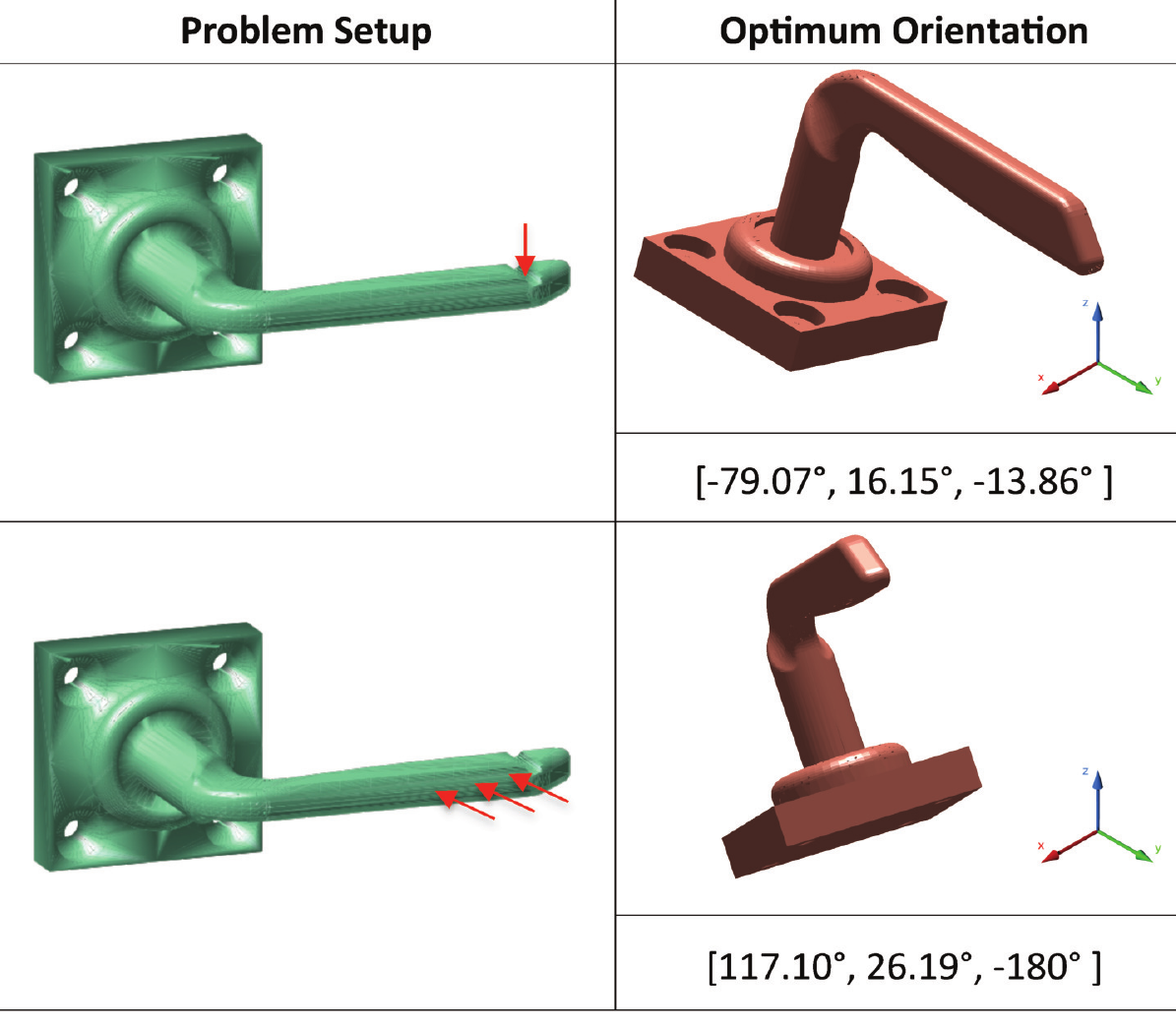} 
\caption{Effect of different loading configurations on the optimum build orientation. Left column shows the problem configuration and right column shows the corresponding optimum build orientations.}
\label{fig:BuildOrientation:Fig11}
\end{figure} 

For a fixed geometry and material, the optimum build orientation may change drastically if the loading configuration changes. In Fig.~\ref{fig:BuildOrientation:Fig11}, optimum orientations are investigated for two different loading configurations  of the door handle (displacement boundary conditions are kept the same). It can be observed that the difference between the optimal build orientations for these two examples is quite distinct.  

\begin{figure*}
\centering
\includegraphics[width=\textwidth]{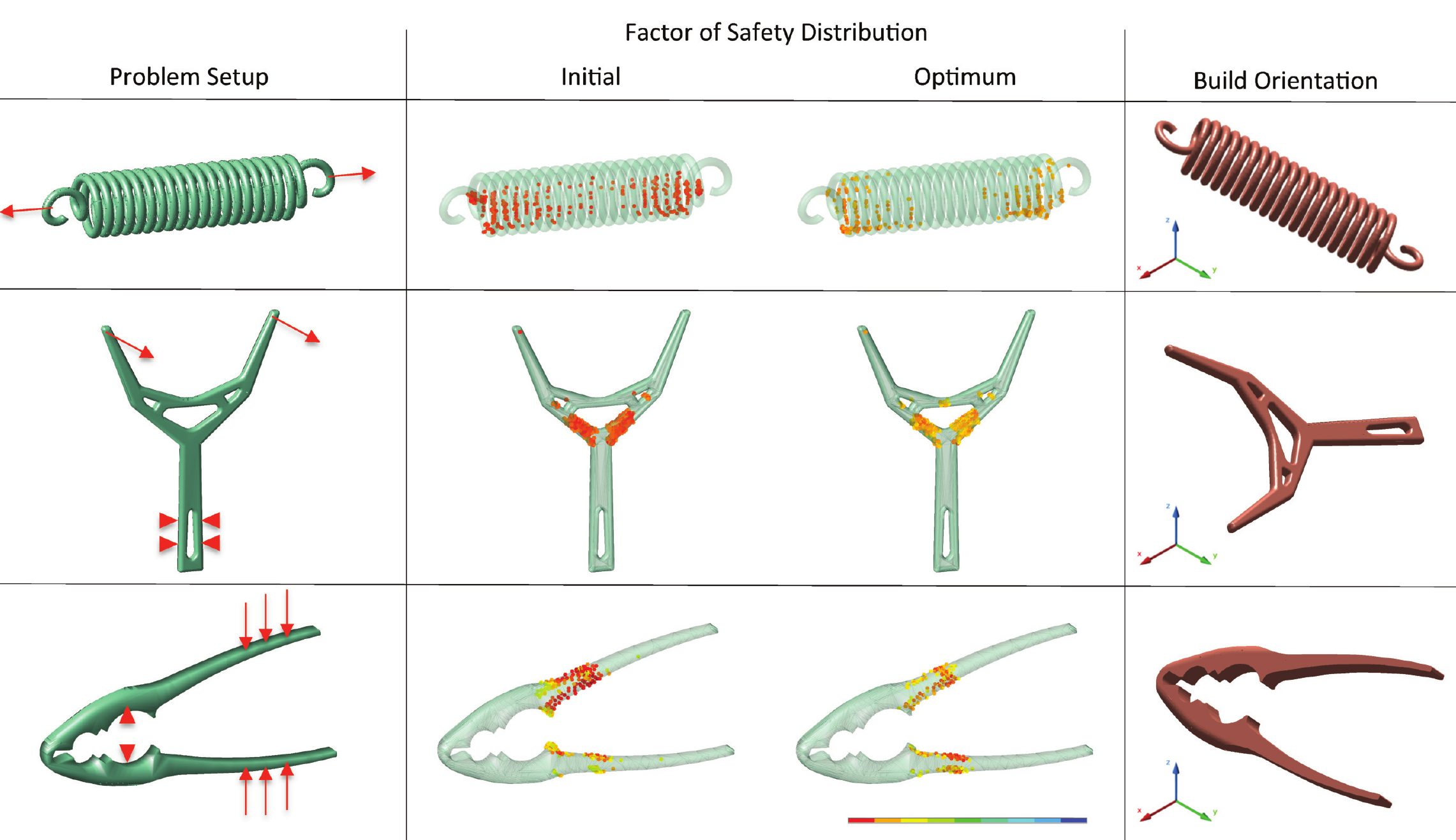} 
\caption{Build orientation optimization results for three different problem configurations. Left column shows the problem settings. Middle column shows the distribution of the lowest 300 FS values over the geometry for the initial and optimum orientations. Right column shows the optimum build orientations.}
\label{fig:BuildOrientation:Fig12}
\end{figure*}

Figure~\ref{fig:BuildOrientation:Fig12} demonstrates the proposed algorithm on different problems. In all cases, we observed significant improvements in the FS when our optimization approach is applied. Table~\ref{tab:BuildOrientation:Tab2} shows the improvement in FS for several problem configurations. Depending on the geometry, loading, boundary conditions and the initial orientation, we were able to achieve up to $90\%$ improvement in the resulting FS values. It is also important to note that in some examples (such as the slingshot and the nut cracker), it is possible to move from unsafe ($FS < 1.0$) to safe ($FS > 1.0$)  using our method without any geometric modification. 

Table~\ref{tab:BuildOrientation:Tab3} shows the computational performance of our approach for problems with different mesh complexities. Because the FE simulations constitute the computational bottleneck, the number of elements directly impact the overall computation time. We observe that the computation time per element is similar for all models and it is approximately $0.1s$. In these problems,  the stopping criterion is the maximum number of objective evaluations (\ie FE simulations) which is selected to be 400. A PC with a 2.4GHz Core CPU and 8GB RAM using MATLAB R2014b is used for surrogate modeling, which drives the FE simulations using a script based ANSYS Mechanical APDL (v14). 

\begin{table*}[t]\small
\caption{Numerical results for several test cases.}
  \centering
  \begin{tabular}{l || c | c | c | c}
      \hline
    \bf Problem Setup & \bf Optimum Orientation & \bf Initial FS & \bf Opt. FS  & \bf Improvement\\ \hline \hline
    Handle (Fig.~\ref{fig:BuildOrientation:Fig11}-Top) & $[-79.07^{\circ}, 16.15^{\circ}, -13.86^{\circ}]$ & 3.3202 & 4.6003 & 38.55$\%$\\
    Handle (Fig.~\ref{fig:BuildOrientation:Fig11}-Bottom) & $[117.10^{\circ}, 26.19^{\circ}, -180^{\circ}]$ & 1.6881 & 2.2259 & 31.85$\%$\\
    Spring (Fig.~\ref{fig:BuildOrientation:Fig12}) & $[56.12^{\circ}, 35.25^{\circ}, -5.33^{\circ}]$ & 9.7441 & 12.0901 & 24.07$\%$\\
    Slingshot (Fig.~\ref{fig:BuildOrientation:Fig12}) & $[-100.12^{\circ}, 128.06^{\circ}, 179.01^{\circ}]$ & 0.9123 & 1.2056 & 32.15$\%$\\
    Nut Cracker (Fig.~\ref{fig:BuildOrientation:Fig12}) & $[7.80^{\circ}, 123.97^{\circ}, -93.32^{\circ}]$ & 0.7815 & 1.4844 & 89.94$\%$\\
    \hline
  \end{tabular}
  \label{tab:BuildOrientation:Tab2}
\end{table*} 

\begin{table}[t]\small
\caption{Computational performance  of our method for several test cases. The maximum number of objective function evaluations are limited to 400.}
  \centering
  \begin{tabular}{l || >{\centering\arraybackslash}m{2cm} | >{\centering\arraybackslash}m{2cm} }
      \hline
    \bf Problem Setup & \bf Number of FEA Elements & \bf Computation Time [s]\\ \hline \hline
    Door Handle (Fig.~\ref{fig:BuildOrientation:Fig11}) & 11290 & 1196.3999\\
    Spring (Fig.~\ref{fig:BuildOrientation:Fig12}) & 94228 & 5571.1827\\
    Slingshot (Fig.~\ref{fig:BuildOrientation:Fig12}) & 37558 & 2655.4636\\
    Nut Cracker (Fig.~\ref{fig:BuildOrientation:Fig12}) & 9145 & 1124.9073\\
    \hline
  \end{tabular}
  \label{tab:BuildOrientation:Tab3}
\end{table}

 \begin{figure*}
\centering
\includegraphics[width=\textwidth]{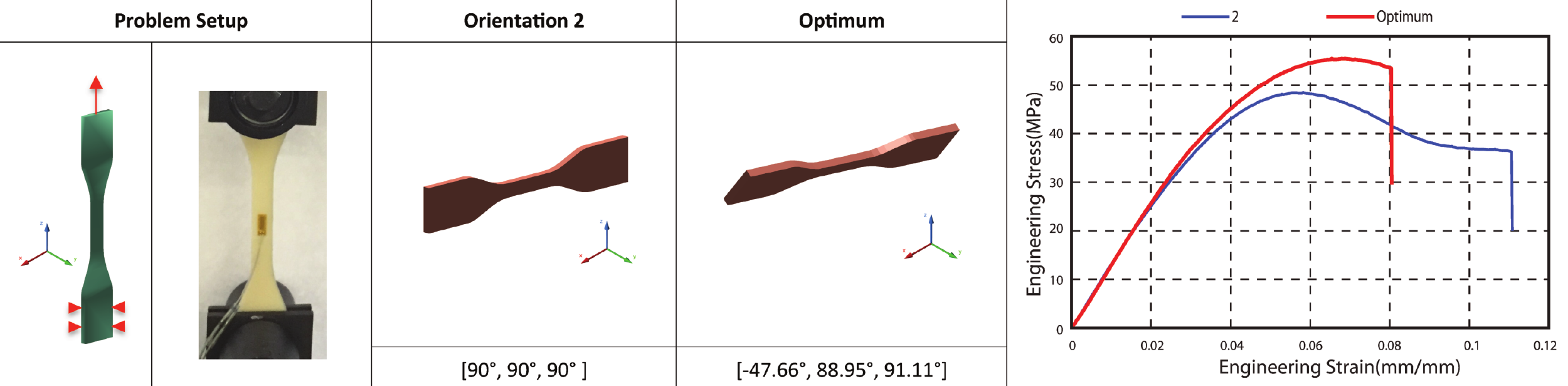}
\caption{Performance evaluation of our algorithm with a standard dog-bone tensile specimen.}
\label{fig:BuildOrientation:Fig13}
\end{figure*}

We conducted two sets of physical experiments to evaluate the performance of our approach. First, for the seven orientations shown in Fig.~\ref{fig:BuildOrientation:Fig4}(a), we computationally determined the best orientation by calculating our objective function for each of the orientations. Direction 2 proved to be the best orientation. For the same geometry and loading configuration, we then computed the optimum orientation using the proposed surrogate modeling approach, printed a new specimen corresponding to this optimum orientation, and conducted a tensile test. Figure~\ref{fig:BuildOrientation:Fig13} compares the stress-strain curves for these two orientations. It is observed that our optimum orientation improves the yield strength by $13\%$ (from 40.14 MPa to 45.56 MPa). 

\begin{figure*}
\centering
\includegraphics[width=\textwidth]{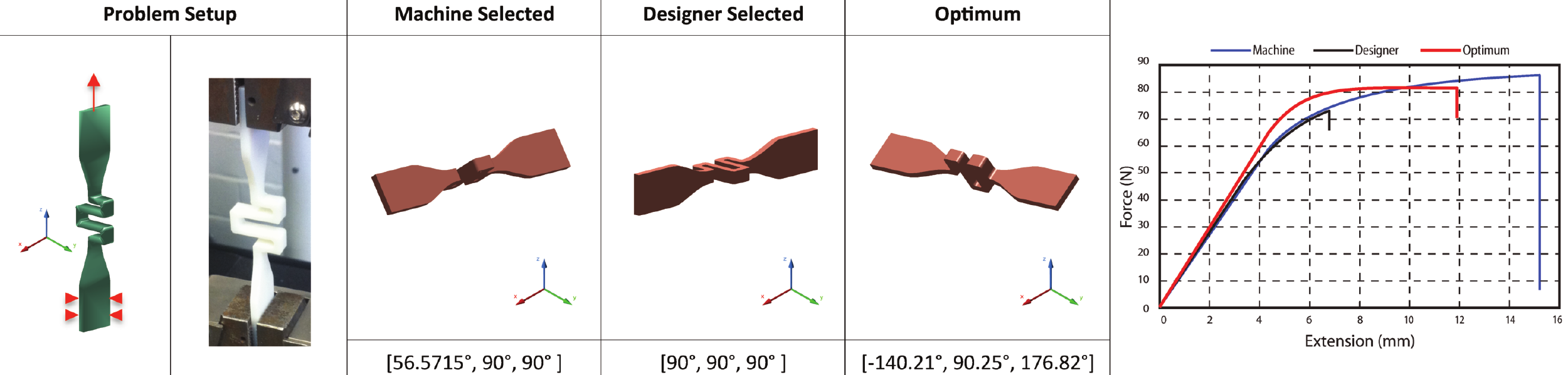}
\caption{Performance evaluation of our algorithm with a custom-designed part.}
\label{fig:BuildOrientation:Fig14}
\end{figure*}

We conducted another test on a custom-designed part shown in Fig.~\ref{fig:BuildOrientation:Fig14}. Note that aside from the conventional tensile specimens, there are not many design alternatives to physically observe and quantify yielding. As a result, we devised an experiment where we  simultaneously acquire the forces and the elongation using a tensile-test machine. We compare the orientation we obtained with our approach against the orientation that minimizes the amount of support structure (machine orientation) and an orientation based on a mechanical engineer's best judgement (Fig.~\ref{fig:BuildOrientation:Fig14}). It is shown that our optimized orientation withstands a higher end force before yielding compared to the other two directions ($12\%$ and $20\%$ better compared to human and machine prediction, respectively). For each of the model and loading configuration shown in this work, we asked several engineers to predict the best orientation to maximize FS. In all cases, our approach outperformed human judgment.

\paragraph{Scope and assumptions} Our analysis is restricted to homogeneous materials and we assume a linear-elastic FE model to successfully simulate the stress and strains in the object. \red{For non-homogeneous material models and higher-order finite elements, stresses may be calculated more accurately at a cost of additional computational complexity. Such modifications in the analysis may result in changes in the resulting optimum orientation.} Our failure criteria is based on maximum stresses and we do not consider strain failure or maximum displacement constraints. However, since our approach is based on FE simulations, it is readily possible to include such criteria in the objective function or the constraints without loss of generality. The yield criterion we use is accurately applicable to ductile materials. For brittle materials, we use the ultimate tensile strength (UTS) as the yield strength. This assumption may cause our analysis for brittle materials to be less accurate compared to ductile materials. 

Our approach assumes that the properties extracted from the tensile test specimens accurately represent the properties of the designed object. In certain printing techniques such as FDM, each layer may involve first a contouring of the layer where the outline of the layer is printed, followed by a raster fill-in. In such cases, the properties will be a function of the object's scale, thus possibly creating a mismatch between the test specimen's and the actual object's properties. Also, we assume that the effects of support structure on the anisotropic mechanical properties of fabricated products are negligible. However, investigation and incorporation of these effects might enhance the performance of the presented method further.

\red{In our approach, we assume that the geometry is known and fixed. Although it is possible to combine our build orientation optimization with the structural optimization methods described in Chapters~\ref{chp:ForceLocationUncertainty} and \ref{chp:ProposedWork}, computational complexity introduced with such an addition would be significant due to the possible increase in the number of local minima.}

\section{Conclusions}
\label{sec:BuildOrientation:Conclusions}

Additively manufactured products exhibit directional dependencies in their structural properties due to the layered nature of the printing process. As a result, the build orientation can significantly affect the structural performance of the resulting objects. In this work, we developed a build orientation optimization algorithm that maximizes the mechanical strength of an additively manufactured object under certain loading/boundary configurations. We start with a set of physical experiments to determine the orthotropic material properties. Then, our optimization approach uses this information to calculate an optimum build orientation. 

Our objective in the optimization problem is formulated based on the factor of safety values obtained using FE simulations. We have shown that a surrogate-based optimization approach can accelerate the optimization process by strategically choosing useful evaluations points. Both our computational and physical experiments show that the optimized build direction can lead to considerable improvements in an object's ability to withstand applied loads. 

\paragraph{Future work} In this work, we explored our build orientation optimization algorithm only for a single polymer material and AM process combination. We expect the proposed formulation to be readily applicable to new polymer-based materials and print technologies. While we have observed a strong consistency among the samples printed in the same orientation, more studies quantifying the sensitivity of the results to the process parameters may be required to further validate the proposed work. We have not tested the proposed method to AM of metals. Recent efforts in process and microstructural modeling/simulation for metals would be critical for a successful extension of the proposed method to metals. Our preliminary discussions reveal that there may be a large number of parameters that affect anisotropy making such properties a strong function of the overall part geometry, spatial position in the print volume, thermal aspects of the process and post processes applied to the part. Moreover, in this work, we optimize the build orientation for a single loading configuration only. Yet, the optimization problem can be extended to ensure that the resulting build orientation is robust to multiple different loading configurations. This may require solving the problem multiple times using different loading configurations and choosing an orientation that jointly maximizes the FS for all considered loading conditions. Likewise, the uncertainty on the loading conditions could be encoded statistically to solve for a more robust build orientation. Finally, other criteria such as creep and fatigue failure provide interesting research opportunities for build direction optimization.

\chapter[Data Driven Structural Optimization]{Data Driven Structural Optimization}
\label{chp:DataDrivenTopOpt}
\blindfootnote{This chapter is based on Ulu \etal, 2014 \cite{ulu2014adata} and Ulu \etal, 2015 \cite{ulu2015adata}.}

Topology optimization problems involving structural mechanics are highly dependent on the design constraints and boundary conditions. Thus, even small alterations in such parameters require a new application of the optimization routine. To address this problem, we examine the use of known solutions for  predicting  optimal topologies under a new set of design constraints. In this context, we explore the feasibility and performance of a data-driven approach to structural topology optimization problems. Our approach takes as input a set of images representing optimal 2-D topologies, each resulting from a random loading configuration applied to a common boundary support condition. These images represented in a high dimensional feature space are projected into a lower dimensional space using component analysis. Using the resulting components, a mapping between the loading configurations and the optimal topologies is learned. From this mapping, we estimate the optimal topologies for novel loading configurations. The results indicate that when there is an underlying structure in the set of existing solutions, the proposed method can successfully predict the optimal topologies in novel loading configurations. In addition, the topologies predicted by the proposed method can be used as effective initial conditions for conventional topology optimization routines, resulting in substantial performance gains. We discuss the advantages and limitations of the presented approach and show its performance on a number of examples.

\section{Introduction} 

Efficient use of material is a key priority for designers in many industries including automotive, aerospace and consumer product industries \cite{bendsoe2004topology,schramm2006recent,richardson2010robust}. Optimizing material layout to satisfy a specific performance criteria, i.e. topology optimization is thus a crucial part of engineering design process. With recent advances in  manufacturing technologies, topology optimization now attracts even more attention \cite{schramm2006recent}. So far, various optimization algorithms including genetic algorithms, method of moving asymptotes, level sets and topological derivatives have been studied for structural topology optimization \cite{bendsoe2004topology,chapman1994genetic,jakiela2000continuum,aage2013parallel,dijk2013levelset,norato2007atopological}. 

Although structural optimization algorithms are becoming computationally more efficient with time, the need for a large number of iterations can rarely be avoided due to the essence of optimization theory. Even with very small alterations in the design constraints, a structurally optimum topology can not be predicted directly by a human from physical principles due to the complex nature of the problem. Based on this observation, we explore how known solutions to topology optimization problems can be exploited to generate a new design for a novel set of  loading configurations. Here, the main challenge is to find a mapping between design constraints and the resulting optimal topologies. With this motivation, we present a data-driven approach to topology optimization involving structural mechanics and explore its feasibility and performance.

\begin{figure}
\centering
\includegraphics[width=1.0\textwidth]{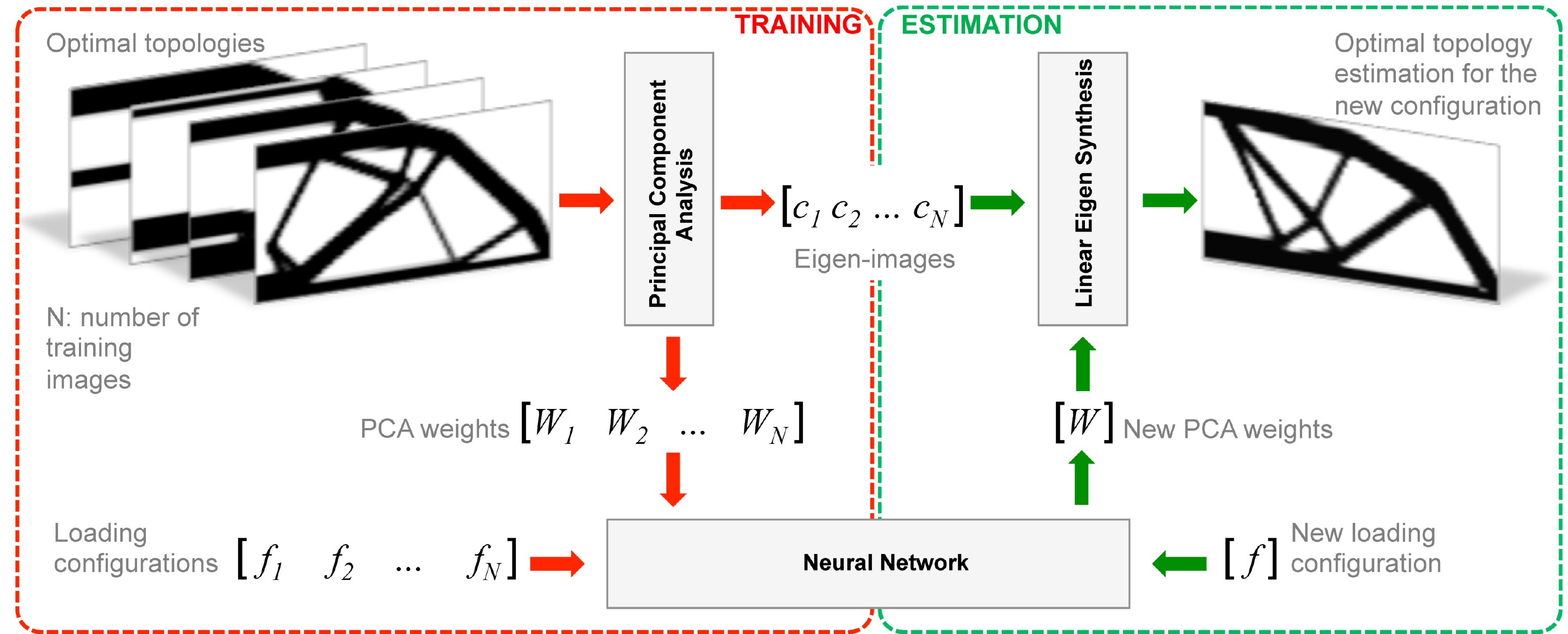}
\caption{Overview of our approach for optimal topology estimation.}
\label{fig:DataDrivenTopOpt:Fig1} 
\end{figure}

Our approach takes as input a set of images representing optimal 2D topologies, each resulting from a conventional optimization method, and generates an optimal topology estimation for a novel set of design constraints (Fig.~\ref{fig:DataDrivenTopOpt:Fig1}). In this study, only the variation in loading configurations is explored under a fixed set of structural boundary conditions. However, the application of the presented method is not limited to this choice, because design constraints can be expanded to accommodate any changes in the boundary conditions as long as the size of the overall design domain is kept the same. In the proposed method, the set of input images (known optimal topologies) which are represented in a high dimensional image space are projected onto a lower dimensional space using Principal Component Analysis (PCA). Once the dimensionality is reduced, a mapping between the loading configurations and the optimal topologies represented as PCA component weights is computed using a feed-forward neural network. Using the trained mapping, we estimate the PCA component weights for a novel loading configuration, and use the resulting estimation to synthesize a solution in the image space. This image represents our estimation of the optimal topology, given a novel loading configuration.

The primary goal of this study is to explore the feasibility and effectiveness of a data-driven approach to structural topology optimization problems. Our results show that the proposed method can successfully predict the optimal topologies in different problem settings, but the results are sensitive to the complexity and the size of the design space dictated by the loading configurations. However, independent of the problem complexity, a practical advantage of the proposed system is that the resulting topology estimations serve as effective initial conditions that facilitate  faster convergence in conventional topology optimization problems. 

Our main contributions are:

\begin{itemize}
\item a novel data-driven approach for structural topology optimization problems,
\item a comparison of mapping methods between the loading configurations and the optimal topologies,
\item a practical estimation of initial topologies to conventional topology optimization approaches.
\end{itemize}

\section{Problem Formulation}

\begin{figure}
\centering
\includegraphics[trim = 0.6in 4.7in 1.2in 0in, clip, width=0.8\textwidth]{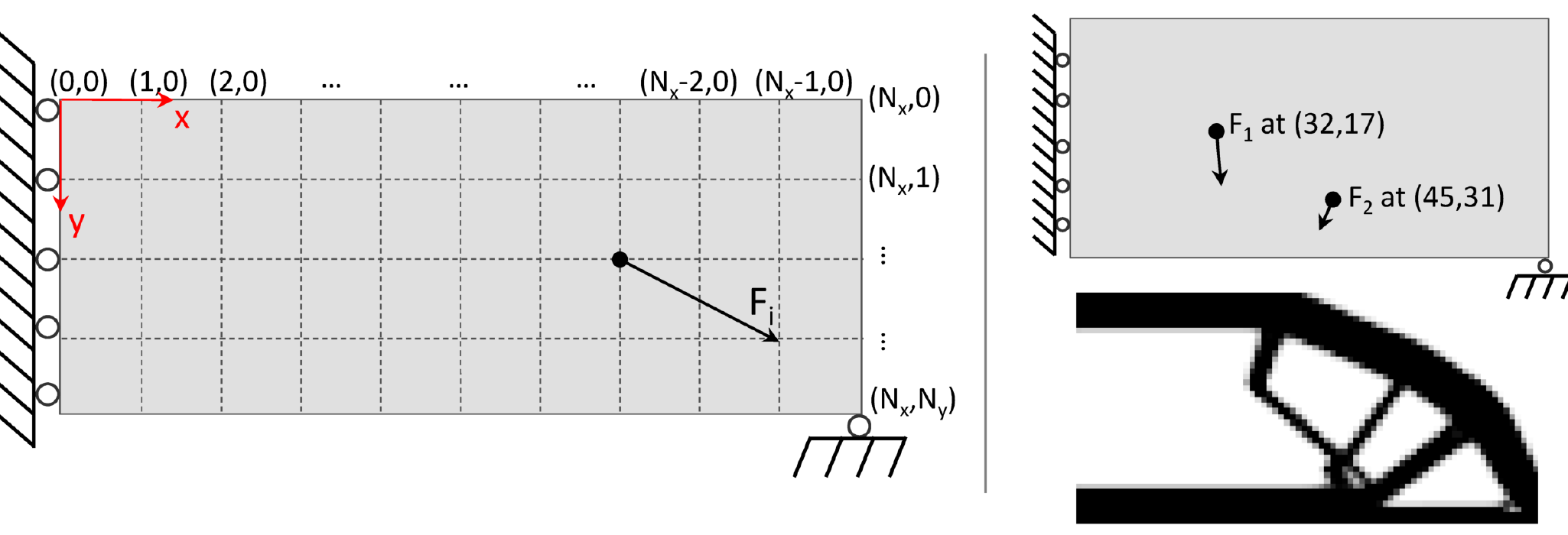}  
\caption{Left: $N_x$-by-$N_y$ design domain for topology optimization problem. Right: Example loading configuration and resulting optimal topology. Multiple external forces, ${F_i}$, can be applied to the beam at the nodes represented by $(x_i,y_i)$ coordinates.}
\label{fig:DataDrivenTopOpt:Fig2} 
\end{figure}

We illustrate our approach using the Messerschmitt-B\"{o}lkow-Blohm (MBB) beam problem, a classical problem in topology optimization. The rectangular beam is represented by an $N_x$-by-$N_y$ image as illustrated in Fig.~\ref{fig:DataDrivenTopOpt:Fig2}. The design domain is discretized by square finite elements each of which corresponds to a pixel in the gray-scale images. A number of external forces, ${F_i}$ $(i=1,...,k)$, can be applied to the beam at the nodes represented by $(x_i,y_i)$ coordinates. Applied forces can be in any direction, i.e. they can have both horizontal and vertical components with magnitudes ranging between $[0,1]$. Boundary support conditions are shown in  Fig.~\ref{fig:DataDrivenTopOpt:Fig2}. Note that this model is used to facilitate discussions; the following sections will demonstrate results on variations of the domain, boundary conditions and loading configurations. 

Our aim is to estimate the optimal topology for such problems when a novel loading condition is prescribed. For this, we generate a pool of training data where each training sample consists of a known loading configuration, and a corresponding optimal topology. To compute the optimal topologies given the loads, we use a density-based topology optimization algorithm given in \cite{andreassen2011efficient}. The method assigns a density value, $\rho_e$, between $[0,1]$ to each pixel $e$ in the domain that dictates the Young's modulus $E_e$ for that particular pixel as: 

\begin{equation}
\label{eq:DataDrivenTopOpt:EqYM}
E_e (\rho_e)=  E_{min} + \rho_e^p (E - E_{min})
\end{equation}

\noindent where $E$ is material stiffness, $E_{min}$ is a very small stiffness value assigned to void regions and $p$ is the penalization factor to attain black and white images. The optimization works toward minimizing the compliance resulting from the generated gray-scale structure. Mathematical formulation for the optimization procedure is given as follows:

\begin{equation}
\label{eq:DataDrivenTopOpt:EqObj}
\begin{aligned}
& \underset{{\boldsymbol \rho}}{\text{minimize}}
& & c({\boldsymbol \rho}) = {\bf U}^{T}{\bf K}{\bf U} \\
& \text{subject to}
& & V({\boldsymbol \rho})/V_0 = r,\\
& & & {\bf K}{\bf U} = {\bf F}, \\
& & & 0 \leq {\boldsymbol \rho} \leq 1.
\end{aligned}
\end{equation}

\noindent where the objective function $c$ is the compliance and ${\bf U}$, ${\bf K}$ and ${\bf F}$ are the global displacement vector, stiffness matrix and force vector, respectively. $r$ is the predetermined fraction of material volume $V({\boldsymbol \rho})$ and design space volume $V_0$. Details of this formulation can be found in \cite{andreassen2011efficient}. 

As a result of the optimization process, resulting images with a varying Young's modulus field represent the optimal topologies for the corresponding loading configurations, where lighter colors represent weaker regions (lower stiffness). The collection of these images establish an input database to our method.

\begin{figure}[t]
\centering
\includegraphics[trim = 0in 3.4in 2.2in 0in, clip, width=0.6\textwidth]{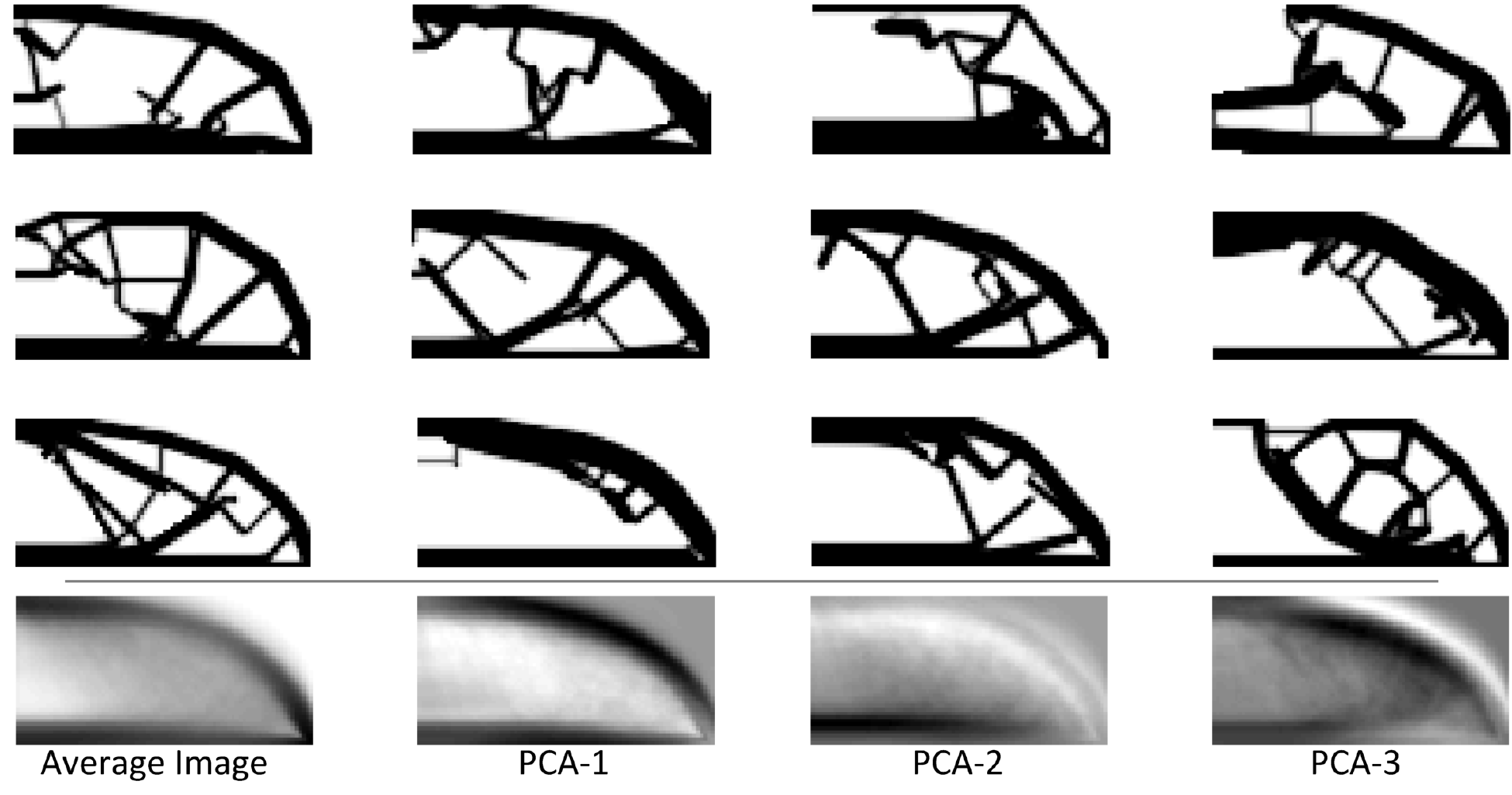} 
\caption{Top: Example training samples. Bottom: Average image and first three PCA images.}
\label{fig:DataDrivenTopOpt:Fig3} 
\end{figure}

\section{Component Analysis}

Principal component analysis is useful for analyzing the input data to identify the significant features inherent in the data, as a way to facilitate dimensionality reduction with minimal information loss. 

Suppose we have $M$ images each with $N_x$-by-$N_y$ resolution in our dataset. Then, gray-scale density values for images are stacked into $M$ column vectors ${\bf t_j}$ $(j=1,...,M)$ of length $l=N_x \times N_y$ to form the high-dimensional image space feature vectors. To mean-shift the data, the average image $\bf{\bar{t}}$ is calculated and subtracted from each sample in the training dataset, i.e. ${\bf t_j}-{\bf \bar{t}}$. In Fig.~\ref{fig:DataDrivenTopOpt:Fig3}, a randomly selected subset of the example training dataset (with 1000 samples), the resulting mean image, and the first three PCA component images are shown.

\begin{figure}[t]
\centering
\includegraphics[trim = 0in 1.3in 2.2in 0.1in, clip, width=1.0\textwidth]{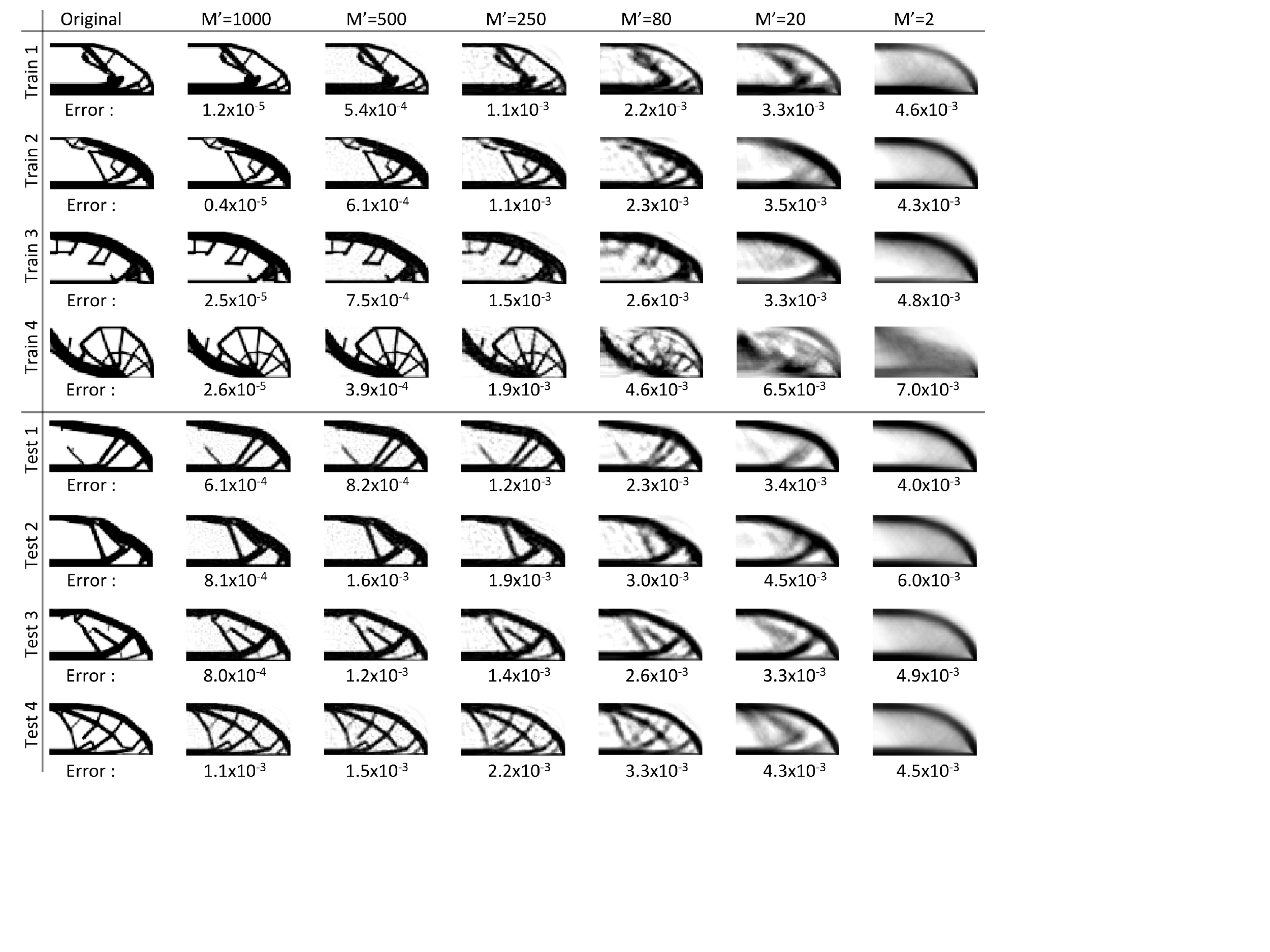} 
\caption{Reconstruction of example samples relative to the different number of eigen-components used. Each row corresponds to a different example. The upper half illustrates reconstructions for sample training images. The lower half shows the same for test images (i.e., images not involved in the construction of  PCA). Difference between each reconstruction image and the original one is evaluated using \eqref{eq:DataDrivenTopOpt:distanceMetric} and given underneath the corresponding image.}
\label{fig:DataDrivenTopOpt:Fig4} 
\end{figure}

\begin{figure}[!t]
\centering
\includegraphics[width=1.0\textwidth]{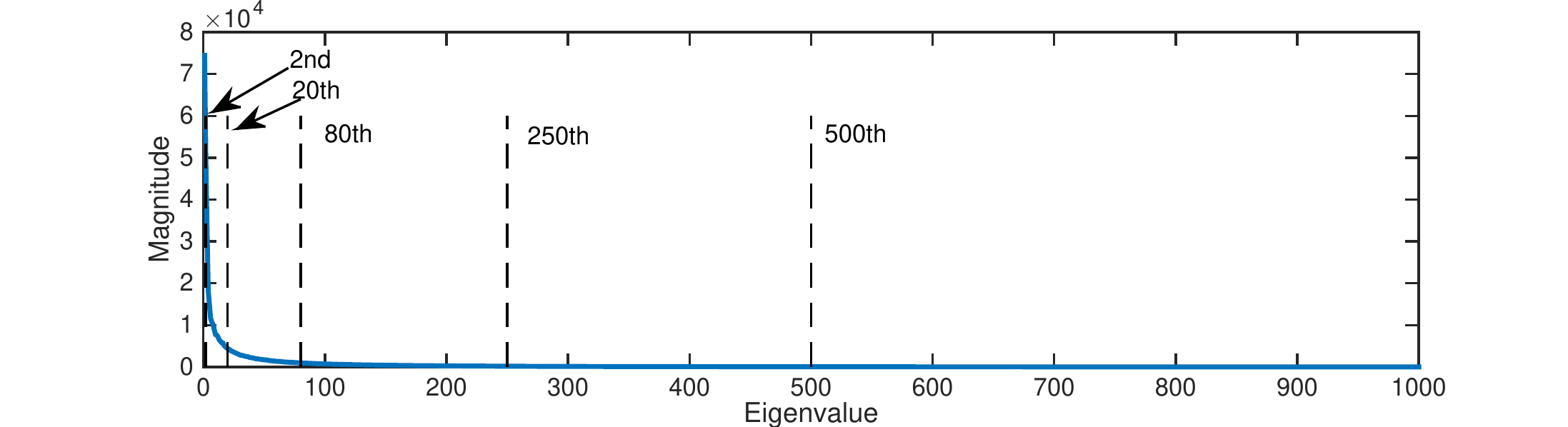} 
\caption{Magnitudes of eigenvalues corresponding to non-zero eigen-images obtained with PCA.}
\label{fig:DataDrivenTopOpt:Fig11} 
\end{figure}

Let the mean centered image be ${\bf p_j} = {\bf t_j}-{\bf \bar{t}}$, we can store our entire data set into $l x M$ matrix ${\bf P}$ to perform principal component analysis. Each column of ${\bf P}$ represents one mean centered image. In order to obtain the eigenvectors (i.e. principal components), the covariance matrix can then be constructed as ${\bf C=PP^{T}}$. However, size of the matrix ${\bf C}$ is $l$-by-$l$ and calculating $l$ eigenvectors may not be practical. As mentioned in \cite{turk1991eigenfaces}, if the number of features is larger than the number of training images $(l>>M)$, there can be at most $(M-1)$ useful eigenvectors (corresponding to non-zero eigenvalues) instead of $l$. These eigenvectors of $l$-by-$l$ ${\bf PP^{T}}$ matrix can be determined from the eigenvectors of $M$-by-$M$ matrix ${\bf P^{T}P}$ as ${\bf c_j=Pv_j}$ where ${\bf v_j}$ is eigenvectors of ${\bf P^{T}P}$. In this work, ${\bf c_j}$'s will be referred to as eigen-images. Each input topology optimization image can then be represented as a linear combination of these $M$ eigen-images, resulting in a PCA weight vector of ${\bf W_i}=[w_1, w_2, ..., w_M]^{T}$. Even using only a few number of eigen-images, $M'$, associated with the largest eigenvalues, a good approximation of an image can be obtained. 

In Fig.~\ref{fig:DataDrivenTopOpt:Fig4}, reconstruction of sample topology optimization images with different number of eigen-images are illustrated. Note that the figure shows example reconstructions of samples that were used during training (train), as well as for novel samples that were not part of the training (test). In this example, images are $80$-by-$40$ pixels and there are $1000$  training images generated by random assignments to the loading configurations and solving for the corresponding optimal topologies. Since PCA is limited by the number of training samples ($1000 < 3200$), the number of non-zero eigen-images is $1000$. In Fig.~\ref{fig:DataDrivenTopOpt:Fig11}, magnitudes of the eigenvalues corresponding to these eigen-images are shown. Here, it can be observed that a remarkably small number of eigen-images are sufficient for a high-fidelity reconstruction of the original images. Based on these observations, we use the first $80$ eigen-images in our examples in remainder of the work, without loss of generality.

\begin{figure}[t]
\centering
\includegraphics[trim = 0.2in 0.2in 0.2in 0.3in, clip, width=0.7\textwidth]{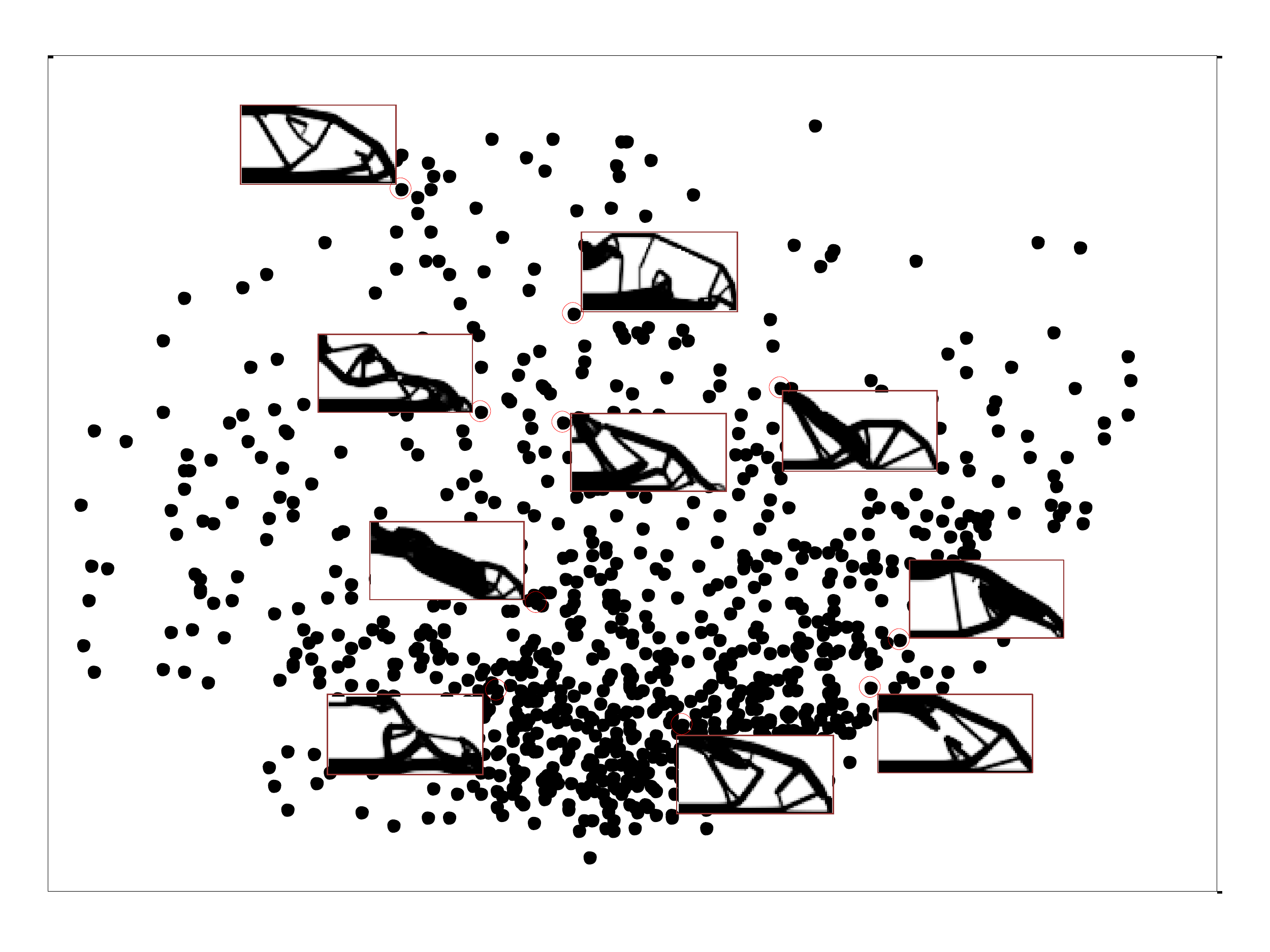} 
\caption{Training images projected into the 2D space created by the first two PCA components. }
\label{fig:DataDrivenTopOpt:Fig5} 
\end{figure}

Fig.~\ref{fig:DataDrivenTopOpt:Fig5} shows the dataset of  $1000$ training images when projected to the space created by the first two eigen-vectors. 

To quantify the mismatch between an original image and its reconstruction, we use the $L_1$ distance between the two images:
   
\begin{equation}
d(t_1, t_2)=\frac{\|t_1-t_2\|_{L1}}{length(t)}
\label{eq:DataDrivenTopOpt:distanceMetric}
\end{equation}  

\begin{figure}
\centering
\includegraphics[trim = 0.3in 6.5in 5in 0.2in, clip, width=0.8\textwidth]{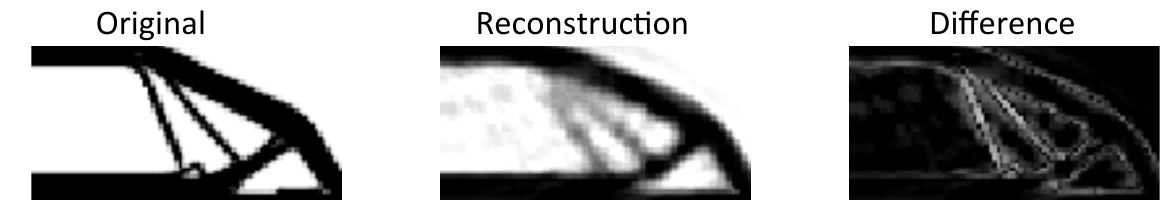} 
\caption{Difference between an example topology and its reconstruction using first 80 eigen-images.}
\label{fig:DataDrivenTopOpt:Fig6} 
\end{figure}

Fig.~\ref{fig:DataDrivenTopOpt:Fig6} illustrates this difference. This metric provides a value between $[0,1]$ where $0$ represents identical images. For this particular example, the difference between the original image and its reconstruction is calculated as $2.8\times10^{-3}$ using \eqref{eq:DataDrivenTopOpt:distanceMetric}. Fig.~\ref{fig:DataDrivenTopOpt:Fig4} also demonstrates the difference values evaluated for various examples underneath each corresponding image.

When properly weighted, eigen-images can be linearly combined to create an approximation to a new image representing a new optimized topology. For this purpose, a set of PCA weights associated with the corresponding loading configuration should be estimated. We address this problem by introducing a mapping function between the loading configurations, ${\bf F}$ and the PCA weights, ${\bf W}$ of the training samples. Details of this process will be described in the next section.

\section{Mapping Load Configurations to Optimal Topologies}

A useful application of PCA decomposition is that with a low dimensional data, a mapping between the original input and the PCA vector space can be created. In this section, we present a neural network approach to generate this mapping, specifically between the force vector indicating load conditions (${\bf F_i}$) and the PCA weights (${\bf W_i}$). In order to show the effectiveness of the neural network mapping, we compare its performance with linear regression and polynomial regression approaches in the following section. 

In our experiments, the mapping functions uses the following input and output configurations:
\begin{enumerate}
 \item The input vector is composed of four real numbers ($x$ and $y$ positions and magnitudes) for each force in the problem.
 \item The output vector is composed of $80$ real numbers corresponding to the PCA weights.
\end{enumerate}

\begin{figure}
\centering
\includegraphics[trim = 0in 0.9in 2.5in 0in, clip, width=0.7\textwidth]{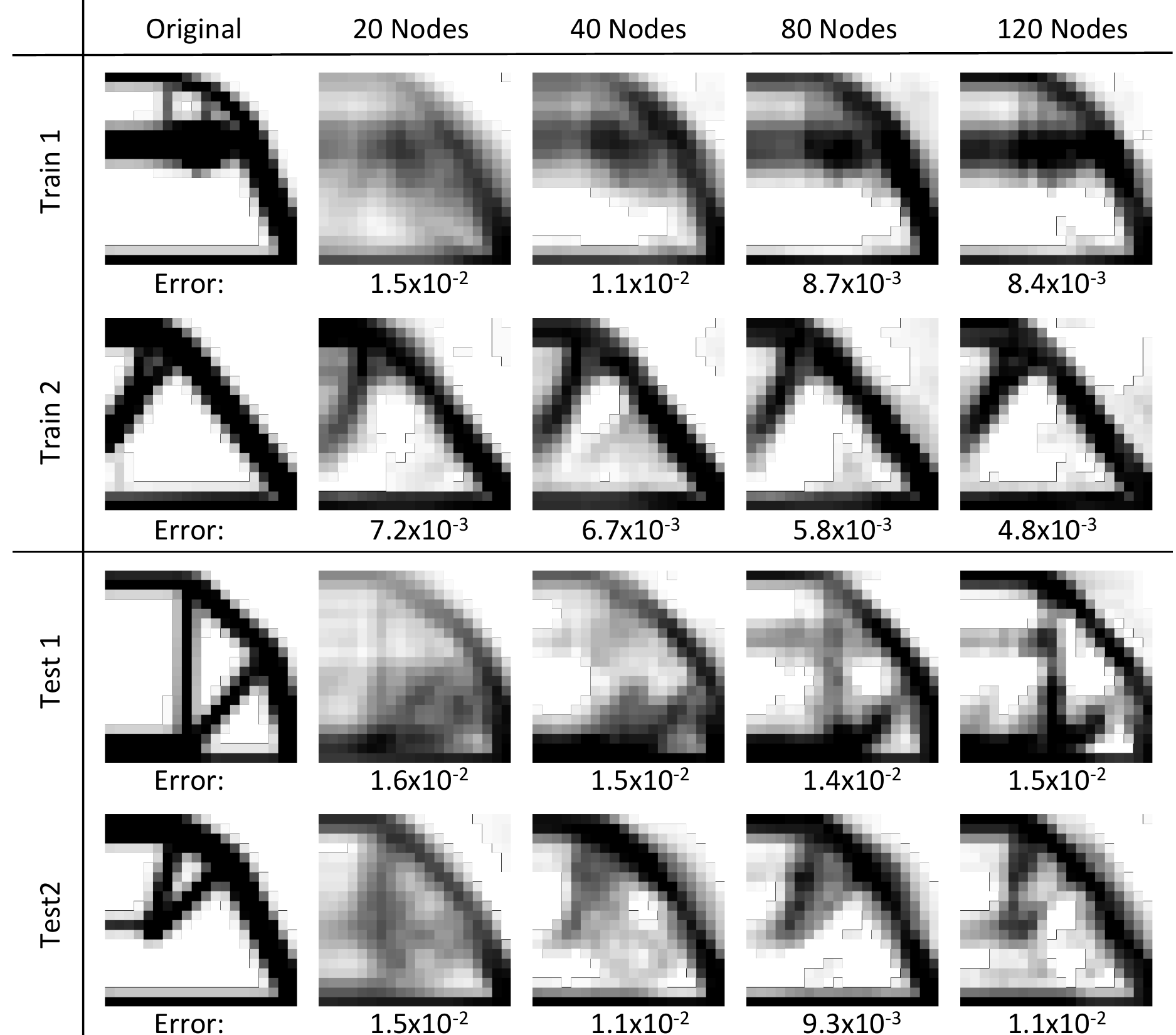} 
\caption{Estimation performance of different neural network configurations with different number of hidden layer nodes. Each row corresponds to a different example. The upper half illustrates estimations for sample training images. The lower half shows the same for test images (i.e., images not involved in the training of neural networks). Difference between each estimation image and the original one is evaluated using \eqref{eq:DataDrivenTopOpt:distanceMetric} and given underneath the corresponding image. }
\label{fig:DataDrivenTopOpt:Fig12} 
\end{figure}

\paragraph{Neural network} Theoretically, a neural network with a sufficient number of nodes and training samples is able to learn any input-output relationship for regression. However, with high dimensional data, such regression would require a considerably large number of training samples and hidden layer nodes resulting in impractical convergence time in the training stage \cite{bishop2006pattern}. We present results that indicate with limited amount of data and empirically determined number of hidden layer nodes, the neural network can appropriately learn the mapping from input force vectors to the output PCA weights for topology optimization.

In our experiments, we utilize a fully-connected feed-forward single hidden layer neural network \cite{bishop2006pattern} as the learner with the aforementioned input and output configurations. We train the resulting neural network with the scaled conjugate gradient algorithm. In order to determine the number of nodes in the hidden layer, we conducted a set of experiments with different number of hidden layer nodes. In Fig.~\ref{fig:DataDrivenTopOpt:Fig12}, a comparison of neural network performances are presented. In these experiments, neural networks are trained with the same dataset created for a specific design domain (middle configuration in Fig.~\ref{fig:DataDrivenTopOpt:Fig9}). This dataset includes 400 training images of size 20-by-20 pixels. Each network is trained until the same level of convergence in network design variables (i.e. adaptive weights) is achieved. The performance of each neural network is then evaluated by using randomly selected samples from both the training and test dataset (a set of randomly generated samples that are not in the training dataset). Expectedly, as the number of hidden layer nodes increases, the neural network performs better for samples in the training dataset. However, the performance gets worse for test samples due to overfitting. Based on this experiment, we use $80$ nodes in the hidden layer for all examples presented in the results section without loss of generality.

\paragraph{Linear regression} As an alternative method, linear regression is one of the most commonly used approaches for learning the mapping between a set of feature vectors \cite{allen2003thespace}. However, the relationship between the loading configurations and the PCA weights in such a high dimensional space may not be accurately predicted by this approach. In this work, we present the linear regression for performance comparison purposes. 

In order to construct the mapping between the input vector (i.e. loading configuration, $\bf F$) and output vector (i.e. PCA weights, $\bf W$), the mutivariate linear regression model can be formulated as follows:

\begin{equation}
\label{eq:DataDrivenTopOpt:linearRegress}
\begin{aligned}
{\bf W} &= {\bf F}{\boldsymbol \beta} \ \ \text{where},\\
{\bf W} = [{\bf W_1}, &{\bf W_2} , ..., {\bf W_i}, ..., {\bf W_{N}}]^T,\\
{\bf F} = [{\bf F_1}, &{\bf F_2} , ..., {\bf F_i}, ..., {\bf F_{N}}, 1]^T.\\
\end{aligned}
\end{equation}

\noindent Here, $N$ is the number of samples in the training dataset (400 in our experiments), vector ${\bf W_i}$ includes $M'$ PCA weights (80 in our experiments) and ${\boldsymbol \beta}$ represents the parameter matrix to be obtained. Equation \eqref{eq:DataDrivenTopOpt:linearRegress} can be solved as ${\boldsymbol \beta} = {\bf F}^{\dagger}{\bf W}$ where ${\bf F}^{\dagger}$ is pseudo-inverse of input matrix ${\bf F}$. 

\paragraph{Polynomial regression} The linear regression method formulated in \eqref{eq:DataDrivenTopOpt:linearRegress} can be extended to polynomial regression in order to account for the non-linearity in the design space. Theoretically, it may be possible to model non-linear behavior in the design space with high order polynomials. However, computational cost increases as the order of the polynomial increases, especially for high dimensional spaces with a large number of training samples due to the (pseudo-)inverse calculation of a large matrix. In this work, we use only quadratic regression for illustration purposes since we do not have any prior knowledge on the complexity of the design space. The mathematical formulation for the quadratic regression  is given as:

\begin{equation}
\label{eq:DataDrivenTopOpt:quadraticRegress}
\begin{aligned}
{\bf W} &= {\bf Q}{\boldsymbol \beta} \ \ \text{where},\\
{\bf W} = [{\bf W_1}, &{\bf W_2} , ..., {\bf W_i}, ..., {\bf W_{N}}]^T,\\
{\bf Q} = [{\bf Q_1}, &{\bf Q_2} , ..., {\bf Q_i}, ..., {\bf Q_{N}}, 1]^T,\\
{\bf Q_i} = [{\bf F_{i1}^2}, {\bf F_{i2}^2}, ..., {\bf F_{ik}^2},&{\bf F_{i1}}{\bf F_{i2}}, {\bf F_{i1}}{\bf F_{i3}}, ..., {\bf F_{i(k-1)}}{\bf F_{ik}}, {\bf F_{i1}}, {\bf F_{i2}}, ...,{\bf F_{ik}}].
\end{aligned}
\end{equation}

\noindent Here, ${\bf F_{ik}}$ is the k'th element of input vector ${\bf F_{i}}$. Different from \eqref{eq:DataDrivenTopOpt:linearRegress}, regressors in ${\bf Q}$ additionally include second order terms. Similar to linear regression, \eqref{eq:DataDrivenTopOpt:quadraticRegress} can be solved as ${\boldsymbol \beta} = {\bf Q}^{\dagger}{\bf W}$ where ${\bf Q}^{\dagger}$ is the pseudo-inverse of input matrix ${\bf Q}$. As seen, when the order of polynomial increases, the size of the matrix to be inverted increases drastically. For example, when there is only one force in the design domain (represented by four elements in ${\bf F_i}$), the size of ${\bf F}$ is 400-by-5 for linear regression with 400 training samples. However, the size of ${\bf Q}$ is 400-by-15 for quadratic regression. This difference is even more substantial for higher order polynomials, especially when the number of input forces or training samples are large. 
  
\begin{figure}
\centering
\includegraphics[trim = 0in 3.9in 4.7in 0in, clip, width=0.7\textwidth]{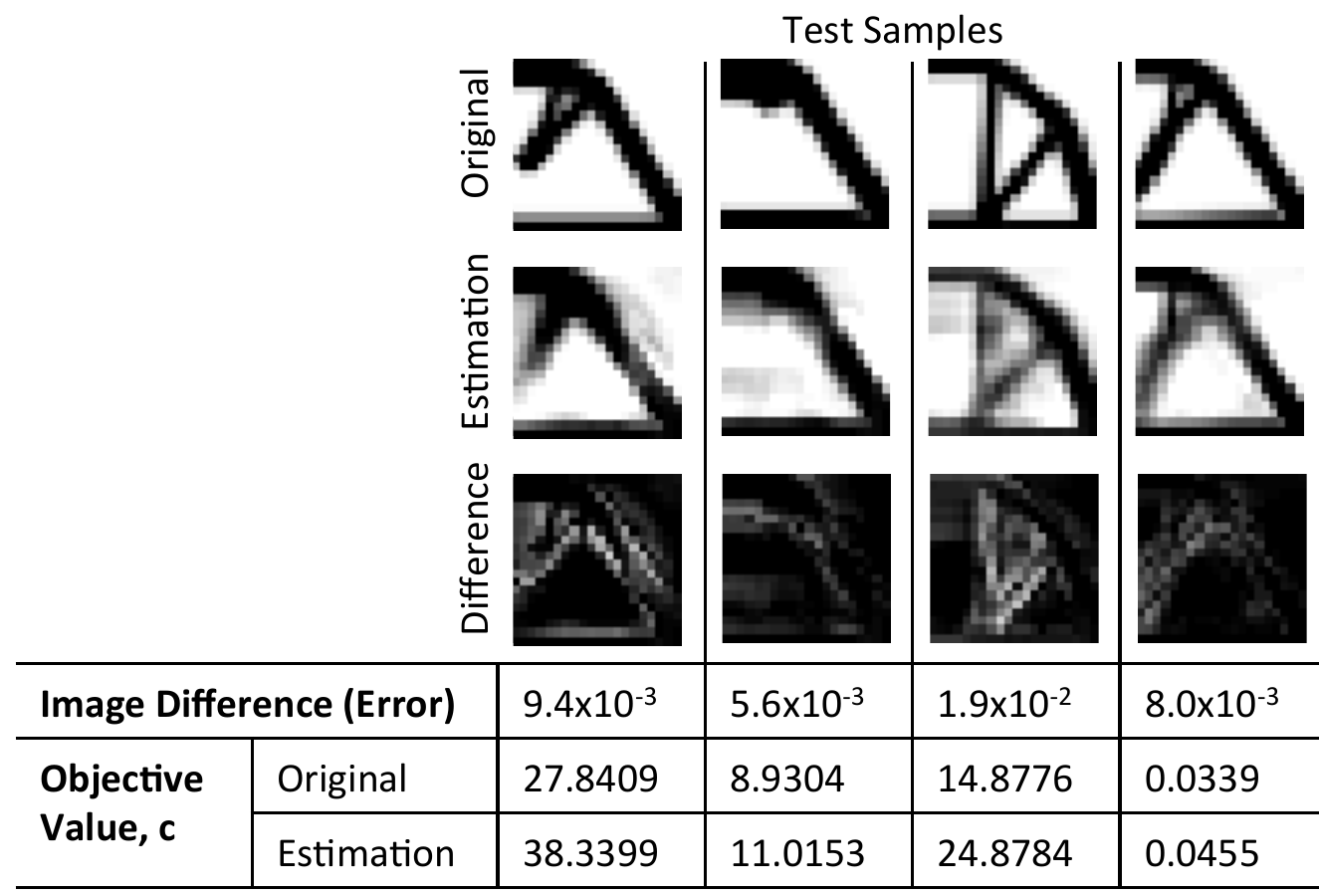} 
\caption{Example topologies and their neural network estimations using the first 80 eigen-images. Difference between original image and estimations (calculated using \eqref{eq:DataDrivenTopOpt:distanceMetric}) and objective values (structural compliance) calculated using \eqref{eq:DataDrivenTopOpt:EqObj} are also shown for numerical comparison.}
\label{fig:DataDrivenTopOpt:Fig7} 
\end{figure}

\begin{figure}
\centering
\includegraphics[trim = 0in 3.5in 0.7in 0in, clip, width=0.9\textwidth]{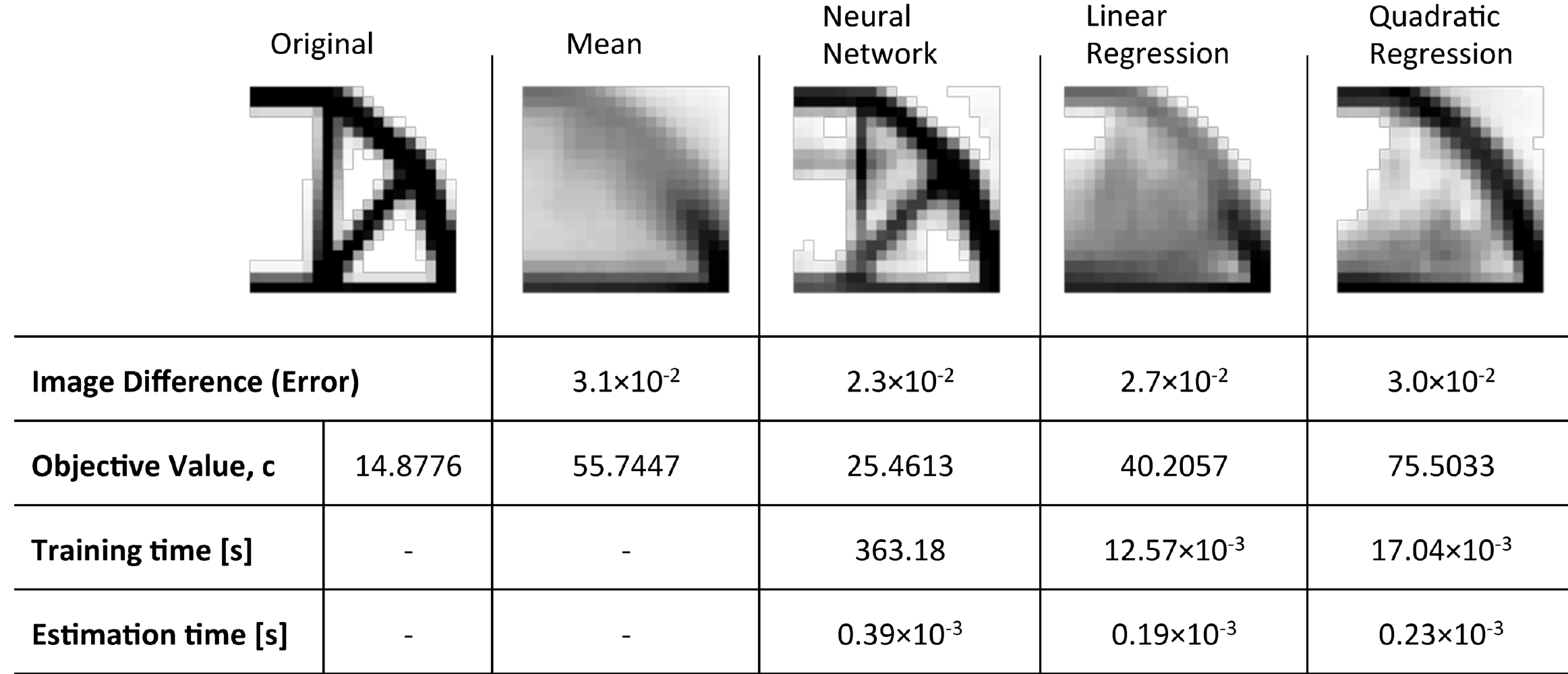} 
\caption{Comparison of neural network estimation with linear regression and quadratic regression estimation of an example topology. Difference between original image and estimations (calculated using \eqref{eq:DataDrivenTopOpt:distanceMetric}), objective values (structural compliance, calculated using \eqref{eq:DataDrivenTopOpt:EqObj}), training and estimation times are also shown for numerical comparison.}
\label{fig:DataDrivenTopOpt:Fig8} 
\end{figure}

\section{Results and Discussion}

\begin{figure}
\centering
\includegraphics[trim = 0in 0.62in 0in 0in, clip, width=1.0\textwidth]{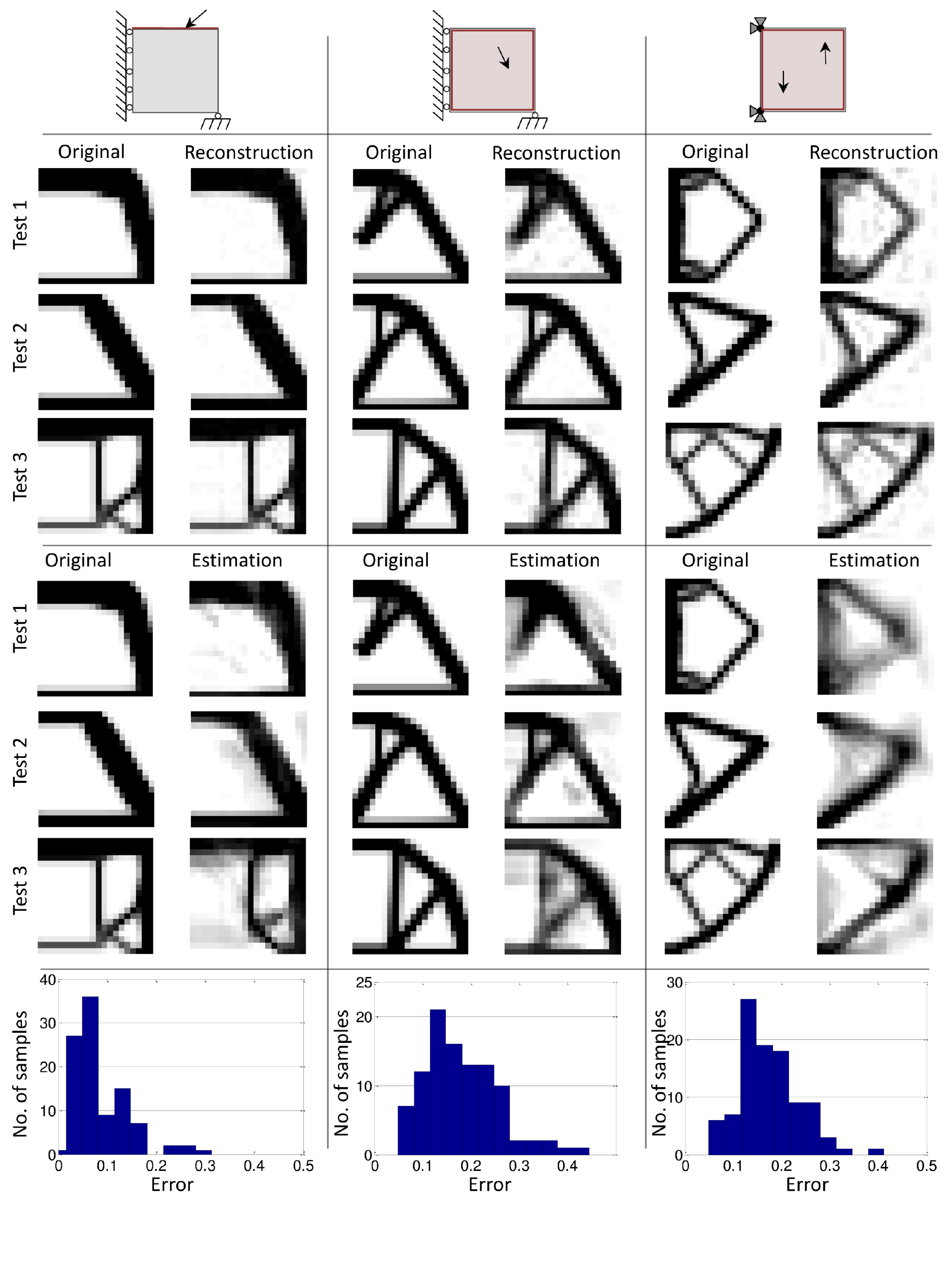} 
\caption{Performance of our method for different design configurations. Left: single force anywhere on the top surface in any direction and magnitude. Middle: single force anywhere in the domain in any direction and magnitude. Right: two vertical forces anywhere in the domain in any magnitude. Reconstruction refers to the PCA reconstruction of the sample using the eigen-images. Estimation uses the proposed neural network, followed by PCA reconstruction. }
\label{fig:DataDrivenTopOpt:Fig9} 
\end{figure}

With a sufficient number of training samples, the neural network can generate a precise mapping between the loading configurations and the corresponding PCA weights. However, the resulting estimation can be affected by the number of eigen-images used to express that image. Fig.~\ref{fig:DataDrivenTopOpt:Fig7} illustrates the performance of our approach on several test samples for a specific design domain (middle configuration in Fig.~\ref{fig:DataDrivenTopOpt:Fig9}). In this configuration, the PCA and neural network are trained using $400$ samples and tested with randomly generated loading configurations. As previously mentioned, only the first $80$ eigen-vectors are used for reconstruction and estimation. Fig.~\ref{fig:DataDrivenTopOpt:Fig7} shows the original samples and their corresponding estimations using our method. Note that estimation involves using the neural network to map the loading configuration into the PCA weight vector, followed by a PCA based reconstruction using $80$ samples. Besides the image differences, compliance value given in \eqref{eq:DataDrivenTopOpt:EqObj} can also be compared to evaluate the performance. In Fig.~\ref{fig:DataDrivenTopOpt:Fig7}, the  table shows the compliance values obtained for regular topology optimization result and our estimation together with the image differences. Here, it can be observed that image difference directly reflects the percent change in the compliance value. Hence, one can judge the performance by only checking the image difference computed with our metric given in \eqref{eq:DataDrivenTopOpt:distanceMetric}. 

In Fig.~\ref{fig:DataDrivenTopOpt:Fig8}, we compare the performance of neural network versus linear regression  and quadratic regression formulated in \eqref{eq:DataDrivenTopOpt:linearRegress} and \eqref{eq:DataDrivenTopOpt:quadraticRegress} on a sample test image. All of the mapping methods are trained with the same $400$ samples. Note that linear and quadratic regression fail to reproduce some of the details in the optimal topology. It can also be observed that the quadratic regression performs even worse than linear regression for this specific test case. Fig.~\ref{fig:DataDrivenTopOpt:Fig8} also demonstrates the numerical values for image differences (from the original), compliance values and computation times for training and estimation. Although neural network performs significantly better than the other methods, there is a time trade-off because of the training process. The training and testing of each mapping method is conducted on a PC with a 2.4GHz Core CPU and 8GB RAM using MATLAB R2014b.  

\begin{table}[h]\small
\renewcommand{\arraystretch}{1.2}
\begin{center}
	\begin{tabular}{c||c|c|c}
	& \multicolumn{3}{ c }{\bf{Configuration}} \\ \cline{2-4} 	
  	\bf{Sample} & \bf{Left} & \bf{Middle} & \bf{Right} \\ \hline \hline
  	\bf{Test 1} & $0.30 \times 10^{-3}$ s & $0.34 \times 10^{-3}$ s & $0.36 \times 10^{-3}$ s \\ \hline
  	\bf{Test 2} & $0.41 \times 10^{-3}$ s & $0.39 \times 10^{-3}$ s & $0.30 \times 10^{-3}$ s \\ \hline
  	\bf{Test 3} & $0.38 \times 10^{-3}$ s & $0.34 \times 10^{-3}$ s & $0.29 \times 10^{-3}$ s \\
  	\hline \hline
  	\bf{Training Time} & 485.09 s & 363.18 s & 625.19 s \\
	\end{tabular}
\end{center}
\captionsetup{justification=centering}
\caption{Computation times for the tests cases demonstrated in Fig.~\ref{fig:DataDrivenTopOpt:Fig9}.}
\label{tab:DataDrivenTopOpt:Table1}
\end{table}

Fig.~\ref{fig:DataDrivenTopOpt:Fig9} illustrates the performance of our algorithm on several test configurations. In the top row, basic representations of the design problems are illustrated. Here, boundary conditions and loading configurations are shown. Red areas represent spaces where forces can be placed. In the following row, reconstructions of several test samples using eigen-images are presented. Since only the first $80$ eigen-images are used to construct an image, there are slight differences from the original optimal topologies. We compare our estimation results with the optimal topologies in the third row. As the design problem becomes more complex, the accuracy of resulting estimations reduces since the number of training samples for neural network may fail to be sufficient. Better estimations can be made using a higher number of training samples for more complex problems. In the last row of Fig.~\ref{fig:DataDrivenTopOpt:Fig9}, a histogram showing the distribution of error between the optimal topology and the estimation result of our approach among $100$ test samples is given for each design configuration. Although we use a limited number of training samples and PCA components, the main structure for optimal topologies can be estimated. Computation times obtained for the same test samples are shown in Table~\ref{tab:DataDrivenTopOpt:Table1} together with the neural network training times.     

\begin{figure}[t]
\centering
\includegraphics[trim = 0in 4.9in 2.1in 0in, clip, width=\textwidth]{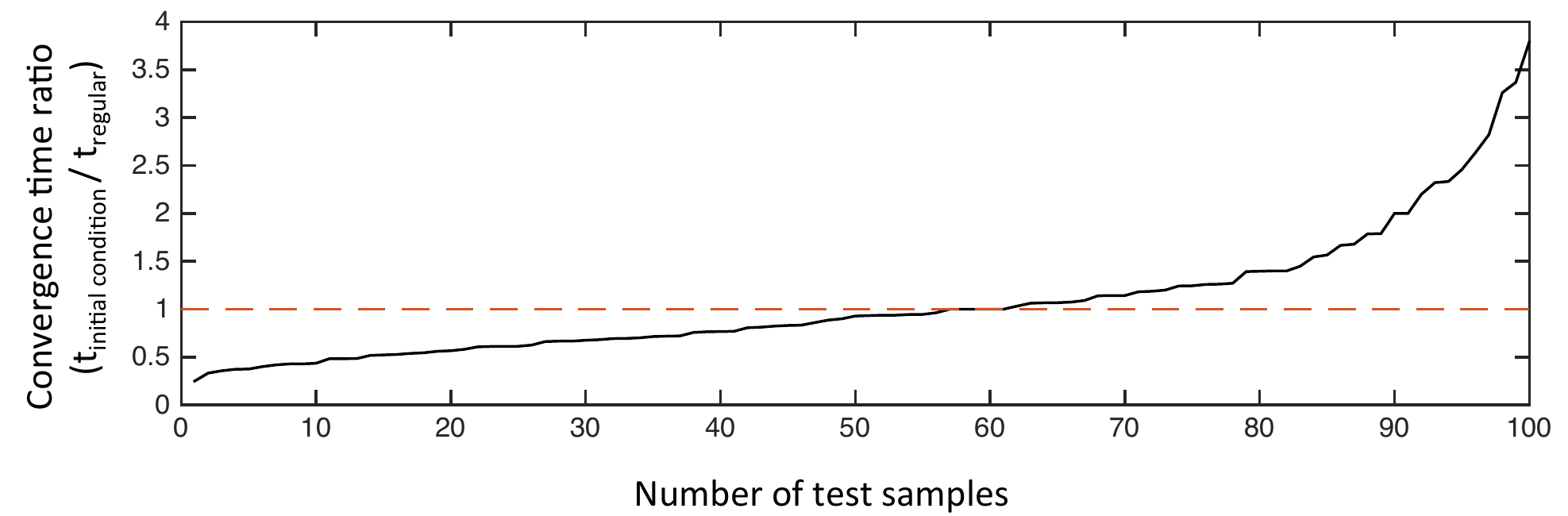} 
\caption{Convergence time comparison. Time of convergence when neural network estimation is used as initial condition ($t_{initial~condition}$) is shorter than that of regular topology optimization ($t_{regular}$) for most of the test samples. Configuration used: \protect{Fig.~\ref{fig:DataDrivenTopOpt:Fig9}} (left).}
\label{fig:DataDrivenTopOpt:Fig10} 
\end{figure}

In design problems involving complex loading configurations, the reconstruction accuracy may decrease visually (e.g., estimations in the last column of Fig.~\ref{fig:DataDrivenTopOpt:Fig9}). However, even in those cases, our optimal topology estimates can be used as an initial condition for a conventional topology optimization algorithm, e.g. \cite{andreassen2011efficient,sigmund2001a99}, to reduce the convergence time. Fig.~\ref{fig:DataDrivenTopOpt:Fig10} shows the effect of using our estimation as an initial condition. For around $70\%$ of 100 test samples, reduction in convergence time is observed for the posed problems. This gain can be even more significant for larger and more complex design domains.   

\section{Conclusions}

We explore the feasibility and performance of a data-driven approach to topology optimization problems involving structural mechanics. We take a set of optimal topology examples for a given configuration, and project them into a lower dimensional space with PCA analysis. We then learn a mapping from loading configurations to optimal topologies using neural networks. Using the trained network, we studied the performance of estimating optimal topologies for novel loading configurations. Our results show that the proposed method can successfully predict the optimal topologies in different problem settings. Moreover, we also prove that the topologies predicted by the proposed method are effective initial conditions for faster convergence in subsequent topology optimization. We believe such time and computational power savings will be greater as the problem size and complexity increase. Thus, a valuable future direction   is the application of the proposed method for 3D topology optimization.

\chapter[Structure Design Under Force Location Uncertainty]{Structure Design Under Force Location Uncertainty}
\label{chp:ForceLocationUncertainty}
\blindfootnote{This chapter is based on Ulu \etal, 2017 \cite{ulu2017lightweight}.}

We introduce a lightweight structure optimization approach for problems in which there is uncertainty in the force locations. Such uncertainty may arise due to force contact locations that change during use or are simply unknown a priori. Given an input 3D model, regions on its boundary where arbitrary normal forces may make contact, and a total force-magnitude budget, our algorithm generates a minimum weight 3D structure that withstands any force configuration capped by the budget. Our approach works by repeatedly finding the most critical force configuration and altering the internal structure accordingly. A key issue, however, is that the critical force configuration changes as the structure evolves, resulting in a significant computational challenge. To address this, we propose an efficient critical instant analysis approach. Combined with a reduced order formulation, our method provides a practical solution to the structural optimization problem. We demonstrate our method on a variety of models and validate it with mechanical tests. 

\section{Introduction} 
\label{sec:ForceLocationUncertainty:Introduction}

\begin{figure*}
\centering
\includegraphics[width = \textwidth]{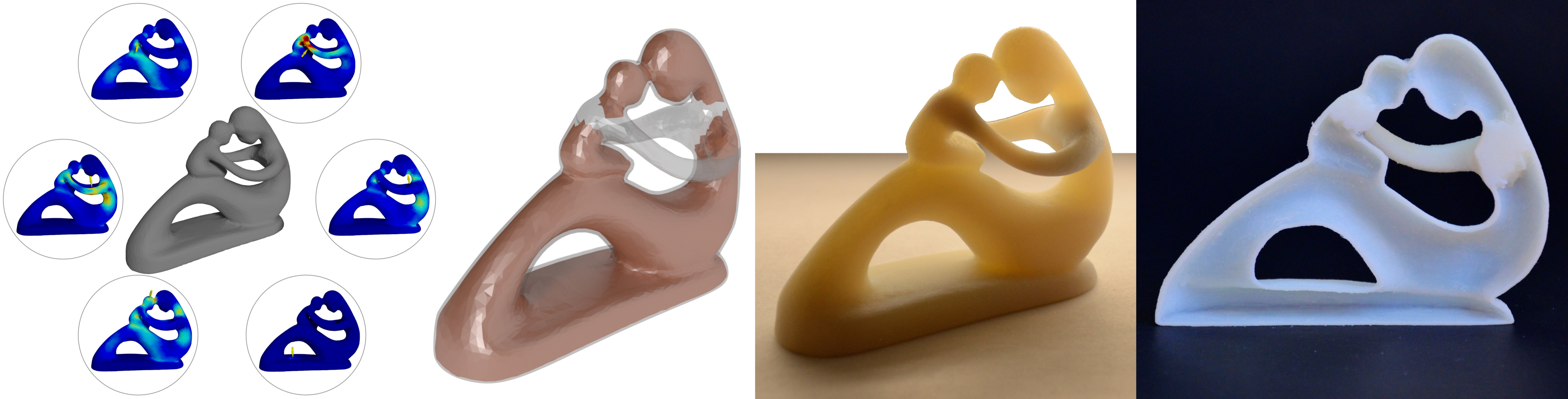}
\caption{We present a method for lightweight structure design for scenarios where external forces may contact an object at a multitude of locations that are unknown a priori. For a given surface mesh (grey), we design the interior material distribution such that the final object can withstand  all external force combinations capped by a budget. The red volume represents the carved out material, while the remaining solid is shown in clear.  Notice the dark material concentration on the fragile regions of the optimum result in the backlit image. The cut-out shows the corresponding  interior structure of the 3D printed optimum.}
\label{fig:ForceLocationUncertainty:teaser}
\end{figure*} 

With the emergence of additive fabrication technologies, structural optimization and lightweighting methods have become increasingly ubiquitous in shape design~\cite{stava2012stress,wang2013cost,lu2014build,christiansen2015combined}. In many such methods, a common approach is to model the external forces as known and fixed quantities. In many real world applications, however, the external forces' contact locations and magnitudes may exhibit significant variations during the use of the object. In such cases, existing techniques are either not directly applicable, or require the designer to make overly conservative simplifications  to account for the uncertainty in the force configurations \cite{choi2002structural}.

We propose a new method for designing minimum weight objects when there exists uncertainty in the external force locations. Such uncertainties may arise in various contexts such as (\textit{i})~multiple force configuration problems where the object experiences a large set of known force configurations such as those arising in machinery, (\textit{ii})~unknown force configuration problems where the location of the contact points may change nondeterministically such as consumer products that are handled in a multitude of ways, or  (\textit{iii})~moving contact problems where a contact force travels on the boundary of an object; such as automated fiber placement manufacturing or cam-follower mechanisms. 

Our approach takes as input (1) a 3D shape represented by its boundary surface mesh, (2) a user-specified \textit{contact region}; a subset of the boundary where external forces may make contact, and (3) a \textit{force-budget}; a maximum cap on the total summed magnitude of the external forces at any given time instance, and produces a minimum weight 3D structure that withstands any force configuration capped by the budget (Figure~\ref{fig:ForceLocationUncertainty:teaser}).

For structural optimization with force location uncertainties, a seemingly reasonable approach would be to compute an optimal structure for every possible force configuration and select the \textit{best} structure at the end. However, this strategy fails to guarantee that the final structure (or any other optimum structure computed along the way) is safe under any force configuration other than the one it was computed for \cite{banichuk2013introduction}. Therefore, at a minimum, finding the best structure requires validating each optimum structure against all possible force configurations. Unfortunately, even this strategy does not guarantee that a solution exists within the set of computed optima.

Our approach overcomes these challenges using a \emph{critical instant analysis} which efficiently determines the most critical force contact location responsible for creating the highest stress within the current shape hypothesis. This capability enables each step of the shape optimization to efficiently determine the maximum possible stress  that can be generated under the force budget and accordingly design the material distribution against failure. Our approach preserves the outer shape through an ingrown boundary shell while optimization removes material from the inside. We do not permit structural alterations to the exterior of the object for strengthening. Hence, our approach is clearly not useful in cases where material failure occurs even in the fully solid version of the object.

Our main contributions are:

\begin{itemize}
\item a novel formulation for  structural optimization problems under force location uncertainty,
\item a method we call \textit{critical instant analysis} that identifies the critical load instant quickly,
\item a practical reduced order lightweighting method using the above two ideas.
\end{itemize}

\section{Problem Formulation}
Our design problem aims to find an optimal material distribution inside the boundary surface mesh $\mathcal{S}_0$ parametrized by the discretized volumetric mesh. Similar to topology optimization \cite{bendsoe2003topology}, material design \cite{skouras2013computational,xu2015interactive} and microstructure design \cite{schumacher2015microstructures} approaches, each element in the discretized domain is associated with a design variable $\rho_e$ representing whether  element $e$ is full ($\rho_e = 1$) or void ($\rho_e = 0$).
To overcome the computational barriers introduced by binary variables, we adopt the common approach of allowing $\rho_e \in [0,1]$ and penalize the intermediate values during  optimization
\cite{bendsoe1989optimal}.
We assume linear isotropic materials and small deformations. The elemental stiffness matrix $\boldsymbol{K}_e$ can be related to $\rho_e$ and the stiffness matrix for base material $\boldsymbol{K}_e^{solid}$ as

\begin{equation}
\boldsymbol{K}_e = \boldsymbol{K}_e^{void} +\rho_e^ \beta (\boldsymbol{K}_e^{solid} - \boldsymbol{K}_e^{void}).
\label{Eq:ForceLocationUncertainty:SIMP}
\end{equation}

Here, $\beta$ is a penalization factor and $\boldsymbol{K}_e^{void} = \epsilon \boldsymbol{K}_e^{solid}$  is the stiffness matrix assigned to the void regions to avoid singularities in FEA. We use $\epsilon = 10^{-8}$ and $\beta = 3$.
In \eqref{Eq:ForceLocationUncertainty:SIMP}, $\boldsymbol{K}_e^{solid}$ is constant for each element and is computed as 

\begin{equation}
\boldsymbol{K}_e^{solid} = V_e \boldsymbol{B}_e^T \boldsymbol{C}_e^{solid} \boldsymbol{B}_e,
\label{Eq:ForceLocationUncertainty:elementStiffness}
\end{equation}

where $V_e$ is volume of the element, $\boldsymbol{B}_e$ is the strain-displacement matrix that depends only on the element's rest shape and $\boldsymbol{C}_e^{solid}$ is the elasticity tensor constructed using the Young's modulus and Poisson's ratio of the base material.
Given a volumetric mesh $\mathcal{V}$ with $m$ elements, one can assemble $\rho_e$ into vector $\boldsymbol{\rho} \in \mathbb{R}^m$ and construct the global stiffness matrix $\boldsymbol{K}(\boldsymbol{\rho})$ in order to determine the displacements $\boldsymbol{u}$ from $\boldsymbol{K} \boldsymbol{u} = \boldsymbol{f}$, where $\boldsymbol{f}$ is the nodal force vector. Then, the stress-displacement relationship can be written as

\begin{equation}
\boldsymbol{\sigma} = \boldsymbol{C_g} \boldsymbol{B} \boldsymbol{u}, 
\label{Eq:ForceLocationUncertainty:stressFormula}
\end{equation}

where $\boldsymbol{\sigma} \in \mathbb{R}^{6m}$ captures the unique six elements of the elemental stress tensor
and $\boldsymbol{B}$ is the assembly of $\boldsymbol{B}_e$ matrices. Block-diagonal matrix $\boldsymbol{C}_g \in \mathbb{R}^{6m \times 6m}$ is constructed with elemental elasticity tensors $\boldsymbol{C}_e(\boldsymbol{\rho})$ on the diagonal.
For each element, $\boldsymbol{C}_e$ can be computed analogous to $\boldsymbol{K}_e$ in \eqref{Eq:ForceLocationUncertainty:SIMP}.
While applicable to different element types, we use linear tetrahedral elements making $\boldsymbol{K}(\boldsymbol{\rho}) \in \mathbb{R}^{3n \times 3n}$, $\boldsymbol{u} \in \mathbb{R}^{3n}$, $\boldsymbol{f} \in \mathbb{R}^{3n}$ and $\boldsymbol{B} \in \mathbb{R}^{6m \times 3n}$ for a volume mesh having $n$ nodes. 

The  approach formulated in~\eqref{Eq:ForceLocationUncertainty:SIMP}-\eqref{Eq:ForceLocationUncertainty:stressFormula} is useful because it preserves the same discretization throughout the optimization.
Additionally, it is amenable to model reduction presented in Section~\ref{sec:ForceLocationUncertainty:ModelReduction} for more efficient iterations (at the expense of reduced degrees freedom). 

\subsection{Force Model} 

\begin{figure}
\centering
\includegraphics[trim = 0in 0in 0in 0in, clip, width = 0.4\textwidth]{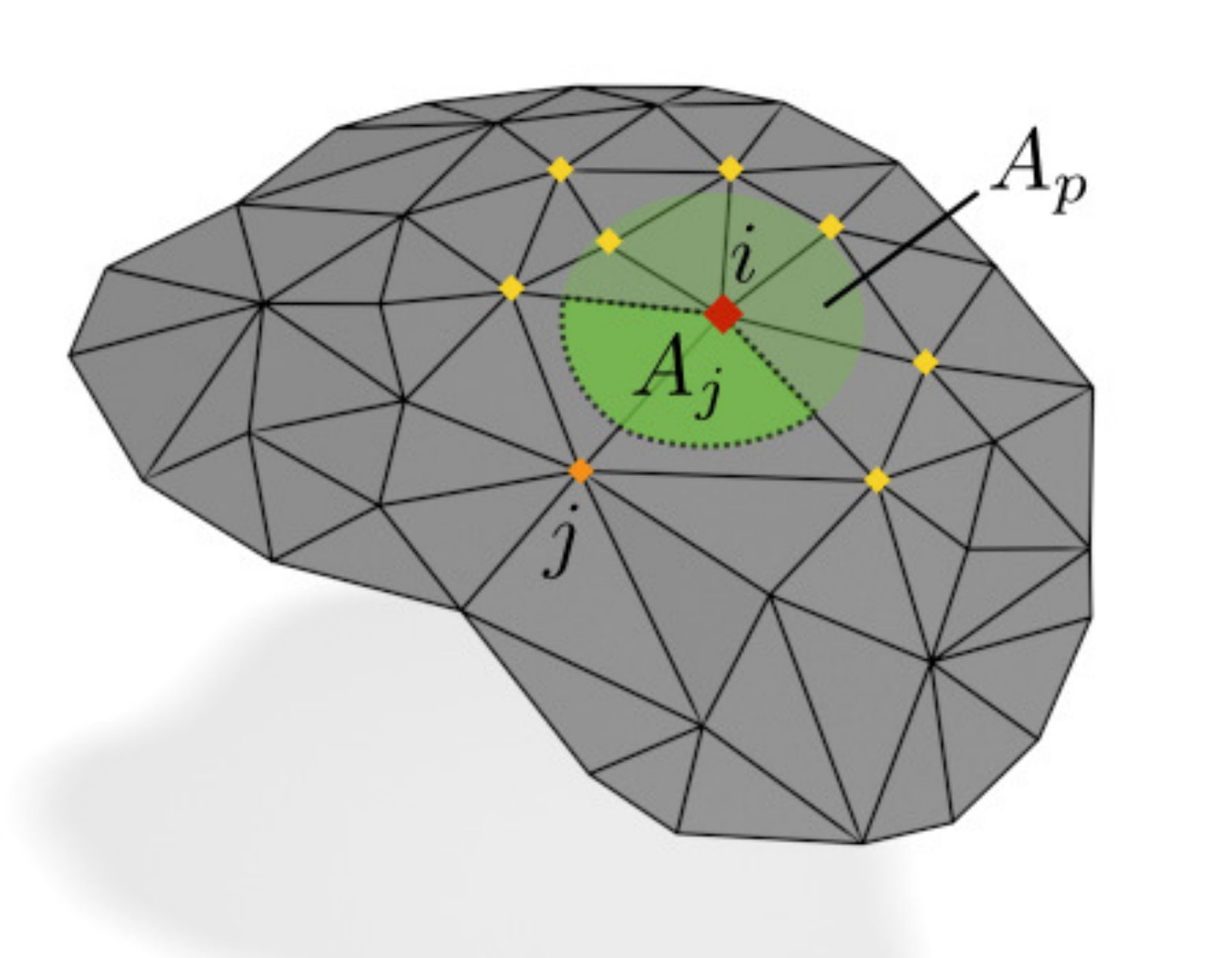}
\caption{The point force on node $i$ is spread in a circular area $A_p$. Nodal forces are computed using \eqref{Eq:ForceLocationUncertainty:areaForce}. Highlighted nodes, including $i$ and $j$, have non-zero nodal forces.}
\label{fig:ForceLocationUncertainty:areaForces}
\end{figure}

External forces are allowed to make contact within a user-specified union of contact regions $\mathcal{S}_L \subseteq \mathcal{S}_0$. To avoid stress singularities (\ie unbounded stresses under a point force), we distribute the force to a small circular area $A_p$ (with radius $r_p$) around the contact point. Then, we construct the force vector $\boldsymbol{f}$ by computing the nodal forces as

\begin{equation}
\boldsymbol{p}_{j} = -P(A_j/3A_p)\boldsymbol{n}_i,
\label{Eq:ForceLocationUncertainty:areaForce}
\end{equation}    

where $\boldsymbol{p}_{j}$ is the force vector at node $j$ when a force with magnitude $P$ is applied to node $i$.
The area $A_j$ is the portion of $A_p$ covered by triangles adjacent to node $j$ and $\boldsymbol{n}_i$ is the surface normal at node $i$ (Figure \ref{fig:ForceLocationUncertainty:areaForces}).
Our approach approximates $A_p$ by intersecting the boundary mesh with a sphere of radius $r_p$ centered at node $i$. Congruent with our earlier problem description of normal contact forces only, \eqref{Eq:ForceLocationUncertainty:areaForce} assumes the force is applied  compressively along the surface normal direction. This formulation thus neglects friction and excludes forces that pull on the surface. Note, however, that most real-world contact scenarios such as handling a part or the contacts within an assembly can be modeled with compressive normal forces developing between interacting bodies.

To anchor the object in space, we require that the mesh is fixed at three or more non-collinear boundary nodes. Boundary constraints remain unchanged during  optimization. 

In this work, we use the von Mises failure criterion. For linear elastic structures, the stress is a linear function of dispacement and the displacement at any given point is a linear function of the force vector. Similarly, the force vector is a linear combination of point normal forces in the contact region. Thus, the stress at a point within the structure can be determined by a superposition of each point force's contribution~\cite{hibbeler2015structural}, thereby making the von Mises stress convex in the applied force. For the force budget $f_B$, the space of allowable forces is defined by $ \| \boldsymbol{p}_{j} \| >0 $ and $ \sum \| \boldsymbol{p}_{j} \| < f_B $, \ie~a simplex $\mathcal{C}$ with vertices of the form $f_B$ for j'th coordinate and zero for the rest and the coordinate origin. By Rockafellar's Theorem 32.2 \cite{rockafellar2015convex}, supremum of a convex function on a convex set $\mathcal{C}$ is attained at one of the points in the set enclosed by the convex hull $\mathcal{C}$. Therefore, von Mises stress at a point will be maximized by spending the entire force budget at a specific point. The same principal holds for the maximum of the von Mises stress over the whole object as a maximum of a set of convex functions is convex. As a result, at the heart of our approach is the search for this most critical contact point given a shape hypothesis (Section~\ref{sec:ForceLocationUncertainty:Critical Instant Analysis})

\subsection{Optimization Problem}

We tackle the following stress-constrained mass minimization problem

\begin{equation}
\begin{aligned}
& \underset{\boldsymbol{\rho}}{\text{minimize}}
& & M(\boldsymbol{\rho})  = \sum_{e=1}^{m} \rho_{e} V_e \\
& \text{subject to}
& & \boldsymbol{K}(\boldsymbol{\rho}) \boldsymbol{u}_i = \boldsymbol{f}_i \quad \forall i \in \mathcal{S}_L, \\
& & & \sigma_{cr}(\boldsymbol{\rho}) \leq \sigma_y,\\
& & & \boldsymbol{0} \leq \boldsymbol{\rho} \leq \boldsymbol{1}.
\end{aligned}
\label{Eq:ForceLocationUncertainty:optimizationProblem}
\end{equation}

Here, $\boldsymbol{f}_i$ and $\boldsymbol{u}_i$ represent the nodal force  and  displacement vectors when the external force is applied to surface node $i$.
The object fails if the maximum stress ever exceeds the yields strength $\sigma_y$.
Hence, we define the critical stress $\sigma_{cr}$ as

\begin{equation}
\sigma_{cr} = \underset{\boldsymbol{i}}{\text{max}}( \underset{\boldsymbol{e}}{\text{max}}(\enskip \sigma_e^{vm}\enskip)) \quad \forall i \in \mathcal{S}_L \enskip \text{and} \enskip \forall e \in \mathcal{V},
\label{Eq:ForceLocationUncertainty:sigmaCritical}
\end{equation}

where $\sigma_e^{vm}$ is the von Mises stress computed for element $e$. 

\begin{figure*}
\centering
\includegraphics[width = \textwidth]{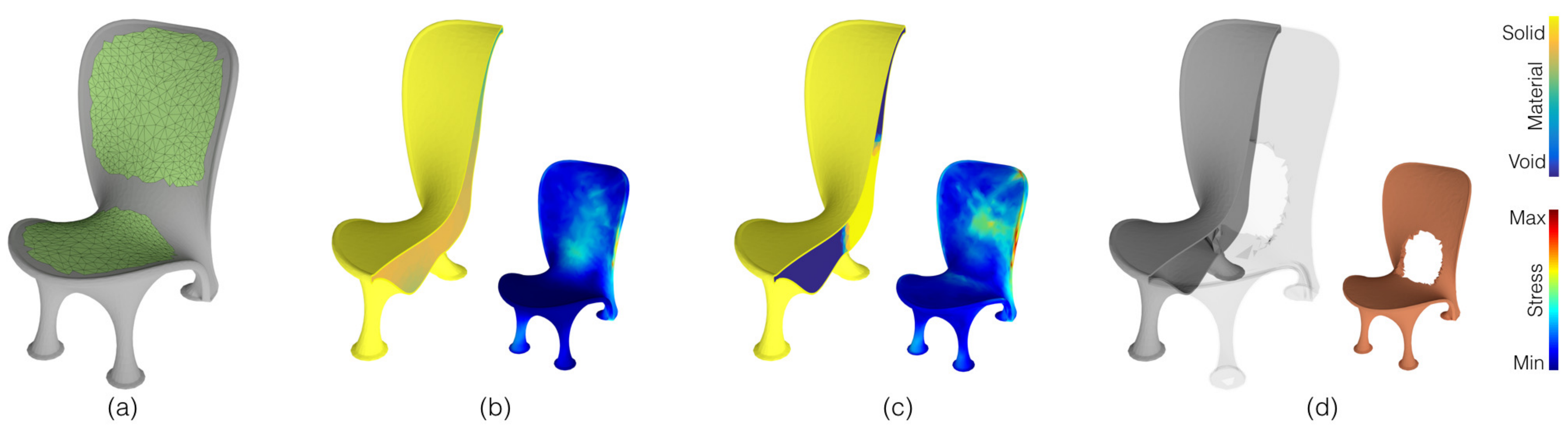}
\caption{Given a contact region (green in a), our algorithm optimizes the interior material distribution (b-c) to find the smallest weight structure (d) that can withstand all possible force configurations. In (b-c), we show the material distribution in two steps of the optimization. Inset figures illustrate the stress distributions for the most critical force instants in (b-c) and the removed material in (d).}
\label{fig:ForceLocationUncertainty:overview}
\end{figure*}

\section{Algorithm}

For force instant $i$, all elemental von Mises stresses $\sigma_e^{vm}$ (hence the maximum) within the object  can be computed using \eqref{Eq:ForceLocationUncertainty:stressFormula} with a single linear solve. However, finding the maximum across all possible instants require as many FEA solves as the number of instants. This number can be large, especially for  structures where $\mathcal{S}_L$ consist of many  nodes.
In such cases, computing the critical stress can be  costly, making shape optimization prohibitively expensive.
We next describe our approach to addressing this problem.

\subsection{Overview}

Figure~\ref{fig:ForceLocationUncertainty:overview} illustrates our approach. From an input 3D shape and prescribed contact regions (Figure~\ref{fig:ForceLocationUncertainty:overview}(a)),
we optimize the material distribution. At each step governed by the current material distribution, we compute the critical stress by efficiently finding the most critical force instant. We call this process \textit{critical instant analysis}. In this analysis, we reduce the search space by computing a set of \textit{force regions} (FR) over the contact region $\mathcal{S}_L$  and \textit{weak regions} (WR) within the entire structure  $\mathcal{V}$.
FRs are a subspace of the surface that are likely to contain the critical force instant. Likewise, WRs are the regions where the maximum stress is likely to occur.  We efficiently find the critical force instant within FRs using a reduced number of FEA evaluations dictated by the number of vertices within FRs. Then, optimization updates the material distribution to minimize mass (Figure~\ref{fig:ForceLocationUncertainty:overview}(b-c)).
At the end, a minimum weight structure satisfying the imposed constraints is obtained (Figure~\ref{fig:ForceLocationUncertainty:overview}(d)). Algorithm~\ref{alg:ForceLocationUncertainty:ourAlgorithm} summarizes our approach. Note that the material distribution is updated only once at each optimization step based on the computed gradients.

\begin{algorithm}
 \SetAlgoLined
 \SetKwInOut{Input}{Input}\SetKwInOut{Output}{Output}
 \Input{$\mathcal{S}_0$ and $\mathcal{S}_L$}
 \Output{Optimized structure}
 \While{Mass is reduced}{
  Compute force regions (FRs)\;  
  Compute weak regions (WRs)\;
  \For{each FR}
  {
  	Perform a hierarchical search to find largest stress at WRs\;
  }
  Choose the maximum stress across all FRs as the critical stress $\sigma_{cr}$\;
  Update material distribution $\boldsymbol{\rho}$\;
 }
 \caption{Our structure optimization algorithm}
 \label{alg:ForceLocationUncertainty:ourAlgorithm}
\end{algorithm}

\subsection{Critical Instant Analysis}
\label{sec:ForceLocationUncertainty:Critical Instant Analysis}

\begin{figure*}
\centering
\includegraphics[width = \textwidth]{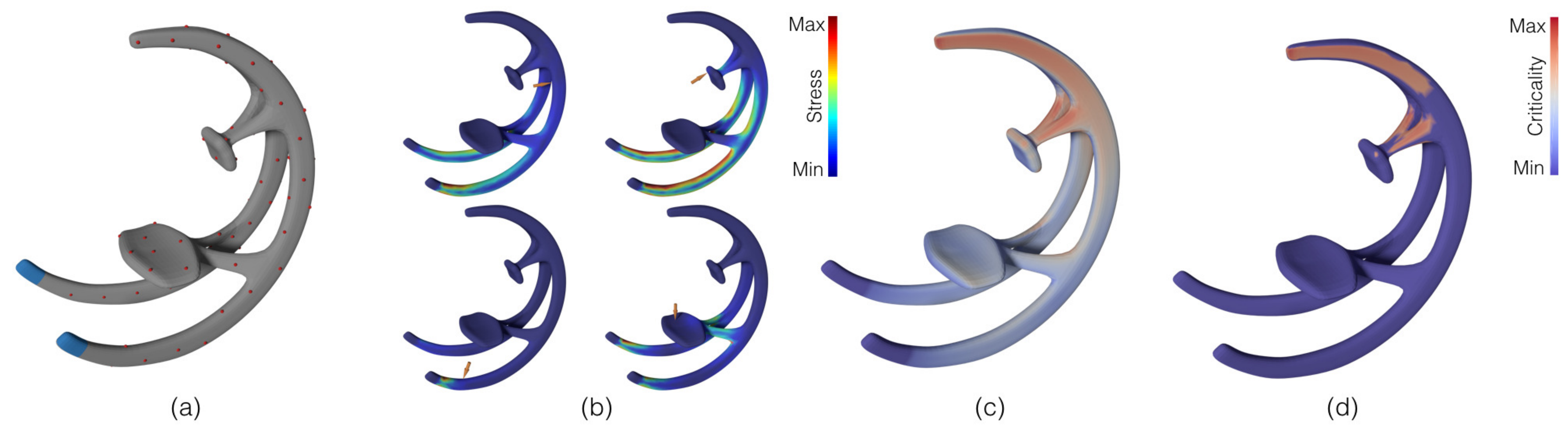}
\caption{
Given a structure represented by the material distribution, we uniformly sample a number of force instants on $\mathcal{S}_L$ (a) and perform FEAs to obtain corresponding the stress distributions (b).
We then use quadratic regression to estimate the stress distributions for the remaining force instants and construct the criticality map (c).
Areas with high criticality constitutes our force regions (d).
The blue regions in (a) represent the fixed boundary condition and the remainder of the boundary surface forms $\mathcal{S}_L$.
}
\label{fig:ForceLocationUncertainty:criticalityMapOverview}
\end{figure*}

Critical instant analysis finds the most critical force instant and the corresponding stress $\sigma_{cr}$ with an order of magnitude fewer FEA evaluations compared to a brute-force approach.

\paragraph{Force Regions} 

For a structure represented by material distribution $\boldsymbol{\rho}$, certain force configurations will cause the largest stresses in the body. We compute these critical force locations as our force regions and restrict the search space for $\sigma_{cr}$ in \eqref{Eq:ForceLocationUncertainty:sigmaCritical} to a smaller space $\mathcal{S}_{fr} \subset \mathcal{S}_L$.

Figure~\ref{fig:ForceLocationUncertainty:criticalityMapOverview} illustrates our approach to compute FRs.
We start by estimating a \textit{criticality map} on $\mathcal{S}_L$, which captures the severity of the force instants. The higher stress a force instant creates, the more critical it is deemed. Thus, the criticality of a force instant is simply the maximum stress it creates in the object.

To efficiently acquire the entire criticality map, we perform FEAs  only for a small number of force instants and estimate the stress distributions for remaining force instants by learning a mapping between the nodal forces and the resulting stress distributions.
We  sample the force instants across $\mathcal{S}_L$ by performing k-means clustering using the farthest-first traversal initialization and selecting the center points of the resulting clusters as our sample instants. We use approximate geodesic distances \cite{crane2013geodesics} as the distance metric in clustering.

Suppose we have $l$ training samples and the boundary mesh $\mathcal{S}_0$ consists of $s$ nodes. In its original form, the sample force instant $\boldsymbol{f_i}$ is represented as a sparse vector of size $3n$ and the corresponding von Mises stress forms a vector of size $m$.
With a small number of training samples ($l \ll 3n$ and $l\ll m$), it is not possible to represent the relationship between two high dimensional data using a simple mapping function.
Moreover, in its sparse form, $\boldsymbol{f_i}$ is devoid of any spatial information relevant to the corresponding force instant. Hence, in this representation, two spatially proximate force instants that likely create similar stress distributions can be as distinct  as two spatially distant force instants. To  reduce the dimensionality of the force space and  establish proximity, we transform and project the sparse force vectors using the surface Laplacian. We stack the magnitudes of forces on boundary surface nodes (\ie $\| \boldsymbol{p} \|$ in \eqref{Eq:ForceLocationUncertainty:areaForce}) into row vectors $\boldsymbol{f'_i}$ of length $s$.
We assemble the mean centered $\boldsymbol{f'_i}$ into a $(l \times s)$ matrix $\boldsymbol{F}$ so that each row is $\boldsymbol{f'_i} - \bar{\boldsymbol{f}}$ where $\mean{\boldsymbol{f}}$ is the average of $\{\boldsymbol{f'_i}\}$.
We then compute the Laplacian basis functions $\boldsymbol{\psi_j}$ as the eigenvectors of the surface graph Laplacian $\boldsymbol{\mathcal{L}}_s \in \mathbb{R}^{s \times s}$.
We assemble the first $q$ eigenvectors to form our lower dimensional basis matrix $\boldsymbol{\Psi} = [ \boldsymbol{\psi_1}, \boldsymbol{\psi_2}, \ldots \boldsymbol{\psi_q} ]$. The lower dimensional representation of the force instants can then be written as 

\begin{equation}
\boldsymbol{F}_L = \boldsymbol{F} \boldsymbol{\Psi},
\label{Eq:ForceLocationUncertainty:forceInLaplacian}
\end{equation} 

where $\boldsymbol{F}_L$ becomes an $(l \times q)$ matrix. 

Similarly, we use principal component analysis (PCA) to project the stress data onto a lower dimensional space.
We assemble the mean centered stress vectors into an $(l \times m)$ matrix $\boldsymbol{T}$.
A PCA on $\boldsymbol{T}$ yields $(l-1)$ principal vectors of size $m$.
Then, each stress vector can be approximated by $(l-1)$ PCA weights through a linear combinations of the principal vectors 

\begin{equation}
\boldsymbol{T}_L = \boldsymbol{T} \boldsymbol{\Phi}
\label{Eq:ForceLocationUncertainty:stressInPCA}
\end{equation}

where $\boldsymbol{T}_L$ is $(l \times l-1)$ matrix storing the PCA weights for each sample in its rows and $\boldsymbol{\Phi}$ is the assembly of principal vectors.

Lower dimensionality in $\boldsymbol{F}_L$ and $\boldsymbol{T}_L$ allows us to learn a simple mapping between the two spaces with a reasonable computational cost. We have found that quadratic regression with L2 regularization performs sufficiently well for capturing the relationship between the PCA weights of the stress vectors and the reduced dimensional force vectors such that $\boldsymbol{T}_L = \hat{\boldsymbol{F}_L} \boldsymbol{\mathcal{W}}$. Here, $\hat{\boldsymbol{F}_L}$ is $(l \times (q^2 + 3q +2)/2)$ matrix including the quadratic terms for $\boldsymbol{F}_L$. In matrix form, the coefficient matrix can be computed as

\begin{equation}
\boldsymbol{\mathcal{W}} = ({\hat{\boldsymbol{F}_L}}^{T}\hat{\boldsymbol{F}_L}+r \boldsymbol{I})^{-1}({\hat{\boldsymbol{F}_L}}^{T}\boldsymbol{T}_L)
\end{equation}

where $r$ is a small number controlling the importance of the regularization term. Using this map, we estimate the criticality of a new force instant by computing the stress distribution it creates through the quadratic map, and extracting its maximum. Note that the estimated stress vectors are only an approximation of the actual values, thus cannot be used directly for $\sigma_{cr}$. However, they provide strong guidance in estimating the location of the most critical force instant. We thus use the synthesized criticality map to determine the force regions (Figure~\ref{fig:ForceLocationUncertainty:criticalityMapOverview}(d)). The connected components of high criticality areas in $\mathcal{S}_L$ comprise our force regions.

\begin{figure}
\centering
\includegraphics[width = \columnwidth]{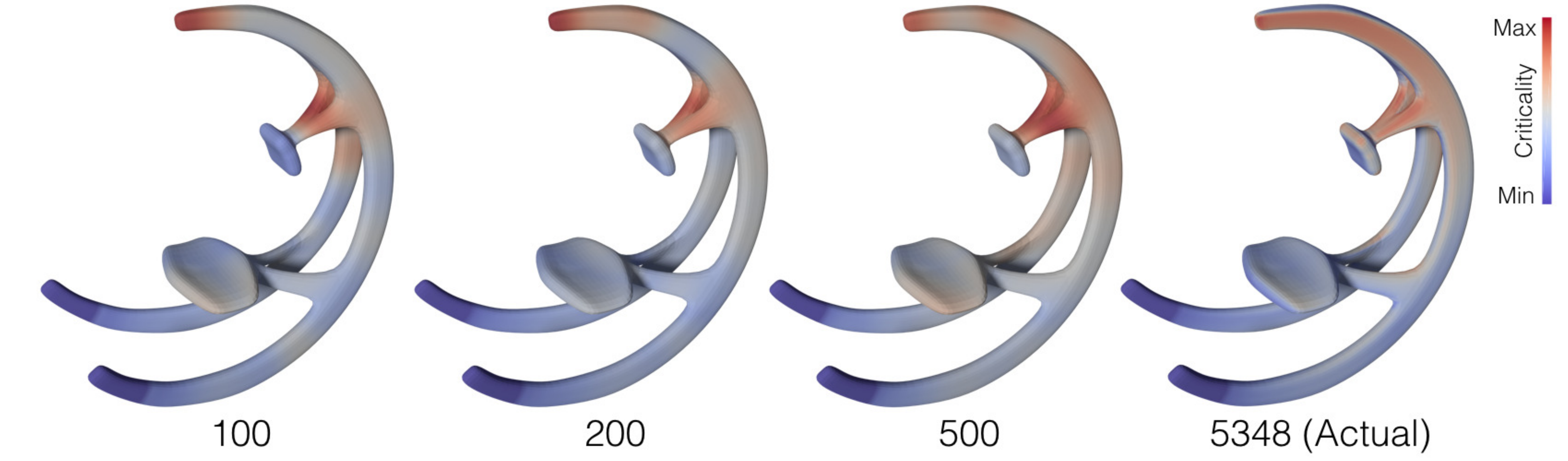}
\caption{Criticality maps as a function of the number of samples. With more samples, the estimated criticality map converges to the actual map, but at an increased cost.}
\label{fig:ForceLocationUncertainty:ActualVSEstimateCriticalityMap}
\end{figure}

\begin{figure}
\centering
\includegraphics[width = 0.7\textwidth]{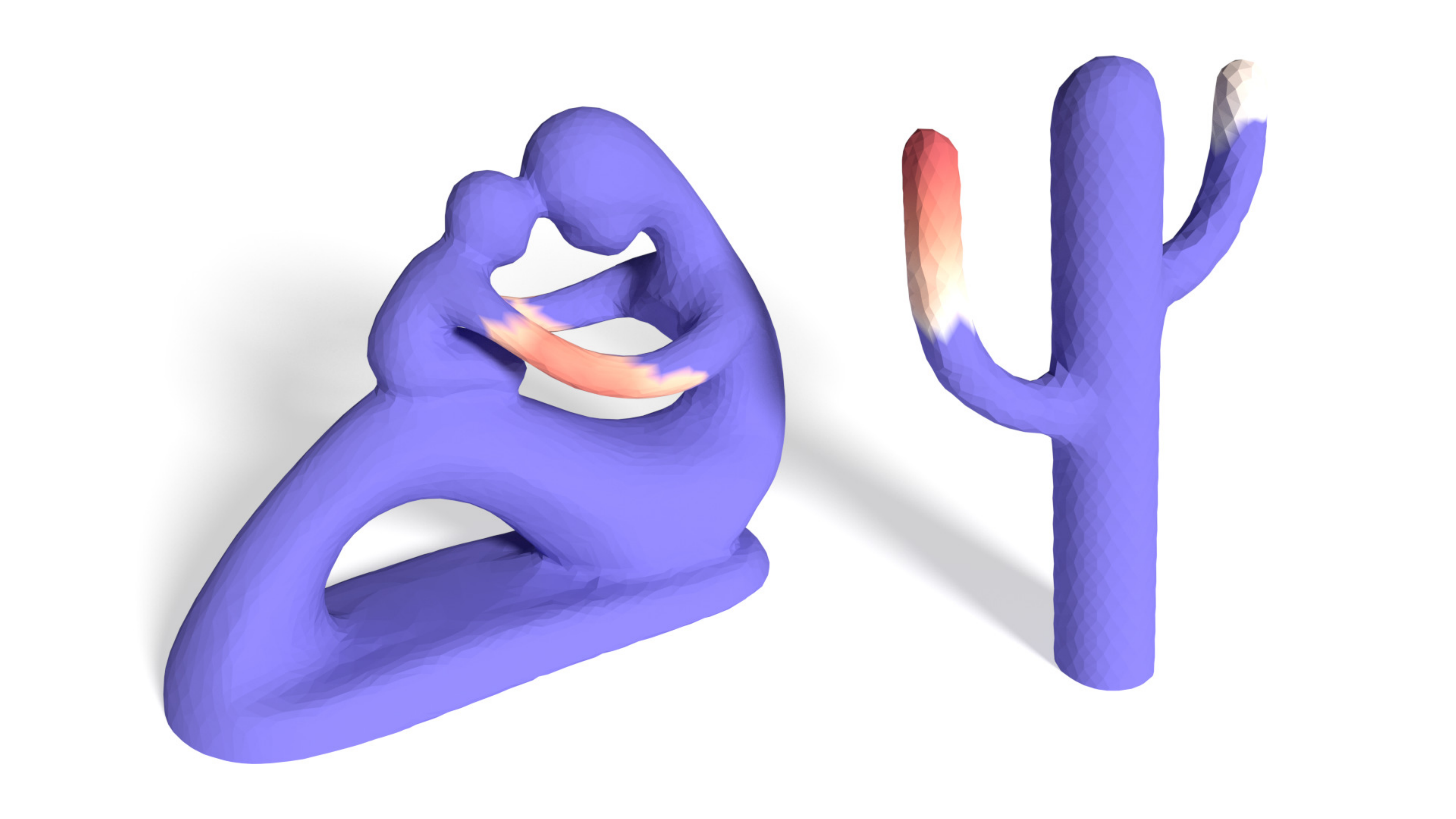}
\caption{Force regions computed for two fully solid models. We take connected components in high criticality regions as our FRs (red-white areas).}
\label{fig:ForceLocationUncertainty:FRs}
\end{figure}

The accuracy of the criticality map depends on the number of training samples $l$. While a large number of samples increases accuracy, the computational cost also increases proportionally. Figure~\ref{fig:ForceLocationUncertainty:ActualVSEstimateCriticalityMap} illustrates the criticality maps obtained with different number of training samples. We observed that using $5\%$ of the nodes in $\mathcal{S}_L$ for training produces an acceptable approximation of the criticality map such that in all of our examples, after the criticality values are estimated, top $10\%$ of the nodes with the largest criticality values always contain the ground truth critical instant. Figure~\ref{fig:ForceLocationUncertainty:FRs} illustrates the force regions we obtained for two different models. 

\paragraph{Weak Regions}

For a structure with material distribution $\boldsymbol{\rho}$, we need to find the maximum stress produced by each force instant to determine $\sigma_{cr}$ in \eqref{Eq:ForceLocationUncertainty:sigmaCritical} and solve \eqref{Eq:ForceLocationUncertainty:optimizationProblem}.
However, failure often occurs at certain regions of the object leaving the remainder safe at all times.
WRs help us constrain the regions of the structure where we seek the maximum stress $\mathcal{V}_{wr} \subset \mathcal{V}$.

In our algorithm, we use an approach similar to \cite{zhou2013worst} to determine the possible failure locations as our weak regions.
We determine WRs using modal analysis that involves solving the generalized eigenvalue problem

\begin{equation}
\lambda_j \boldsymbol{M}_g(\boldsymbol{\rho}) \boldsymbol{u}_j = -\boldsymbol{K}(\boldsymbol{\rho}) \boldsymbol{u}_j, \quad j=1,2,\ldots,m
\label{Eq:ForceLocationUncertainty:modalAnalysis}
\end{equation}  

where $\boldsymbol{u}_j$ is $j$'th eigenmode, $\lambda_j$ is the corresponding eigenvalue and $\boldsymbol{M}_g(\boldsymbol{\rho})$ is the mass matrix for the tetrahedral mesh.
Note that WRs are structure dependent, hence need to be updated at each step of the optimization.
To reduce computational cost, we use a lumped mass matrix and distribute each element's mass to its nodes equally, thus creating a sparse diagonal matrix $\boldsymbol{M}_g~\in~\mathbb{R}^{3n \times 3n}$. 

\begin{figure}
\centering
\includegraphics[width = 0.7\textwidth]{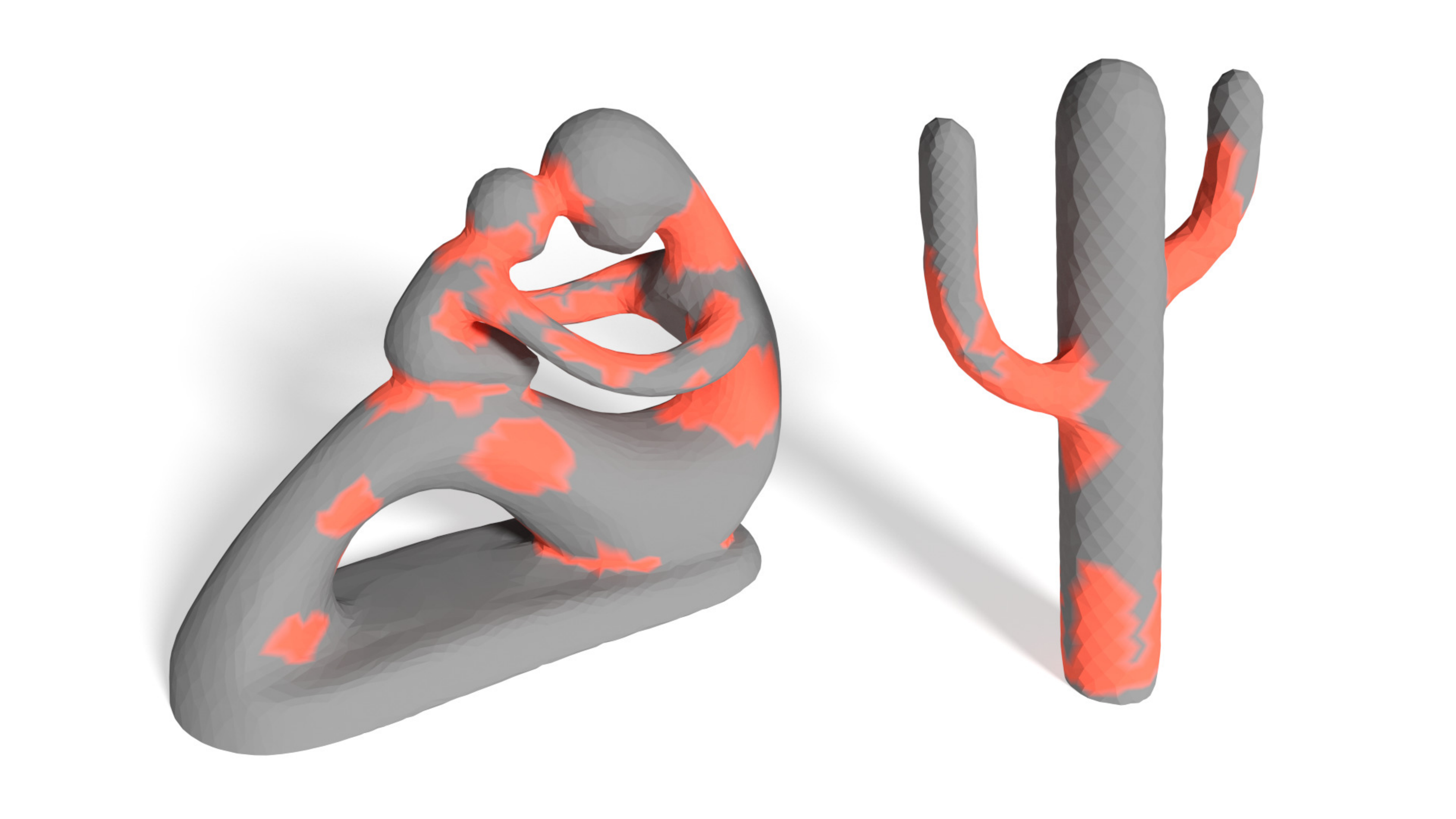}
\caption{Weak-regions are found around thin parts or sharp edges where stress concentration is likely to occur. Both are fully solid models and are fixed at the bottom.}
\label{fig:ForceLocationUncertainty:WRs}
\end{figure}

WRs can be extracted by computing the low frequency eigenmodes in \eqref{Eq:ForceLocationUncertainty:modalAnalysis} and identifying the nodes that experience large stresses under these deformations. In our examples, we use the first $15$ vibration modes and $2.5\%$ of the most stressed nodes to form our WRs. Different from \cite{zhou2013worst}, we combine all unique nodes obtained from modal analysis to construct the WRs. Figure \ref{fig:ForceLocationUncertainty:WRs} shows the WRs we obtained for two fully solid models ($\rho_e = 1 \enskip \forall e$). Note that WRs are found around possible stress concentration points such as thin parts and crease edges.

\paragraph{Critical Instant}

We solve \eqref{Eq:ForceLocationUncertainty:sigmaCritical} for $\sigma_{cr}$ in a much smaller domain defined by the FRs $\mathcal{S}_{fr} \subset \mathcal{S}_{L}$ and WRs $\mathcal{V}_{wr} \subset \mathcal{V}$. In particular, only the force instants captures with FRs are used, and maximum stresses are only sought in WRs.

We use a simple greedy hierarchical search to find the force instant creating the largest stress in the structure. For each FR island, we partition it into four segments and perform FEA by applying the force to their central nodes. We then further select and partition the segment that produces the highest stress within WRs, and repeat this process until converging to a single node. After repeating this process for all FR islands, we choose the maximum stress across all FR islands as $\sigma_{cr}$ for that particular optimization step. The stiffness matrix $\boldsymbol{K}$ and $\boldsymbol{C}_g$ need to be computed only once during these evaluations as the structure remains unchanged.  We thus factorize $\boldsymbol{K}$ once for each optimization step, and determine displacements and stresses using efficient forward and backward substitutions.

\subsection{Stress Singularity}
\label{sec:ForceLocationUncertainty:Stress Singularity}

Stress constrained mass minimization problems are prone to singularity issues \cite{sved1968structural,kirsch1990on,lee2012stress}.
For instance, in  \eqref{Eq:ForceLocationUncertainty:optimizationProblem}, the global optimum is obtained when $\boldsymbol{\rho}=\boldsymbol{0}$, making all elements void. To mitigate this problem, we establish a layer of fully solid boundary shell that is excluded from optimization. This approach overcomes the singularity problem by coercing the  optimization to employ material in the remaining inner volume as a way relieve the high stresses generated on the boundary shell. Enforcing a boundary shell also preserves the original outer surface.

\begin{figure}
\centering
\includegraphics[width = 0.8\textwidth]{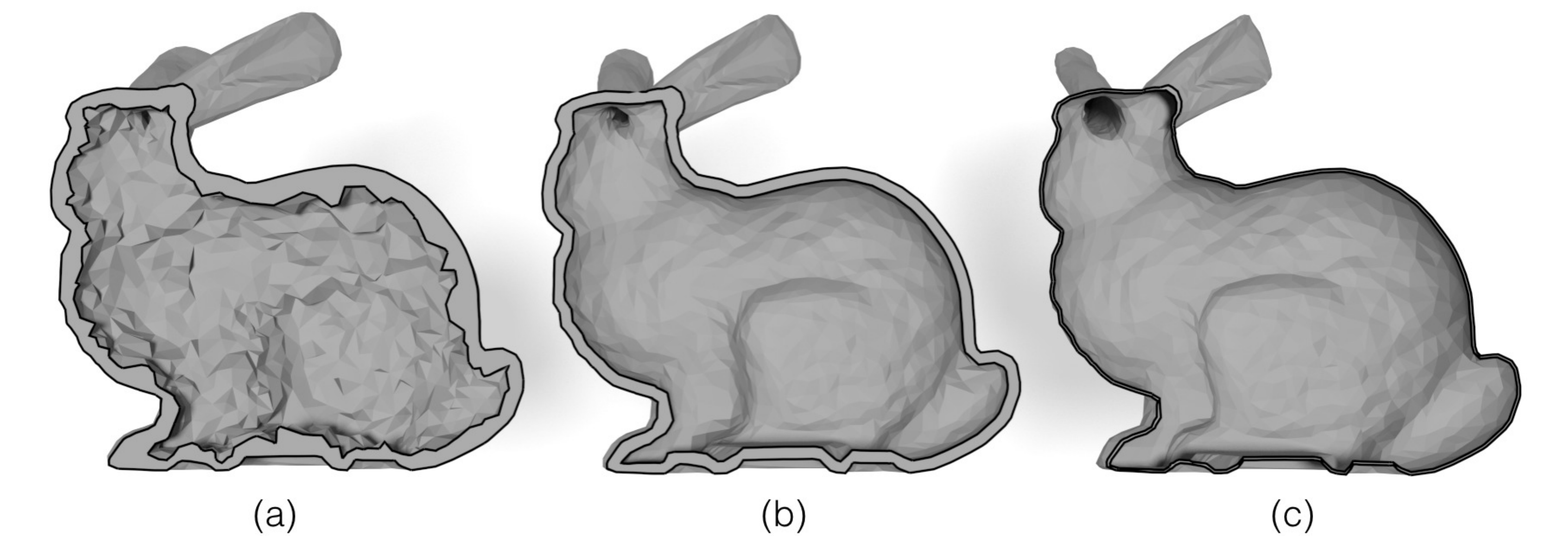}
\caption{To prevent undulations in the shell (a), we use offset surfaces to create a smooth boundary shell with prescribed thicknesses (b-c). In all three cases, the volumetric mesh has $29k$ elements.}
\label{fig:ForceLocationUncertainty:offsetSurfaces}
\end{figure}

All elements that contribute one or more vertices to the outer boundary $\mathcal{S}_0$ could serve as the boundary elements. For an arbitrary volumetric mesh, however, forming a solid shell using only these elements may introduce stress concentrations (Figure~\ref{fig:ForceLocationUncertainty:offsetSurfaces}(a)). To create a smooth and uniform thickness shell, we first generate an inner offset surface $\mathcal{S}_i$ using the method presented in \cite{musialski2015reduced}.
Then, we tetrahedralize the entire domain, which results in a smooth shell layer $\mathcal{V}_s \subset \mathcal{V}$ sandwiched by $\mathcal{S}_i$ and $\mathcal{S}_0$  (Figure~\ref{fig:ForceLocationUncertainty:offsetSurfaces}(b-c)). We use a uniform shell thickness prescribed by the user, which can be adjusted based on a 3D printer's minimum print thickness. Although we do not optimize the shell thickness, we discuss its effects in Section~\ref{sec:ForceLocationUncertainty:resultsAndDiscussion}.

\section{Model Reduction}
\label{sec:ForceLocationUncertainty:ModelReduction}

\begin{figure}
\centering
\includegraphics[width = 0.8\textwidth]{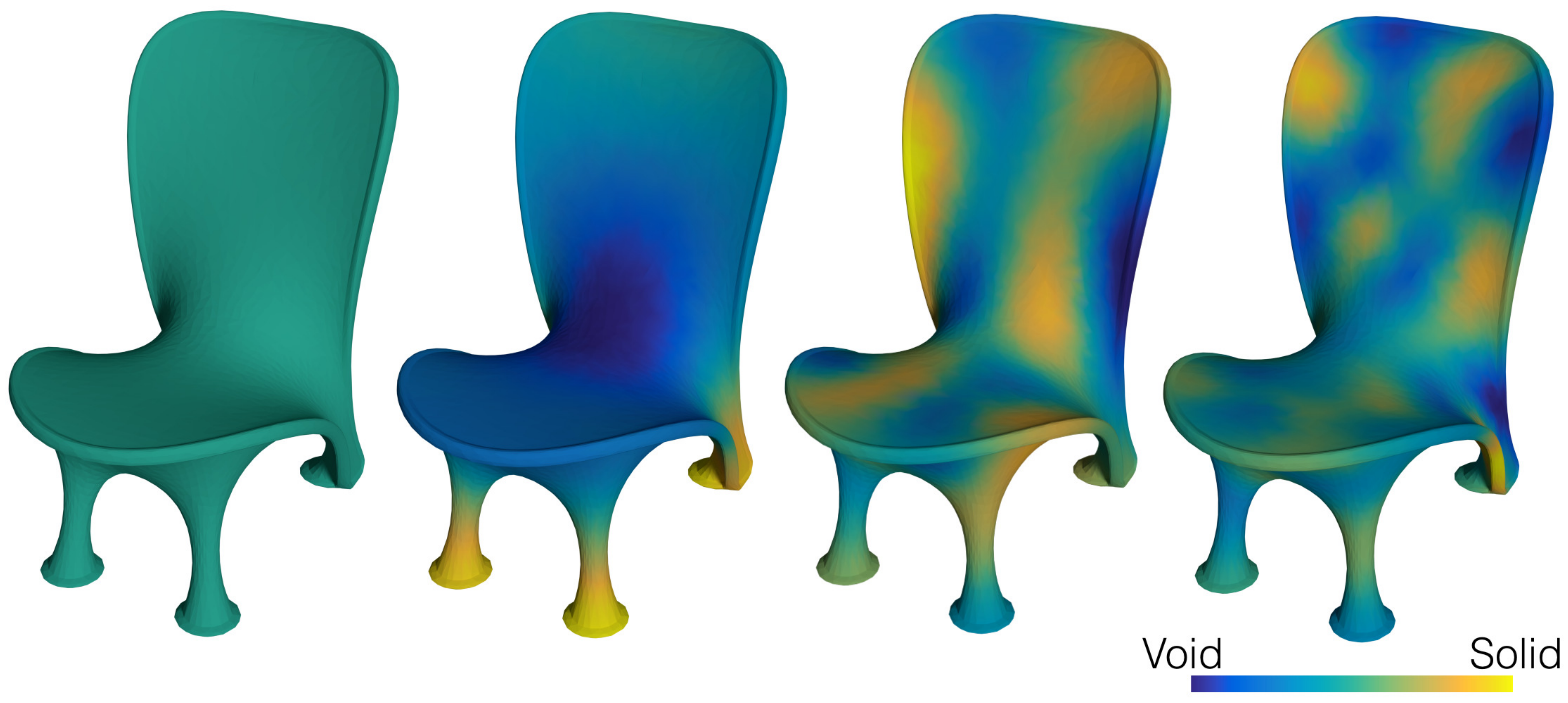}
\caption{Left to right: material distributions corresponding to $1st$, $5th$, $25th$ and $50th$ modes. Lower modes control global material distribution while higher modes enable local details.}
\label{fig:ForceLocationUncertainty:materialModes}
\end{figure}

The optimization problem \eqref{Eq:ForceLocationUncertainty:optimizationProblem} is typically very high-dimensional as the number of design variables is equal to the number elements in the volumetric mesh $\mathcal{V}$. This, in turn, has a significant impact on the computational performance. To accelerate optimization, we compute a set of material modes \cite{xu2015interactive}, which helps control the material distribution using only a small number of variables.  Material modes can be computed as the eigenvectors of the element-based graph Laplacian $\boldsymbol{\mathcal{L}} \in \mathbb{R}^{m \times m}$ defined on $\mathcal{V}$ by solving the generalized eigenvalue problem

\begin{equation}
\mu_j \boldsymbol{V} \boldsymbol{\gamma}_j = -\boldsymbol{\mathcal{L}} \boldsymbol{\gamma}_j, \quad j=1,2,\ldots,m
\label{Eq:ForceLocationUncertainty:materialModes}
\end{equation}

where $\mu_j$ are non-negative eigenvalues, $\boldsymbol{\gamma}_j$ are corresponding eigenvectors and $\boldsymbol{V} \in \mathbb{R}^{m \times m}$ is a diagonal matrix composed of $V_e$s.
Eigenvectors $\boldsymbol{\gamma}_j$ are orthogonal and smooth scalar functions that spectrally decompose the material distribution \cite{zhou2005large,zhang2010spectral}. The first  mode represents the homogeneous material distribution and while the level of detail control increases with higher frequencies (Figure~\ref{fig:ForceLocationUncertainty:materialModes}).
We assemble the first $k$ eigenvectors to form the reduced order basis  $\boldsymbol{\Gamma} = [\boldsymbol{\gamma}_1, \boldsymbol{\gamma}_2, \ldots \boldsymbol{\gamma}_k]$ so that the material distribution can be written as

\begin{equation}
\boldsymbol{\rho} = \boldsymbol{1} + \boldsymbol{\Gamma} \boldsymbol{\alpha},
\label{Eq:ForceLocationUncertainty:reducedOrderDensity}
\end{equation}

where $\boldsymbol{\alpha} = [\alpha_1, \alpha_2, \ldots \alpha_k]^T$ is the design vector for the reduced order problem.
This formulation allows us to trivially enforce fully solid material on the boundary shell elements by by setting the corresponding rows in the reduced  basis matrix $\boldsymbol{\Gamma}_e$ to be $\boldsymbol{0}$. This way the entries in $\boldsymbol{\alpha}$ can take on any value during the optimization without violating the geometrical constraints. 

\paragraph{Logistic Function} 
In our reduced order formulation, we use a logistic function $G(x)$ to penalize the intermediate values of $\rho_e$ by modifying \eqref{Eq:ForceLocationUncertainty:reducedOrderDensity} as

\begin{subequations}
\label{Eq:ForceLocationUncertainty:logistic}
\begin{align}
\boldsymbol{\rho} &= G(\boldsymbol{\Gamma} \boldsymbol{\alpha}) \label{Eq:ForceLocationUncertainty:logisticDensity}\\
G(x) &= 1/[1+e^{(\kappa(x-x_0))}]. \label{Eq:ForceLocationUncertainty:logisticFunction}
\end{align}
\end{subequations}

Here, $\kappa$ and $x_0$ determine the steepness and inflection point of the logistic function.
Note that $x_0$ should be adjusted to satisfy $G(\boldsymbol{0}) \approx \boldsymbol{1}$ to ensure that the elements on the boundary shell are solid.
While increasing $\kappa$  intensifies binarization by pushing the intermediate densities  toward $0$ and $1$, it also hampers convergence. We use $\kappa=5$ for all of our examples.   

In addition to binarizing the intermediate densities, the use of logistic function in \eqref{Eq:ForceLocationUncertainty:logisticDensity} guarantees $\rho_e \in [0,1]~\forall e \in \mathcal{V}$ for $-\infty<\boldsymbol{\alpha}<\infty$.
This allows us to remove a large number of constraints $\boldsymbol{0} \leq \boldsymbol{\rho} \leq \boldsymbol{1}$ from our reduced order optimization problem.

\subsection{Reduced Order Problem}
Applying \eqref{Eq:ForceLocationUncertainty:logistic} to our optimization formulation \eqref{Eq:ForceLocationUncertainty:optimizationProblem}, the reduced order optimization problem can be stated as

\begin{equation}
\begin{aligned}
& \underset{\boldsymbol{\alpha}}{\text{minimize}}
& & M(\boldsymbol{\alpha})  = G(\boldsymbol{\Gamma} \boldsymbol{\alpha}) \cdot \boldsymbol{V} \\
& \text{subject to}
& & \boldsymbol{K}(\boldsymbol{\alpha}) \boldsymbol{u}_i = \boldsymbol{f}_i \quad \forall i \in \mathcal{S}_{fr}, \\
& & & \sigma_{cr}(\boldsymbol{\alpha}) \leq \sigma_y,\\
\end{aligned}
\label{Eq:ForceLocationUncertainty:reducedOptimizationProblem}
\end{equation}
where 
\begin{equation}
\sigma_{cr} = \underset{\boldsymbol{i}}{\text{max}}( \underset{\boldsymbol{e}}{\text{max}}( \enskip \sigma_e^{vm} \enskip)) \quad \forall i \in \mathcal{S}_{fr} \enskip \text{and} \enskip \forall e \in \mathcal{V}_{wr},
\label{Eq:ForceLocationUncertainty:reducedSigmaCritical}
\end{equation}
for $\boldsymbol{\Gamma}_e = \boldsymbol{0} \quad \forall e \in \mathcal{V}_s$.

A benefit of the reduced order formulation is that the number of new design variables $k$ can be markedly small compared to the number of original variables $m$ in \eqref{Eq:ForceLocationUncertainty:optimizationProblem}. This number is independent of the input mesh and needs to be prescribed by the user. 
Because the structural optimization algorithm involves a large number of costly FEA evaluations per iteration, we found $k \leq 15$ to provide a favorable tradeoff between speed and expressiveness. Hence, we use $15$ material modes in all of our examples, unless otherwise stated. The optimization starts with a fully solid model ($\boldsymbol{\alpha} = \boldsymbol{0}$). Our approach is predicated on the assumption that this starting solution is feasible, hence amenable to lightweighting through optimization.

At the end of the optimization, the resulting material distribution may still contain elements with intermediate densities. In such cases, we  threshold the gray scale material distribution  to obtain a fully binarized solution. In our examples, we use $\rho=0.5$ as the threshold; elements with lower densities are set to void. Nonetheless, there exists more sophisticated thresholding methods \cite{hsu2005interpreting}. Another positive byproduct of the reduced order approach is that the resulting material distribution after binarization typically does not suffer from a checkerboard effect, as the reduction leads to smooth material modes in $\mathcal{V}$, especially for $k \ll m$. This, in turn, helps alleviate exhaustive post-processing.

Note that the material modes are precomputed as they depend only on $\mathcal{V}$. Because only the first $k$ modes are used in the reduced order formulation, they can be computed efficiently using iterative methods. We use ARPACK for this purpose \cite{lehoucq1998arpack}.
In order to solve \eqref{Eq:ForceLocationUncertainty:reducedOptimizationProblem}, we use sequential quadratic programming \cite{nocedal2006numerical}.
Our code solves linear systems using the Eigen library's SimplicialLDLT sparse solver \cite{eigenweb}.

Our hierarchical search method and density based shape representation allow us to compute gradients analytically. We compute gradients of the critical stress with respect to the design variables using the adjoint method \cite{Paris2010stress}. We use p-norm approximations ($p=15$) for the max functions. Details are given in Appendix~\ref{sec:ForceLocationUncertainty:Appendix}.

\begin{figure}
\centering
\includegraphics[width = 0.8\textwidth]{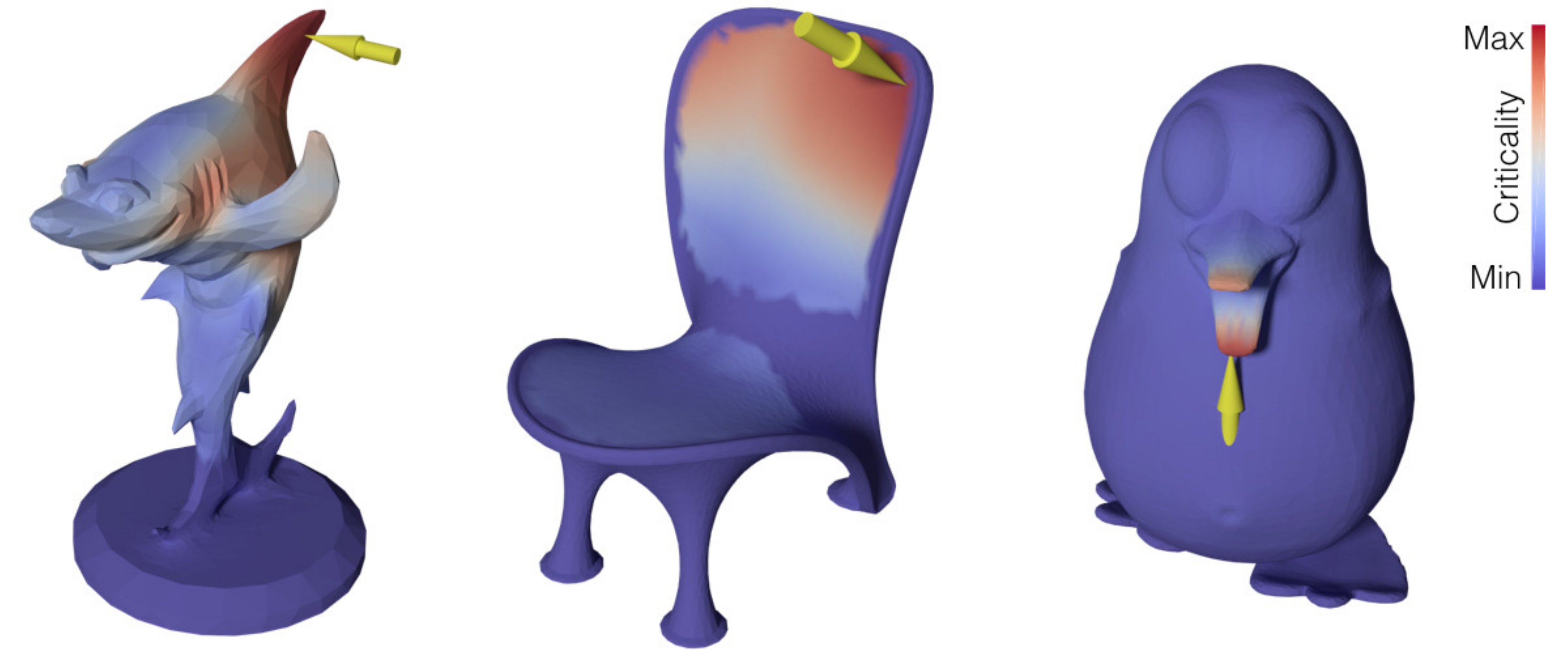}
\caption{Critical force instants are shown on estimated criticality maps. All models are fully solid and are fixed at the bottom. }
\label{fig:ForceLocationUncertainty:Analysis}
\end{figure}

\section{Results and Discussion}
\label{sec:ForceLocationUncertainty:resultsAndDiscussion}
   
\subsection{Criticality Analysis}

\begin{figure}
\centering
\includegraphics[width = 0.8\textwidth]{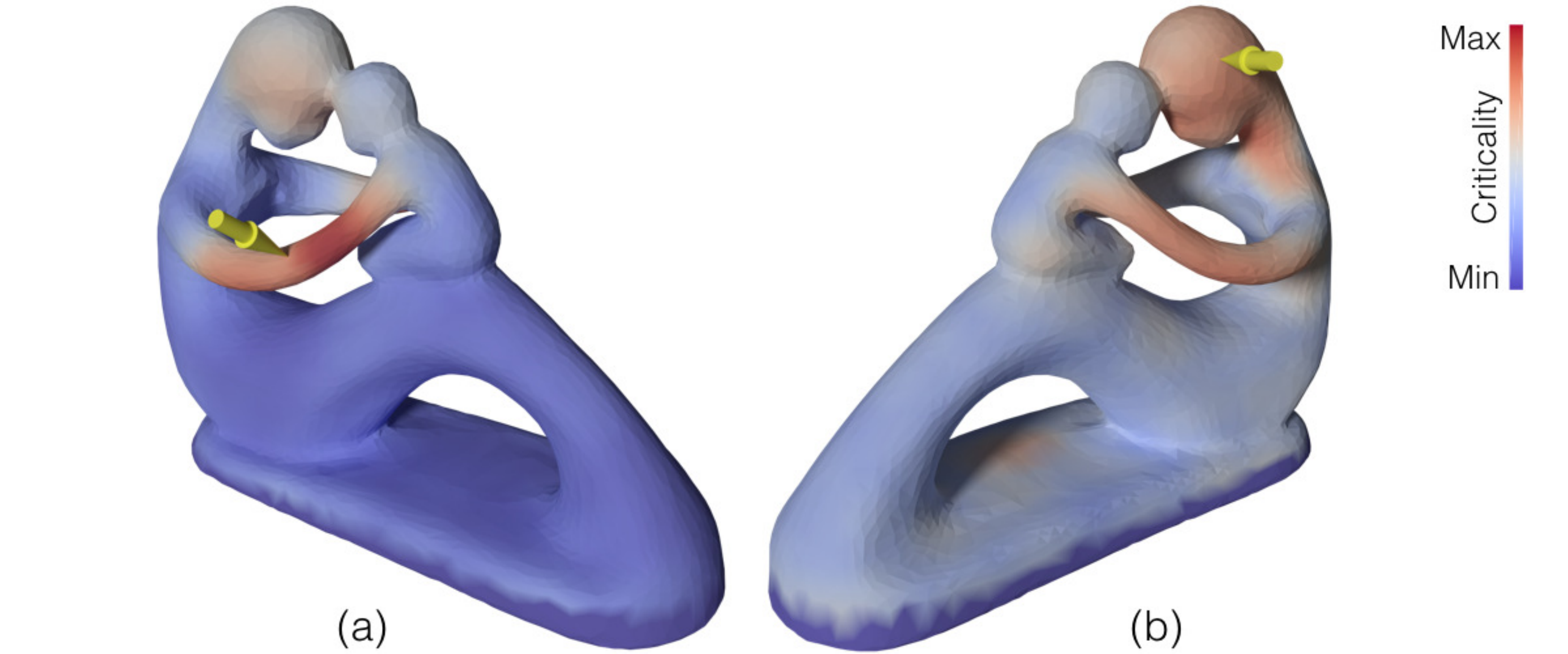}
\caption{Critical instant can change significantly during optimization. While forces around the arms are critical for the fully solid model (a), the critical instant shifts to the mother's temple as the structure is hollowed (b).}
\label{fig:ForceLocationUncertainty:AnalysisFertility}
\end{figure}

Figure~\ref{fig:ForceLocationUncertainty:Analysis} illustrates the results of our critical instant analysis  on a collection of fully solid models.
In all cases presented in this work, our approach involving criticality map estimation followed by a hierarchical search is able to determine the true critical force instant (yellow arrows). We verified this match using an expensive brute force search method. As shown in Figure~\ref{fig:ForceLocationUncertainty:Analysis}, our estimated criticality map captures the true critical instant quite well in that it finds the most critical force to be at the point with the highest estimated criticality (dark red). Nonetheless, our analysis can tolerate inaccuracies in the estimated criticality map. Figure~\ref{fig:ForceLocationUncertainty:AnalysisFertility}(a) shows such a case. Although the criticality map estimates the most critical point to be on the arm closer to the baby (dark red), our algorithm subsequently finds the true critical instant in that vicinity by searching an expanded area forming the force region.

\begin{figure}
\centering
\includegraphics[width = 0.7\textwidth]{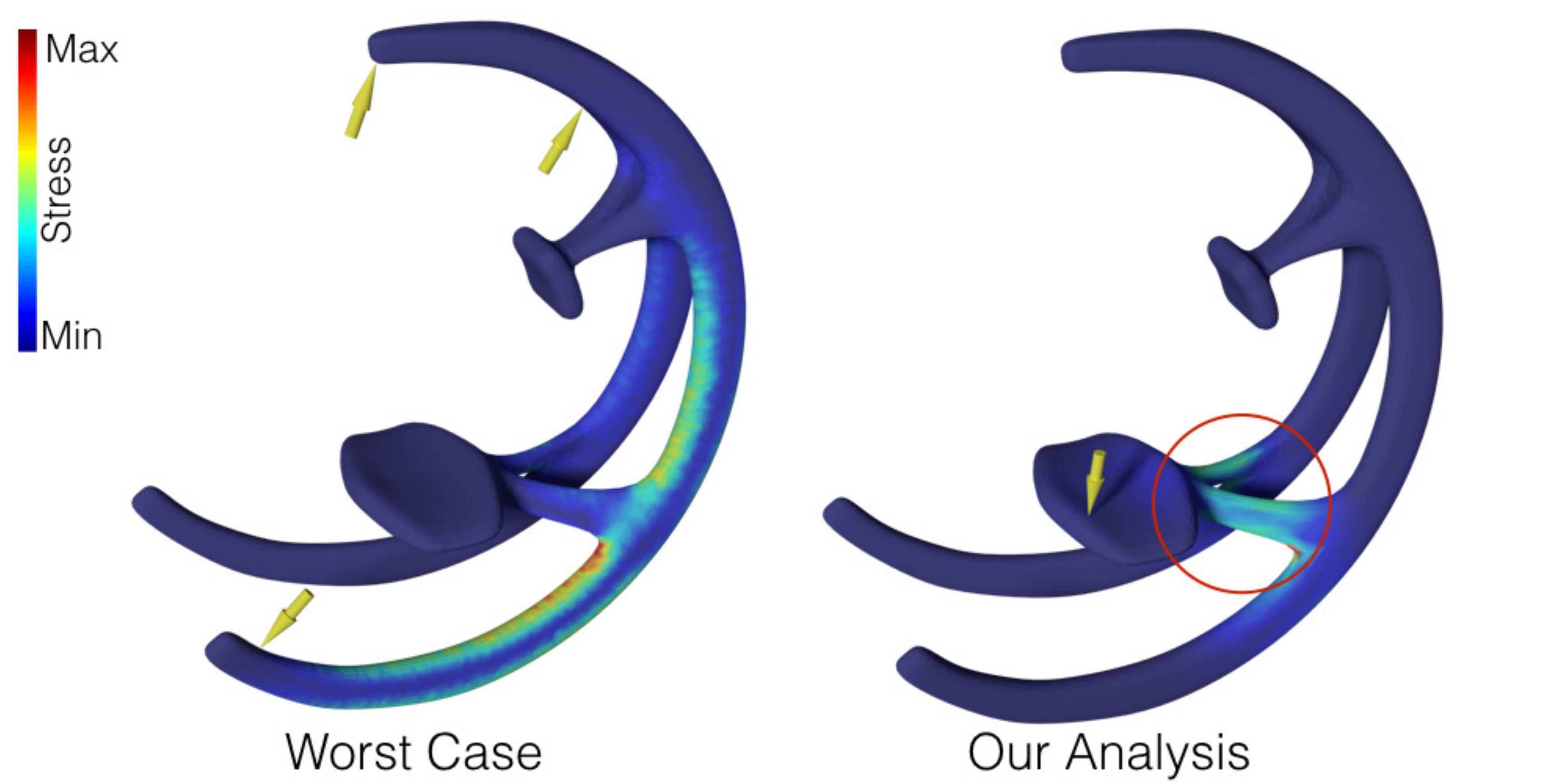}
\caption{A comparison between worst case structural analysis \protect\cite{zhou2013worst} and our method. 
While worst case analysis predicts the critical forces well for handling scenarios, it may miss contextually relevant critical instants during actual use cases. Our analysis, with delineated contact regions and boundary conditions,  predicts the critical instants to be around the seat of the rocker.}
\label{fig:ForceLocationUncertainty:WorstCaseVSOurs}
\end{figure}

Figure~\ref{fig:ForceLocationUncertainty:AnalysisFertility} illustrates the critical instants for two different material distributions.
Because we construct the criticality map for each step of the optimization, the force regions are updated and the change in the critical instant is captured  well. In all of our examples, the critical force instant is always contained in the identified force regions. 

\begin{figure}
\centering
\includegraphics[width = \columnwidth]{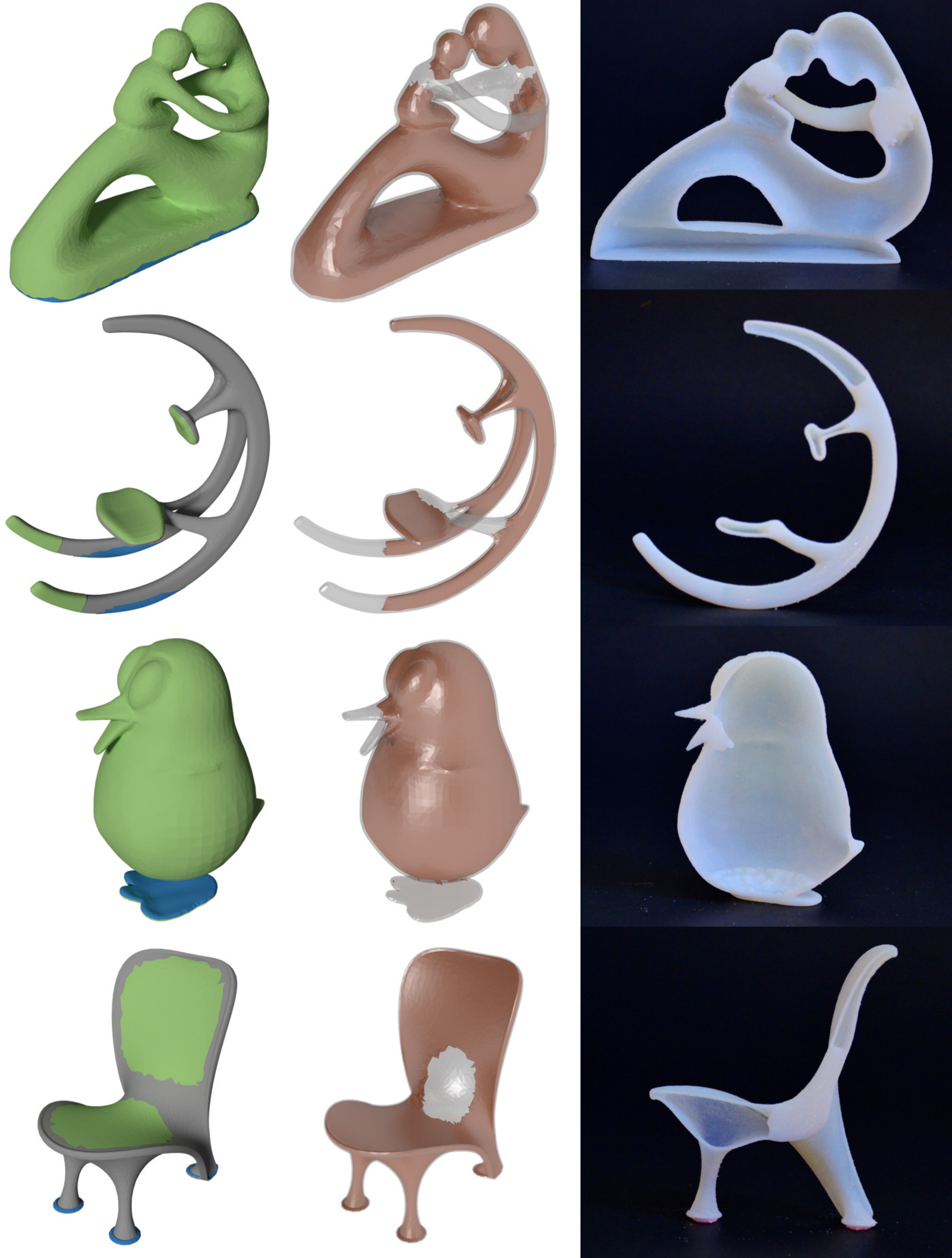}
\caption{Structural optimization results. Left-to-right, problem setups with fixed boundary conditions (blue) and contact regions (green), our optimized structures and 3D printed cut-outs of optimal results. Red shows the removed material.}
\label{fig:ForceLocationUncertainty:Results}
\end{figure}

\paragraph{Comparison} Figure~\ref{fig:ForceLocationUncertainty:WorstCaseVSOurs} compares our critical instant analysis with the worst-case structural analysis of Zhou \etal \cite{zhou2013worst}.
The worst-case structural analysis method is designed to predict the critical force configurations that develop during handling an object.
Thus, it may have a tendency to overpredict the weakness, resulting in overengineered solutions if used for structural optimization.
In our method, because we delineate the contact regions and the displacement constraints to reflect the knowledge of actual use, our approach captures the critical forces that are more likely to be encountered during a product's nominal use. For instance, the region encircled in Figure~\ref{fig:ForceLocationUncertainty:WorstCaseVSOurs} for our analysis shows the high stress region worst-case analysis fails to capture. The boundary conditions and the contact regions used in our analysis are shown in Figure~\ref{fig:ForceLocationUncertainty:Results}.

\subsection{Structural Optimization}

Figure~\ref{fig:ForceLocationUncertainty:Results} illustrates the results of our method on various 3D models. The displacement-constrained regions are shown in blue. The contact regions are shown in green. Our reduced order  optimization  detects the parts of the objects where high stresses may develop and distributes material accordingly. For objects with small weak regions such as the fertility and penguin models, our optimization allows us to preserve the same structural strength that their fully solid versions possess, while shedding a large portion of the mass.
Similarly, our algorithm performs well for models with thin elements such as the chair by utilizing mass around the stress carrying regions of the object.
Notice the arms supporting the seat in the rocking chair and the back support in the chair model.

In these examples, we  achieved $50\%$ to $90\%$ mass reduction. Table~\ref{tab:ForceLocationUncertainty:Results} summarizes the weight reduction together with various other metrics relevant to these models.

\begin{figure}
\centering
\includegraphics[width = 0.8\textwidth]{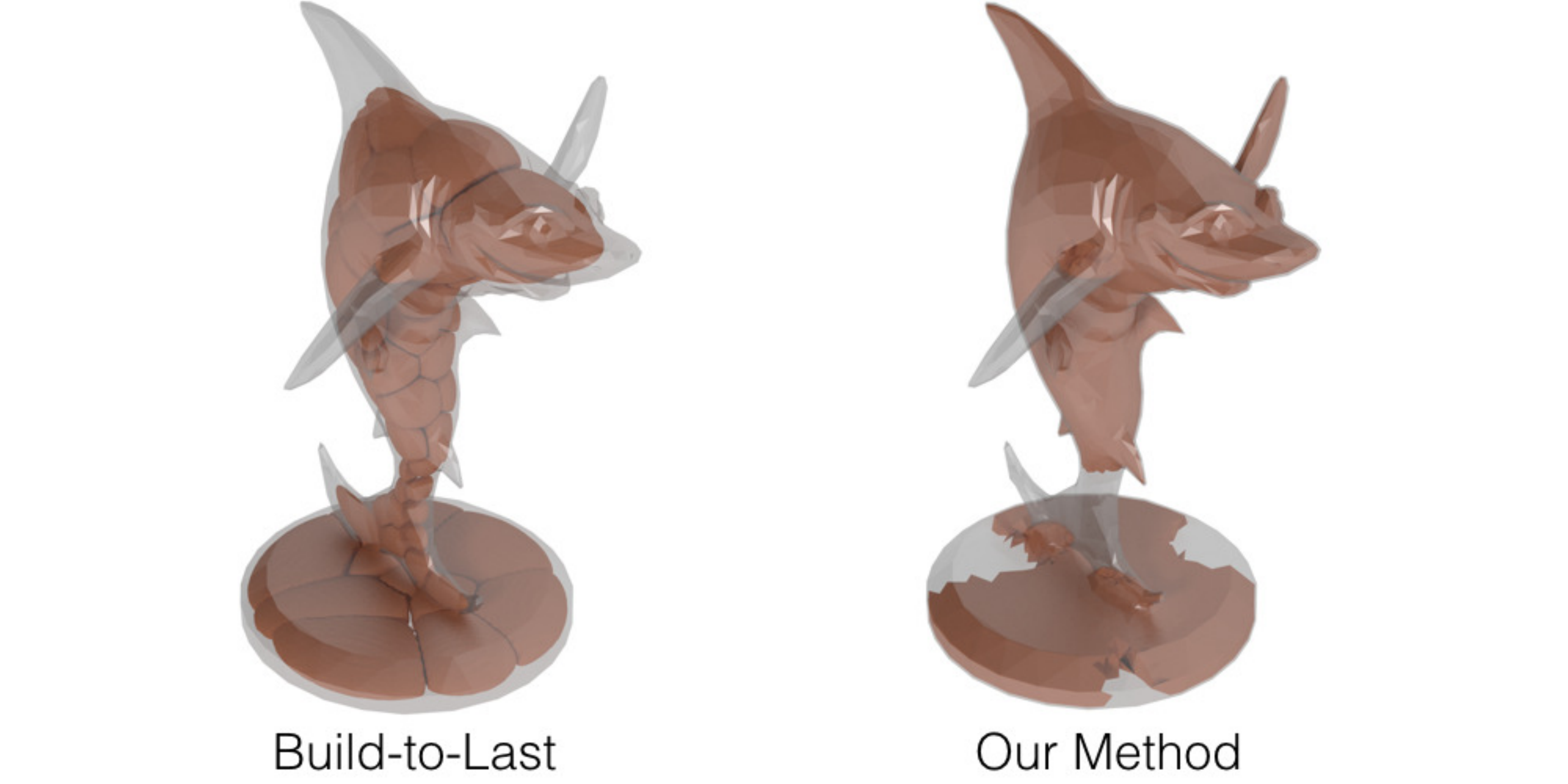}
\caption{A comparison between build-to-last \protect\cite{lu2014build} and our method. Our optimization approach produces a lighter structure while sustaining any possible force applied on the boundary. Build-to-last optimizes the structure for  a single static force.}
\label{fig:ForceLocationUncertainty:BuildToLastVSOurs}
\end{figure}

\paragraph{Comparison}
In Figure~\ref{fig:ForceLocationUncertainty:BuildToLastVSOurs}, we compare our approach with the build-to-last method  \cite{lu2014build}.
We impose a force budget of $20N$ applied anywhere on the surface of the shark. Our optimum result weighs $33\%$ less than the build-to-last structure that takes a prescribed force location as input\footnote{In \cite{lu2014build}, we could not identify the force magnitude being used.}.
Unlike build-to-last, our algorithm hollows the fins and the nose where high stresses cannot be generated in any force configuration.
Also, our method generates a three-pronged rib structure at the base, possibly to accommodate the forces that can be applied laterally in all arbitrary directions. 

\begin{sidewaysfigure}
\centering
\includegraphics[width = \textwidth]{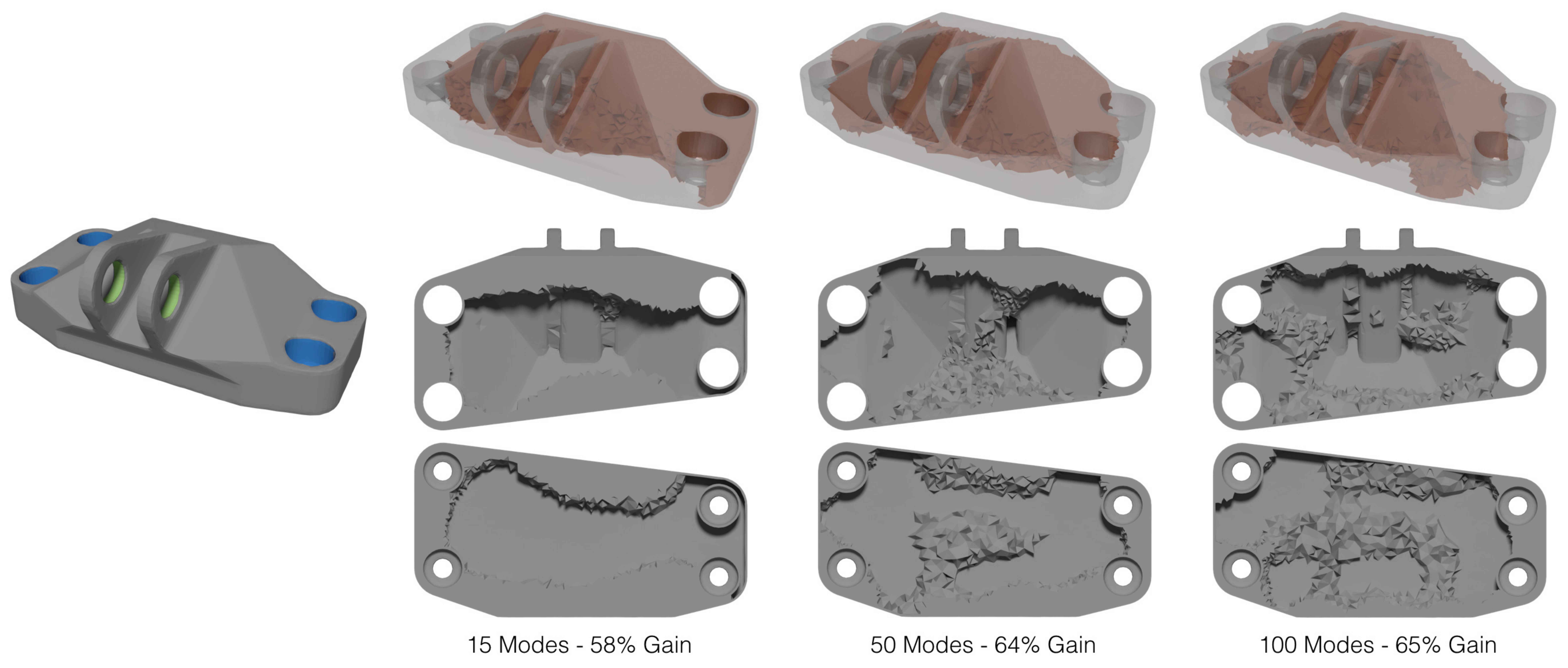}
\caption{Bracket model (left) is optimized using three different number of material modes. As the number of material modes increases, optimization can perform more localized alterations resulting in smaller mass structures at a cost of higher computational complexity.}
\label{fig:ForceLocationUncertainty:GEBracketDifferentModes}
\end{sidewaysfigure}

\paragraph{Number of Material Modes} Figure~\ref{fig:ForceLocationUncertainty:GEBracketDifferentModes} illustrates the effect of the number of material modes.
We optimized the bracket using $15$, $50$ and $100$  modes for the same force budget and boundary conditions.
As shown, higher material modes allow for a finer local shape control. Hence, a larger mass reduction is obtained with increasing number of material modes (see Table~\ref{tab:ForceLocationUncertainty:Results}). Although larger number of modes allow our algorithm to remove more material through local control, computational  cost also increases. In Table~\ref{tab:ForceLocationUncertainty:Results}, per iteration computation times are given for the bracket model using different number of material modes. As the number of modes increases, computational cost increases significantly, while only a minor improvement in further mass reduction is achieved.  

\begin{figure}
\centering
\includegraphics[width = \columnwidth]{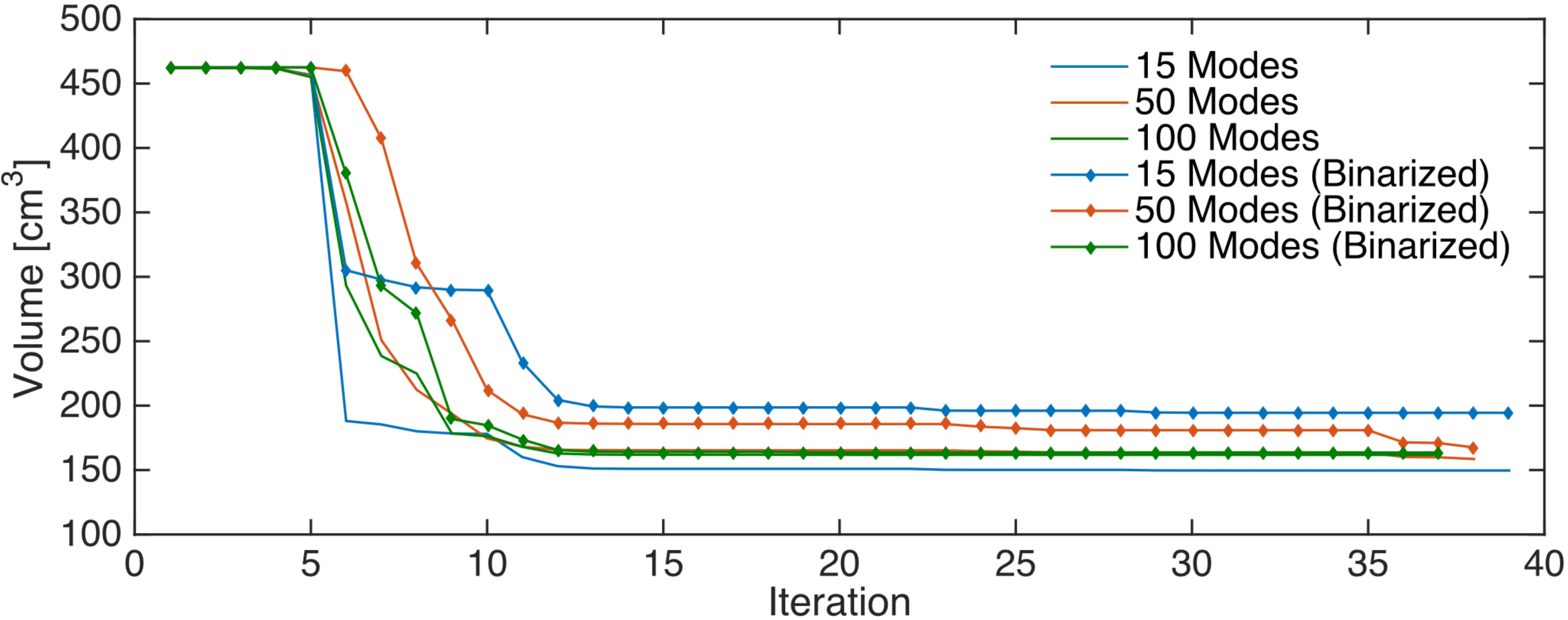}
\caption{Convergence plot for the bracket model using three different material modes. Binarized values are also shown.}
\label{fig:ForceLocationUncertainty:Convergence}
\end{figure}

\paragraph{Convergence} Figure~\ref{fig:ForceLocationUncertainty:Convergence} shows the convergence profiles for the  bracket model. While convergence is achieved after a similar number of iterations, the smaller number of material modes tend to leave more intermediate density elements in the optimized distribution, resulting in a larger mass when binarized.

\begin{figure}
\centering
\includegraphics[width = \columnwidth]{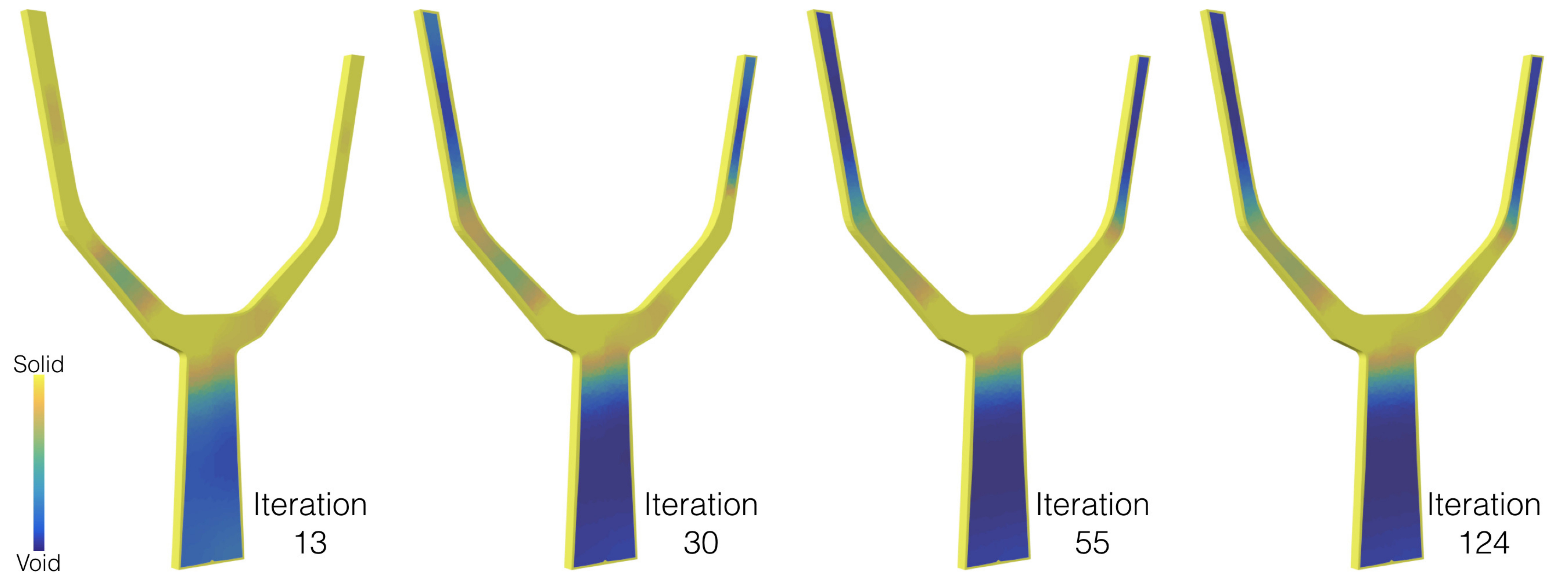}
\caption{Evolution of material distribution throughout the optimization process for a symmetrical slingshot model. Optimization is initialized with fully solid object. Left to right  material distribution converges to the resulting state. The model is fixed at the bottom.}
\label{fig:ForceLocationUncertainty:ConvergenceSymmetric}
\end{figure}

One might worry that in a symmetrical object, the critical point could jump from side to side, with every incremental improvement to one side causing a symmetric worsening of the other side, and convergence never being reached. We attempted to trigger this potential failure by creating a test case with a carefully-constructed perfectly symmetric boundary and tetrehedral mesh (Figure~\ref{fig:ForceLocationUncertainty:ConvergenceSymmetric}). Our algorithm nonetheless converged in $124$ steps. We conjecture that the homogenous (first) material mode tends to absorb enough of the change to avoid oscillation; though it could also be that low-level numerical asymmetries, \eg ordering effects in linear algebra routines, account for the convergence.

\paragraph{Boundary shell thickness}

\begin{figure}
\centering
\includegraphics[width = 0.8\textwidth]{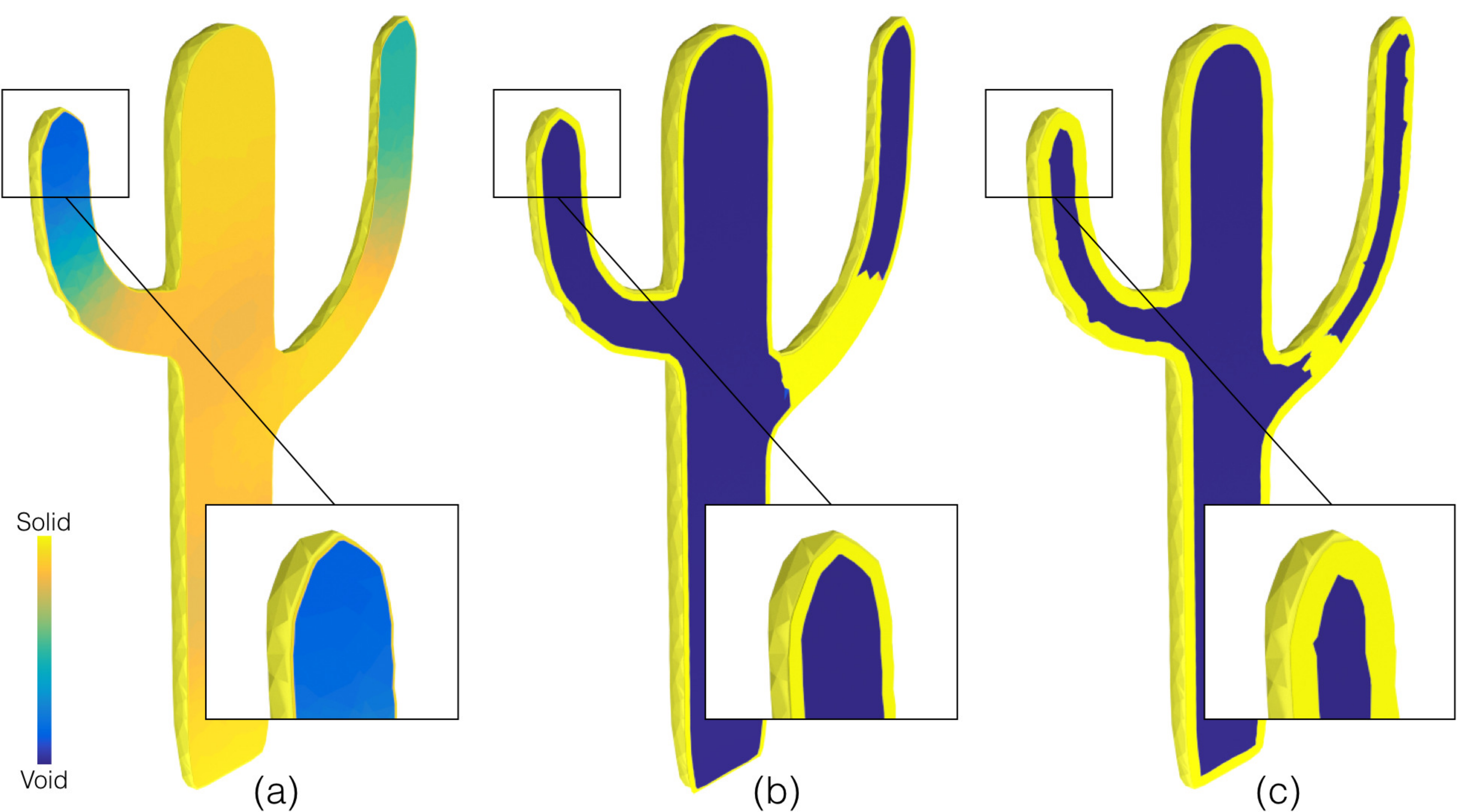}
\caption{Effect of boundary shell thickness on material distribution. From (a) to (c), shell thickness is increased while all other parameters are kept constant. For very small thickness values, our algorithm fails to converge to a binary material distribution using only few material modes due to high local compressive stresses. Close-up images show the shell thickness for each case.}
\label{fig:ForceLocationUncertainty:boundaryShellThickness}
\end{figure}

There is a trade-off between the shell thickness and the flexibility of our algorithm in generating an internal structure.
As the boundary shell gets thicker, its contribution to the structure's strength becomes more prominent, thereby shrinking the design space for our algorithm.
On the other hand, too thin boundary shells can lead to large local compressive stresses that renders optimization infeasible.
In such cases, our method fails to converge to a varied solid versus void material distribution (Figure~\ref{fig:ForceLocationUncertainty:boundaryShellThickness}(a)).
The main reason is that high local compressive stresses encourages the optimization to perform local thickening around the force application points. However, low frequency modes can alter the material distribution only in large chunks.
Therefore, fine-level local modifications cannot be achieved unless a large number of material modes are used.
Figure~\ref{fig:ForceLocationUncertainty:boundaryShellThickness} illustrates the effect of the shell thickness on the resulting material distribution.
With a proper choice of the thickness (Figure~\ref{fig:ForceLocationUncertainty:boundaryShellThickness}(b)), our method is able to reduce mass by $72\%$ compared to fully solid model, while the reduction was only $17\%$ and $54\%$ for (a) and (c), respectively.

\subsection{Validation and Performance}

\paragraph{Fabrication} We 3D printed our optimum results on an OBJET Connex printer using inkjet printing technology.
We use VeroWhitePlus material with a yield strength of $50 MPa$, a Young's modulus of $2.1 GPa$,  and  a Poisson's ratio of $0.3$  \cite{polyjetDatasheet}.

When soluble support material is used, the boundary shell can be pierced by small holes to empty the internal support material. We observed that our reduced order method has a tendency to create only a small number of inner void regions (especially for a small number of material modes) and thus the support material can be removed with minimal alterations. To avoid trapping support material, resulting models can be printed in several pieces and glued together after cleaning.

\begin{figure}
\centering
\includegraphics[width = \columnwidth]{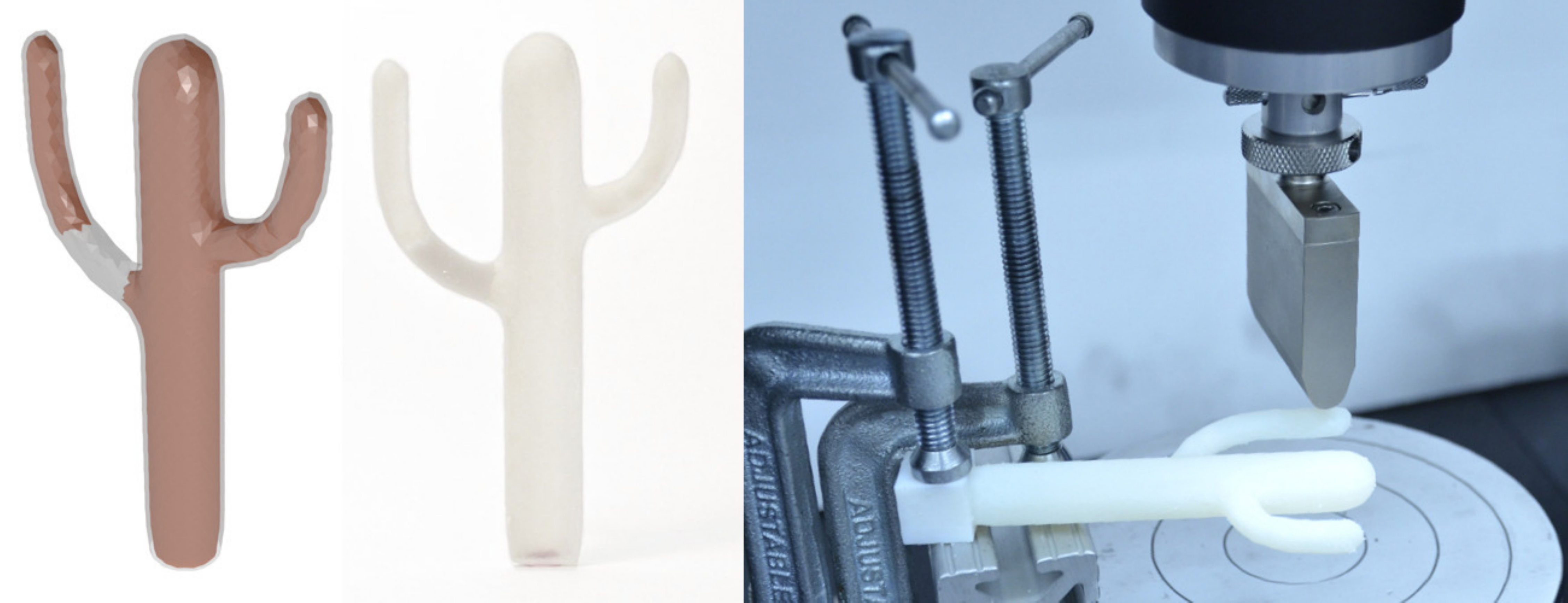}
\caption{Compression tests on the optimized cactus. Left and middle images show the resulting model and a 3D printed cutout.}
\label{fig:ForceLocationUncertainty:CactusPhysicalExperiment_1}
\end{figure}

\paragraph{Physical Tests} We performed compression tests on our optimized cactus model to physically evaluate the strength of the 3D printed models, as shown in Figure~\ref{fig:ForceLocationUncertainty:CactusPhysicalExperiment_1}. We used an INSTRON universal testing machine and ran compression tests on the optimized cactus model. For comparison, we chose an identically weighing uniform thickness cactus. We performed the same compression test on the long arm and measured the failure load. For our optimized model, we measured the failure force to be $36.82N$. The uniform thickness model snapped at $20\%$ less force of $29.69N$. 

\begin{figure}
\centering
\includegraphics[width = 0.8\textwidth]{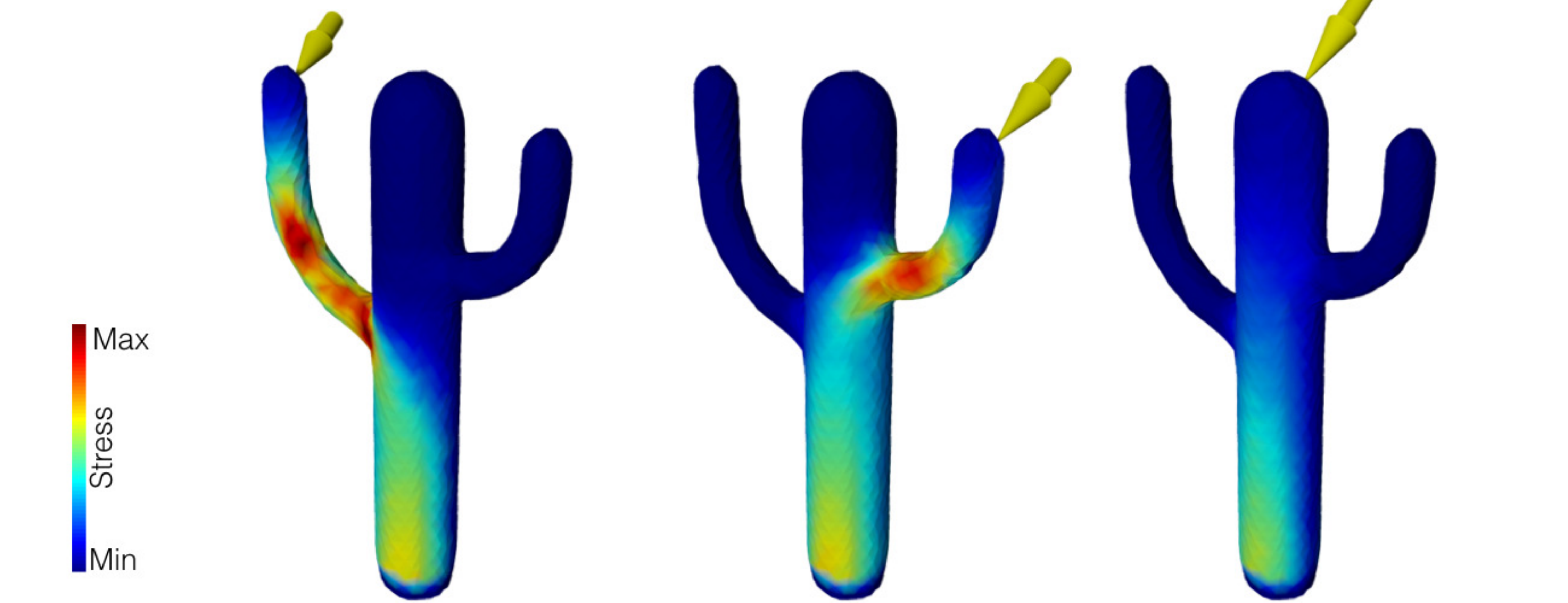}
\caption{Stress distributions when the same magnitude force is applied to three different locations on the optimized cactus. Analysis results match  the compression tests in that failure occurs on the long arm, short arm and trunk, in order. }
\label{fig:ForceLocationUncertainty:CactusPhysicalExperiment_2}
\end{figure}

For our optimized model, we measured the failure force of $39.62N$ for the short arm, and  $74.22N$ for the trunk.
The  test results agree well with our critical instant analysis in that the tip of the long arm turns out to be physically the most critical force point in the optimum model and there is no need to add material to either the short arm or the trunk.
Figure~\ref{fig:ForceLocationUncertainty:CactusPhysicalExperiment_2} illustrates the stress distribution when a $35N$ force is applied to the same three points as the physical tests. It can be observed that the short arm and the trunk are quite safe, while very high stresses are present on the long arm.

\begin{figure}
\centering
\includegraphics[width = 0.8\textwidth]{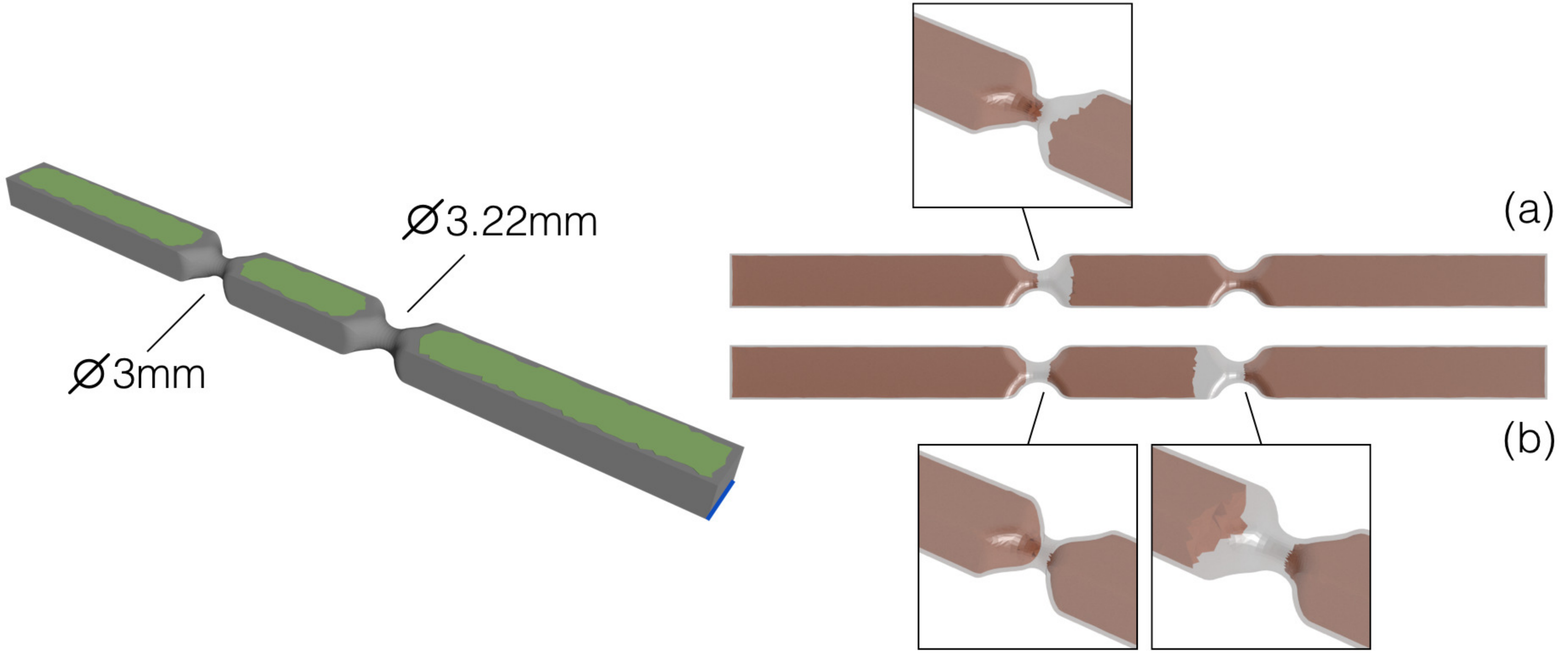}
\caption{A model with two failure-prone regions optimized for two different force budgets. For a force budget of $4N$ (a), material is only placed around the thinnest region while both critical regions are beefed up  for a larger force budget of $5.5N$ (b).}
\label{fig:ForceLocationUncertainty:TestExample}
\end{figure}

\begin{figure}
\centering
\includegraphics[width = \columnwidth]{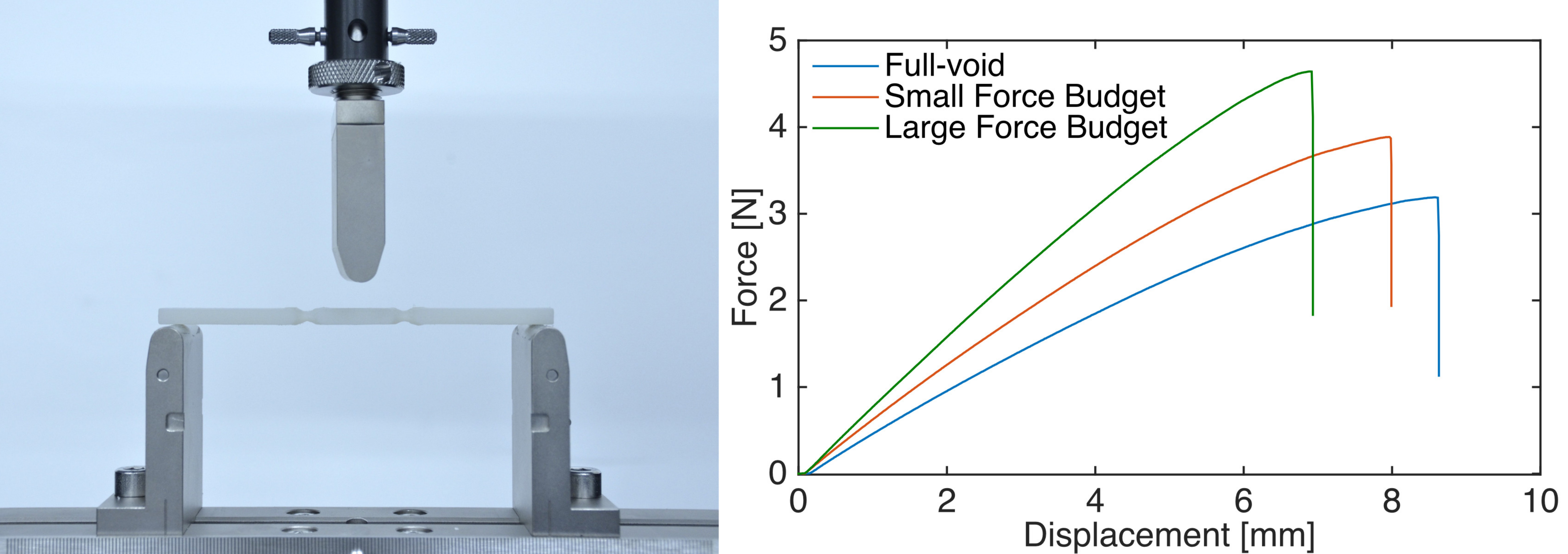}
\caption{Three point bend tests on optimized test models. The model optimized for larger force budget performs best while empty model breaks at a small force value. }
\label{fig:ForceLocationUncertainty:TestExperiment}
\end{figure}

Figure~\ref{fig:ForceLocationUncertainty:TestExample} illustrates a test model we designed to observe the effect of the force budget.
The model has two thin regions with slightly different dimensions.
We set the contact region to be the entire top surface while fixing it only at the bottom right and left edges to simulate a simply supported beam.
We then optimized the structure for two different force budgets of $4N$ and $5.5N$.
For the smaller force budget, the optimization focuses material around the thinnest part only.
However, for the larger force budget, material is distributed around both of the failure prone regions.
Notice that only a portion of the thinnest neck is filled for the small force budget while it is entirely filled for the larger force budget.
To validate the optimization results of the test model, we performed a set of three-point bend tests.
For benchmarking, we also tested a completely empty shell model (full-void). Figure~\ref{fig:ForceLocationUncertainty:TestExperiment} shows the experimental setup together with the force plots we obtained. The test model optimized for the larger force budget performs best while the empty model breaks at the smallest force magnitude.
Our structural optimization method strengthens the model up to $46\%$ while increasing its mass by only $15\%$.
The  discrepancies  between the input force budget  and the measured failure forces can be due to the FEA modeling of the problem as well as the possible anisotropic behavior of the 3D printed material \cite{ulu2015enhancing}.

\begin{sidewaystable}
\small
\caption{Performance of our algorithm on a variety of models. The number of FEAs and analysis times in columns 4 and 6 are average and given per step of critical instant analysis. Performance of the brute force approach is also shown for benchmarking. Columns 7 and 8 are number of iterations to fully optimize the structures and average times per iteration of our structural optimization, respectively.} 
\centering 
\begin{tabular}{l ccccccccc} 
\hline\hline 
\multirow{2}{*}{Model} & \multirow{2}{*}{Elements} & \multicolumn{2}{c}{\# of FEA} & \multicolumn{2}{c}{Analysis Time [s]}  & \multirow{2}{*}{Iteration} & \multirow{2}{*}{Time[s]} &\multicolumn{2}{c}{Volume [cm\textsuperscript{3}]}\\ \cline{3-4} \cline{5-6} \cline{9-10}
& & Brute Force  & Our Method  & Brute Force & Our Method & & & Initial & Optimized\\ [0.5ex]
\hline 
Cactus     				& 25229		& 1658	& 216	& 17.69		& 4.35		& 49	&	39.61	& 68.990	& 18.800\\
Slingshot				& 54290		& 2192	& 148	& 89.02			& 10.66		& 124	&	72.52	& 17.822	& 9.451\\
Fertility    			& 57006		& 3914	& 248	& 131.85	& 20.68		& 122	&	83.33	& 54.023 	& 16.350\\
Test (Small Force)		& 58555		& 822	& 69	& 26.53		& 7.52		& 180	&	59.15	& 4.443		& 1.216\\
Test (Large Force)		& 58555		& 822	& 69	& 26.53		& 7.52		& 223	&	59.15	& 4.443		& 1.300\\
Chair   				& 58848		& 663	& 52	& 27.96		& 13.28		& 104	&	75.32	& 17.076	& 6.192\\
Penguin        		& 68615		& 3035	& 304	& 149.80	& 45.13		& 31	&	132.06	& 91.643 	& 10.211\\
Shark			 		& 70397		& 4282	& 273	& 178.93	& 18.10		& 72	&	115.07	& 66.454	& 14.997\\ 
Rocking Chair  		& 78025		& 1531	& 103	& 85.48		& 11.17		& 160	&	154.32	& 13.015	& 6.709\\ 
Bracket (k = 15)   	& 111498	& 408	& 55	& 70.85		& 48.87		& 39	&	301.01	& 462.520	& 194.434\\ 
Bracket (k = 50)  		& 111498	& 408	& 55	& 70.85		& 48.87		& 38	&	382.49	& 462.520	& 167.552\\ 
Bracket (k = 100) 		& 111498	& 408	& 55	& 70.85		& 48.87		& 37	&	487.44	& 462.520	& 163.581\\ 

\hline 
\end{tabular}
\label{tab:ForceLocationUncertainty:Results}
\end{sidewaystable}

\paragraph{Performance} Table~\ref{tab:ForceLocationUncertainty:Results} shows the performance of our algorithm.
We tested our method on a 3.2GHz Intel Core i5 computer with 8GB of memory.
We selected various 3D models and optimized under different force configurations.
Although the major computational cost comes from the critical instant analysis, we achieve $5\times$ acceleration on average  over a brute force approach. This acceleration becomes more significant as the contact region (hence the number of force instants) grows in relation to the total boundary surface. Shark (large contact region) vs. Bracket (small contact region) in Table~\ref{tab:ForceLocationUncertainty:Results} highlights this difference. The main reason is that the stiffness matrix is assembled and factorized only once at each optimization step and $\sigma_{cr}$ is computed by only performing a back-substitution for each force instant. For a large number of force instants, the cost of a single assembly and factorization becomes much smaller compared to the number of linear solves performed. 

\subsection{Limitations and Future Work}

Our critical instant analysis is based on an approximation to the relationship between input forces and resulting stresses. We found this approximation to work well when stresses due to bending and torsion are dominant compared to local compressive stresses. For shapes where local compressive stresses play a dominant role in failure,  an efficient approximation of stress and an accurate estimation of the criticality map  remains an open problem. For models with many small protruding features, our geodesic force instant sampling may fail to sample such features, thus causing the estimated criticality map to miss potential critical instants. 

In our approach, we benefit from the boundary shell in preserving the external shape and solving the singularity problems in the optimization. However, it can also serve as a main structural component as its thickness is increased. In this work, we do not optimize the boundary shell thickness. A natural extension of our approach would be to efficiently determine the boundary shell thickness as a preprocessing step to our algorithm.

In the future, our analysis could be extended nonlinear and/or anisotropic material models. One of the advantages of using material modes is that the resulting density is smoothly varying, which makes the results easier to fabricate. This work focuses on making objects as safe at all times, but in some cases one may want to incorporate \textit{weak} points to enable fail-safe designs. Our critical instant analysis might be able to handle this case by using a criticality map construction that takes into account a spatially-varying thresholds.

\section{Conclusions}
We present a lightweight structure optimization method for 3D objects under force location uncertainty.
We propose a novel critical instant analysis method to efficiently determine the force instant creating the highest stress in the structure.
With this method, we show that an approximation to the relationship between the force configurations and  resulting stress distributions can be captured using only small number of FEA evaluations.
Combined with a reduced order formulation, we demonstrate that our method provides a practical solution to this computationally demanding  optimization problem.
We evaluate the performance of our algorithm on a variety of 3D models.
Our results show that significant mass reduction can be achieved by optimizing the material distribution while ensuring that the object is structurally sound against a wide range of force configurations capped by a force budget.

\chapter{Conclusion}
\label{chp:conclusion}

In this thesis, computational methods to improve structural performance of additively manufactured objects are investigated. A geometry preserving method as well as data-driven shape optimization approaches to the structural design problems are presented. 

We start by introducing a new build orientation selection method for AM processes that aims to maximize an input object's resistance to failure under prescribed external loads. Additively manufactured objects often exhibit directional dependencies in their structure due to the layered nature of the printing process. Our optimization algorithm exploits this anisotropy in the material properties and identifies the build orientation that maximizes the factor of safety of an input object under prescribed loading and boundary configurations. We use a surrogate-based optimization method that starts with a small number of finite element solutions corresponding to different build orientations. The initial solutions are progressively improved with the addition of new solutions until the optimum orientation is computed.

Second, we present a data-driven approach to topology optimization involving structural mechanics. In topology optimization problems, even small alterations in design constraints or boundary conditions require a new application of the optimization routine. Our algorithm addresses this problem by using known solutions for predicting optimal topologies under a new set of design constraints. Our approach takes as input a set of images representing optimal 2D topologies, each resulting from a random loading configuration applied to a common boundary support condition. 
Using these sample images, a mapping between the loading configurations and the optimal topologies is learned. From this mapping, we estimate the optimal topologies for novel loading configurations.

We, then, introduce a structure optimization approach for a more general class of problems in which there is uncertainty in the force locations. Such uncertainty may arise due to force contact locations that change during use or are simply unknown a priori. Given an input 3D model and regions on its boundary where arbitrary normal forces may make contact, our algorithm generates a minimum weight 3D structure that withstands any force configuration possible. Our approach works by repeatedly finding the most critical force configuration and altering the internal structure accordingly. 
Combined with a reduced order formulation, our method provides a practical solution to the structural optimization problem.

Finally, we present a method for designing shell structures for worst-case loading scenarios. Topology optimization, when geometrically unconstrained, tends to create very complex internal structures. For closed geometries, such internal structures prevent easy removal of support material encapsulated inside as well as disrupt the inner cavity that could be used for functional purposes. To address this, given an input 3D shape, our algorithm alters the model's shell thickness and generates a minimum weight 3D structure that withstands a wide range of force configurations capped by a prescribed force budget. Resulting geometry is hollow with a single inner cavity and preserves the appearance of the input model.

Methods presented in this thesis greatly enhance the value of AM technology by taking advantage of the design space enabled by it for a broad class of problems involving complex in-use loads.

\appendix\pagestyle{plain}

\chapter{Analytic Gradients}
\label{sec:ForceLocationUncertainty:Appendix}

Following the final formulation in \eqref{Eq:ForceLocationUncertainty:reducedOptimizationProblem}, the gradient of mass with respect to the reduced order design variables $\boldsymbol{\alpha}$ can be calculated as

\begin{equation}
 \frac{\partial{M}}{\partial{\boldsymbol{\alpha}}} = \frac{\partial{M}}{\partial{\boldsymbol{\rho}}} \frac{\partial{\boldsymbol{\rho}}}{\partial{\boldsymbol{\alpha}}}
\end{equation}

where the first term is simply the elemental volume vector $\boldsymbol{V}$ and the second term can be obtained by following \eqref{Eq:ForceLocationUncertainty:logistic} as

\begin{equation}
 \frac{\partial{\boldsymbol{\rho}}}{\partial{\boldsymbol{\alpha}}}  = \frac{\partial{G}}{\partial{\boldsymbol{x}}} \frac{\partial{\boldsymbol{x}}}{\partial{\boldsymbol{\alpha}}}.
\end{equation}

Here, $\boldsymbol{x} = \boldsymbol{\Gamma} \boldsymbol{\alpha}$ and its derivative with respect to $\boldsymbol{\alpha}$ is simply the reduced order material basis matrix $\boldsymbol{\Gamma}$. 

The gradient of the critical stress $\sigma_{cr}$ with respect to $\boldsymbol{\alpha}$ can be obtained following the formulation in \eqref{Eq:ForceLocationUncertainty:reducedSigmaCritical} as

\begin{equation}
\label{Eq:ForceLocationUncertainty:dsigmacr_dalpha}
 \frac{\partial{\sigma_{cr}}}{\partial{\boldsymbol{\alpha}}} = \frac{\partial{\sigma_{cr}}}{\partial{\boldsymbol{\rho}}} \frac{\partial{\boldsymbol{\rho}}}{\partial{\boldsymbol{\alpha}}}.
\end{equation}

We use p-norm approximations ($p=15$) for the max functions. For $H(\boldsymbol{\sigma}^{vm}) = \| \boldsymbol{\sigma}^{vm} \|_{p}$ where $\boldsymbol{\sigma}^{vm}$ is composed of $\sigma_e^{vm} \quad \forall e \in \mathcal{V}_{wr}$, the derivative of $\sigma_{cr}$ with respect to $\boldsymbol{\rho}$ can be computed as 

\begin{equation}
\label{Eq:ForceLocationUncertainty:dsigmacr_dro}
 \frac{\partial{\sigma_{cr}}}{\partial{\boldsymbol{\rho}}} = \frac{\partial{H}}{\partial{\boldsymbol{\sigma}^{vm}}} \frac{\partial{\boldsymbol{\sigma}^{vm}}}{\partial{\boldsymbol{\sigma}}} \left( \frac{\partial{\boldsymbol{\sigma}}}{\partial{\boldsymbol{u}}} \frac{\partial{\boldsymbol{u}}}{\partial{\boldsymbol{\rho}}} +  \frac{\partial{\boldsymbol{\sigma}}}{\partial{\boldsymbol{\rho}}} \right) 
\end{equation}

where the derivative of $\boldsymbol{u}$ with respect to $\boldsymbol{\rho}$ can be obtained from the equilibrium equation $\boldsymbol{K}\boldsymbol{u} = \boldsymbol{f}$ as

\begin{equation}
\label{Eq:ForceLocationUncertainty:du_dro}
 \frac{\partial{\boldsymbol{u}}}{\partial{\boldsymbol{\rho}}} = \boldsymbol{K}^{-1} \left( -\frac{\partial{\boldsymbol{K}}}{\partial{\boldsymbol{\rho}}} \boldsymbol{u} \right). 
\end{equation}

Applying the adjoint variable method, \eqref{Eq:ForceLocationUncertainty:dsigmacr_dro} can be re-written as

\begin{equation}
\label{Eq:ForceLocationUncertainty:dsigmacr_dro_adjoint}
 \frac{\partial{\sigma_{cr}}}{\partial{\boldsymbol{\rho}}} = \boldsymbol{\xi}^T \left(- \frac{\partial{\boldsymbol{K}}}{\partial{\boldsymbol{\rho}}} \boldsymbol{u} \right)  + \frac{\partial{H}}{\partial{\boldsymbol{\sigma}^{vm}}} \frac{\partial{\boldsymbol{\sigma}^{vm}}}{\partial{\boldsymbol{\sigma}}}\frac{\partial{\boldsymbol{\sigma}}}{\partial{\boldsymbol{\rho}}} 
\end{equation}

where $\boldsymbol{\xi}$ is the adjoint variable and defined as 

\begin{equation}
\label{Eq:ForceLocationUncertainty:adjointVariable}
 \boldsymbol{\xi}^T = \frac{\partial{H}}{\partial{\boldsymbol{\sigma}^{vm}}} \frac{\partial{\boldsymbol{\sigma}^{vm}}}{\partial{\boldsymbol{\sigma}}} \frac{\partial{\boldsymbol{\sigma}}}{\partial{\boldsymbol{u}}}\boldsymbol{K}^{-1}. 
\end{equation}

All the terms in equations \eqref{Eq:ForceLocationUncertainty:dsigmacr_dalpha} and \eqref{Eq:ForceLocationUncertainty:dsigmacr_dro_adjoint} can be directly obtained and the adjoint variable can be computed by solving the system of linear equations in \eqref{Eq:ForceLocationUncertainty:adjointVariable}.

\bibliographystyle{IEEEtran}

\bibliography{content/references}

\begin{thebibliography}{100}
\providecommand{\url}[1]{#1}
\csname url@samestyle\endcsname
\providecommand{\newblock}{\relax}
\providecommand{\bibinfo}[2]{#2}
\providecommand{\BIBentrySTDinterwordspacing}{\spaceskip=0pt\relax}
\providecommand{\BIBentryALTinterwordstretchfactor}{4}
\providecommand{\BIBentryALTinterwordspacing}{\spaceskip=\fontdimen2\font plus
\BIBentryALTinterwordstretchfactor\fontdimen3\font minus
  \fontdimen4\font\relax}
\providecommand{\BIBforeignlanguage}[2]{{%
\expandafter\ifx\csname l@#1\endcsname\relax
\typeout{** WARNING: IEEEtran.bst: No hyphenation pattern has been}%
\typeout{** loaded for the language `#1'. Using the pattern for}%
\typeout{** the default language instead.}%
\else
\language=\csname l@#1\endcsname
\fi
#2}}
\providecommand{\BIBdecl}{\relax}
\BIBdecl

\bibitem{zhou2013worst}
\BIBentryALTinterwordspacing
Q.~Zhou, J.~Panetta, and D.~Zorin, ``Worst-case structural analysis,''
  \emph{ACM Trans. Graph.}, vol.~32, no.~4, pp. 137:1--137:12, Jul. 2013.
  [Online]. Available: \url{http://doi.acm.org/10.1145/2461912.2461967}
\BIBentrySTDinterwordspacing

\bibitem{lu2014build}
\BIBentryALTinterwordspacing
L.~Lu, A.~Sharf, H.~Zhao, Y.~Wei, Q.~Fan, X.~Chen, Y.~Savoye, C.~Tu,
  D.~Cohen-Or, and B.~Chen, ``Build-to-last: Strength to weight 3d printed
  objects,'' \emph{ACM Trans. Graph.}, vol.~33, no.~4, pp. 97:1--97:10, Jul.
  2014. [Online]. Available: \url{http://doi.acm.org/10.1145/2601097.2601168}
\BIBentrySTDinterwordspacing

\bibitem{scott2012additive}
J.~Scott, N.~Gupta, C.~Weber, S.~Newsome, T.~Wohlers, and T.~Caffrey,
  ``Additive manufacturing: status and opportunities,'' \emph{Science and
  Technology Policy Institute, Washington, DC}, pp. 1--29, 2012.

\bibitem{vaezi2013areview}
M.~Vaezi, H.~Seitz, and S.~Yang, ``A review on 3d micro-additive manufacturing
  technologies,'' \emph{The International Journal of Advanced Manufacturing
  Technology}, vol.~67, no. 5-8, pp. 1721--1754, 2013.

\bibitem{gibson2010additive}
I.~Gibson, D.~W. Rosen, B.~Stucker \emph{et~al.}, \emph{Additive manufacturing
  technologies}.\hskip 1em plus 0.5em minus 0.4em\relax Springer, 2010, vol.
  238.

\bibitem{lee2012stress}
E.~Lee, K.~A. James, and J.~R. R.~A. Martins, ``Stress-constrained topology
  optimization with design-dependent loading,'' \emph{Structural and
  Multidisciplinary Optimization}, vol.~46, no.~5, pp. 647--661, 2012.

\bibitem{stava2012stress}
\BIBentryALTinterwordspacing
O.~Stava, J.~Vanek, B.~Benes, N.~Carr, and R.~M\v{e}ch, ``Stress relief:
  Improving structural strength of 3d printable objects,'' \emph{ACM Trans.
  Graph.}, vol.~31, no.~4, pp. 48:1--48:11, Jul. 2012. [Online]. Available:
  \url{http://doi.acm.org/10.1145/2185520.2185544}
\BIBentrySTDinterwordspacing

\bibitem{wang2013cost}
\BIBentryALTinterwordspacing
W.~Wang, T.~Y. Wang, Z.~Yang, L.~Liu, X.~Tong, W.~Tong, J.~Deng, F.~Chen, and
  X.~Liu, ``Cost-effective printing of 3d objects with skin-frame structures,''
  \emph{ACM Trans. Graph.}, vol.~32, no.~6, pp. 177:1--177:10, Nov. 2013.
  [Online]. Available: \url{http://doi.acm.org/10.1145/2508363.2508382}
\BIBentrySTDinterwordspacing

\bibitem{banichuk2013introduction}
N.~V. Banichuk, \emph{Introduction to optimization of structures}.\hskip 1em
  plus 0.5em minus 0.4em\relax Springer Science \& Business Media, 2013.

\bibitem{musialski2015reduced}
\BIBentryALTinterwordspacing
P.~Musialski, T.~Auzinger, M.~Birsak, M.~Wimmer, and L.~Kobbelt,
  ``Reduced-order shape optimization using offset surfaces,'' \emph{ACM Trans.
  Graph.}, vol.~34, no.~4, pp. 102:1--102:9, Jul. 2015. [Online]. Available:
  \url{http://doi.acm.org/10.1145/2766955}
\BIBentrySTDinterwordspacing

\bibitem{christiansen2015combined}
\BIBentryALTinterwordspacing
A.~N. Christiansen, J.~A. B{\ae}rentzen, M.~Nobel-J{\o}rgensen, N.~Aage, and
  O.~Sigmund, ``Combined shape and topology optimization of 3d structures,''
  \emph{Comput. Graph.}, vol.~46, no.~C, pp. 25--35, Feb. 2015. [Online].
  Available: \url{http://dx.doi.org/10.1016/j.cag.2014.09.021}
\BIBentrySTDinterwordspacing

\bibitem{bendsoe2003topology}
M.~P. Bends{\o}e and O.~Sigmund, \emph{Topology optimization: theory, methods,
  and applications}.\hskip 1em plus 0.5em minus 0.4em\relax Springer-Verlag,
  2003.

\bibitem{zhang2015efficient}
P.~Zhang, J.~Toman, Y.~Yu, E.~Biyikli, M.~Kirca, M.~Chmielus, and A.~C. To,
  ``Efficient design-optimization of variable-density hexagonal cellular
  structure by additive manufacturing: theory and validation,'' \emph{Journal
  of Manufacturing Science and Engineering}, vol. 137, no.~2, p. 021004, 2015.

\bibitem{huang2017cost}
R.~Huang, E.~Ulu, L.~B. Kara, and K.~S. Whitefoot, ``Cost minimization in metal
  additive manufacturing using concurrent structure and process optimization,''
  in \emph{ASME 2017 International Design Engineering Technical Conferences and
  Computers and Information in Engineering Conference}.\hskip 1em plus 0.5em
  minus 0.4em\relax American Society of Mechanical Engineers, 2017, pp.
  V02AT03A030--V02AT03A030.

\bibitem{ulu2018concurrent}
E.~Ulu, R.~Huang, L.~B. Kara, and K.~S. Whitefoot, ``Concurrent structure and
  process optimization for minimum cost metal additive manufacturing,''
  \emph{Journal of Mechanical Design}, 2018.

\bibitem{wang2018efficient}
Y.~Wang, E.~Ulu, A.~Singh, and L.~B. Kara, ``Efficient load sampling for
  worst-case structural analysis under force location uncertainty,'' in
  \emph{ASME 2018 International Design Engineering Technical Conferences and
  Computers and Information in Engineering Conference}.\hskip 1em plus 0.5em
  minus 0.4em\relax American Society of Mechanical Engineers, 2018.

\bibitem{ulu2017lightweight}
\BIBentryALTinterwordspacing
E.~Ulu, J.~Mccann, and L.~B. Kara, ``Lightweight structure design under force
  location uncertainty,'' \emph{ACM Trans. Graph.}, vol.~36, no.~4, pp.
  158:1--158:13, Jul. 2017. [Online]. Available:
  \url{http://doi.acm.org/10.1145/3072959.3073626}
\BIBentrySTDinterwordspacing

\bibitem{arisoy2017methods}
E.~B. Arisoy, S.~Musuvathy, E.~Ulu, and N.~G. Ulu, ``Methods and system to
  predict hand positions for multi-hand grasps of industrial objects,'' Patent
  WO2\,017\,132\,134 A1, 2017.

\bibitem{arisoy2016data}
E.~B. Arisoy, G.~Ren, E.~Ulu, N.~G. Ulu, and S.~Musuvathy, ``A data-driven
  approach to predict hand positions for two-hand grasps of industrial
  objects,'' in \emph{ASME 2016 International Design Engineering Technical
  Conferences and Computers and Information in Engineering Conference}.\hskip
  1em plus 0.5em minus 0.4em\relax American Society of Mechanical Engineers,
  2016, pp. V01AT02A067--V01AT02A067.

\bibitem{ulu2015adata}
\BIBentryALTinterwordspacing
E.~Ulu, R.~Zhang, and L.~B. Kara, ``A data-driven investigation and estimation
  of optimal topologies under variable loading configurations,'' \emph{Computer
  Methods in Biomechanics and Biomedical Engineering: Imaging \&
  Visualization}, vol.~4, no.~2, pp. 61--72, 2016. [Online]. Available:
  \url{http://dx.doi.org/10.1080/21681163.2015.1030775}
\BIBentrySTDinterwordspacing

\bibitem{ulu2015enhancing}
E.~Ulu, E.~Korkmaz, K.~Yay, O.~B. Ozdoganlar, and L.~B. Kara, ``Enhancing the
  structural performance of additively manufactured objects through build
  orientation optimization,'' \emph{Journal of Mechanical Design}, vol. 137,
  no.~11, p. 111410, 2015.

\bibitem{ulu2014adata}
\BIBentryALTinterwordspacing
E.~Ulu, R.~Zhang, M.~E. Yumer, and L.~B. Kara, \emph{A Data-Driven
  Investigation and Estimation of Optimal Topologies under Variable Loading
  Configurations}.\hskip 1em plus 0.5em minus 0.4em\relax Cham: Springer
  International Publishing, 2014, pp. 387--399. [Online]. Available:
  \url{http://dx.doi.org/10.1007/978-3-319-09994-1_38}
\BIBentrySTDinterwordspacing

\bibitem{xu1999considerations}
F.~Xu, H.~Loh, and Y.~Wong, ``Considerations and selection of optimal
  orientation for different rapid prototyping systems,'' \emph{Rapid
  Prototyping Journal}, vol.~5, no.~2, pp. 54--60, 1999.

\bibitem{thrimurthulu2004optimum}
K.~Thrimurthulu, P.~M. Pandey, and N.~V. Reddy, ``Optimum part deposition
  orientation in fused deposition modeling,'' \emph{International Journal of
  Machine Tools and Manufacture}, vol.~44, no.~6, pp. 585 -- 594, 2004.

\bibitem{ahn2007fabrication}
D.~Ahn, H.~Kim, and S.~Lee, ``Fabrication direction optimization to minimize
  post-machining in layered manufacturing,'' \emph{International Journal of
  Machine Tools and Manufacture}, vol.~47, no. 3-4, pp. 593 -- 606, 2007.

\bibitem{canellidis2009genetic}
V.~Canellidis, J.~Giannatsis, and V.~Dedoussis, ``Genetic-algorithm-based
  multi-objective optimization of the build orientation in stereolithography,''
  \emph{The International Journal of Advanced Manufacturing Technology},
  vol.~45, no. 7-8, pp. 714--730, 2009.

\bibitem{alexander1998part}
P.~Alexander, S.~Allen, and D.~Dutta, ``Part orientation and build cost
  determination in layered manufacturing,'' \emph{Computer-Aided Design},
  vol.~30, no.~5, pp. 343 -- 356, 1998.

\bibitem{vanek2014clever}
J.~Vanek, J.~A.~G. Galicia, and B.~Benes, ``Clever support: Efficient support
  structure generation for digital fabrication,'' \emph{Computer Graphics
  Forum}, vol.~33, no.~5, pp. 117--125, 2014.

\bibitem{ahn2002anisotropic}
S.-H. Ahn, M.~Montero, D.~Odell, S.~Roundy, and P.~K. Wright, ``Anisotropic
  material properties of fused deposition modeling abs,'' \emph{Rapid
  Prototyping Journal}, vol.~8, no.~4, pp. 248--257, 2002.

\bibitem{bagsik2011mechanical}
A.~Bagsik and V.~Sch{\"o}ppner, ``Mechanical properties of fused deposition
  modeling parts manufactured with ultem* 9085,'' \emph{ANTEC 2011}, 2011.

\bibitem{elgizawy2011process}
A.~S. El-Gizawy, S.~Corl, and B.~Graybill, ``Process-induced properties of fdm
  products,'' in \emph{Proceedings of The ICMET}, 2011.

\bibitem{hill2014deposition}
N.~Hill and M.~Haghi, ``Deposition direction-dependent failure criteria for
  fused deposition modeling polycarbonate,'' \emph{Rapid Prototyping Journal},
  vol.~20, no.~3, pp. 221--227, 2014.

\bibitem{barclift2012examining}
M.~W. Barclift and C.~B. Williams, ``Examining variability in the mechanical
  properties of parts manufactured via polyjet direct 3d printing,'' in
  \emph{International Solid Freeform Fabrication Symposium, August}, 2012, pp.
  6--8.

\bibitem{kesy2010Mechanical}
A.~Kesy and J.~Kotli{\'n}ski, ``Mechanical properties of parts produced by
  using polymer jetting technology,'' \emph{Archives of civil and mechanical
  engineering}, vol.~10, no.~3, pp. 37--50, 2010.

\bibitem{galeta2013influence}
T.~Galeta, I.~Kladaric, and M.~Karakasic, ``Influence of processing factors on
  the tensile strength of 3d-printed models,'' \emph{Material Technology
  ({MTAEC9})}, vol.~47, no.~6, pp. 781--788, 2013.

\bibitem{hildebrand2013orthogonal}
K.~Hildebrand, B.~Bickel, and M.~Alexa, ``Orthogonal slicing for additive
  manufacturing,'' \emph{Computers and Graphics}, vol.~37, no.~6, pp. 669 --
  675, 2013.

\bibitem{umetani2013cross}
\BIBentryALTinterwordspacing
N.~Umetani and R.~Schmidt, ``Cross-sectional structural analysis for 3d
  printing optimization,'' in \emph{{SIGGRAPH} Asia 2013 Technical Briefs},
  ser. SA '13.\hskip 1em plus 0.5em minus 0.4em\relax New York, NY, USA: ACM,
  2013, pp. 5:1--5:4. [Online]. Available:
  \url{http://doi.acm.org/10.1145/2542355.2542361}
\BIBentrySTDinterwordspacing

\bibitem{Luo2012Chopper}
L.~Luo, I.~Baran, S.~Rusinkiewicz, and W.~Matusik, ``Chopper: Partitioning
  models into 3d-printable parts,'' \emph{ACM Trans. Graph.}, vol.~31, no.~6,
  pp. 129:1--129:9, Nov. 2012.

\bibitem{suh1995integration}
Y.~S. Suh and M.~J. Wozny, ``Integration of a solid freeform fabrication
  process into a feature-based {CAD} system environment,'' in \emph{Solid
  Freeform Fabrication Symposium}, H.~L.~M. et~al., Ed., University of Texas,
  Austin, August 1995.

\bibitem{thompson1995optimizing}
D.~C. Thompson and R.~H. Crawford, ``Optimizing part quality with
  orientation,'' in \emph{Proceedings of the Solid Freeform Fabrication
  Symposium}, vol.~6, 1995, pp. 362--368.

\bibitem{bendsoe2004topology}
M.~P. Bendsoe and O.~Sigmund, \emph{Topology Optimization: Theory, Methods and
  Applications}, 2nd~ed.\hskip 1em plus 0.5em minus 0.4em\relax Springer, 2004.

\bibitem{dijk2013levelset}
N.~P. Dijk, K.~Maute, M.~Langelaar, and F.~Keulen, ``Level-set methods for
  structural topology optimization: A review,'' \emph{Structural and
  Multidisciplinary Optimization}, vol.~48, no.~3, pp. 437--472, 2013.

\bibitem{norato2007atopological}
J.~A. Norato, M.~P. Bendsoe, R.~B. Haber, and D.~A. Tortorelli, ``A topological
  derivative method for topology optimization,'' \emph{Structural and
  Multidisciplinary Optimization}, vol.~33, no. 4-5, pp. 375--386, 2007.

\bibitem{bendsoe1988generating}
M.~P. Bendsoe and N.~Kikuchi, ``Generating optimal topologies in structural
  design using a homogenization method,'' \emph{Computer Methods in Applied
  Mechanics and Engineering}, vol.~71, no.~2, pp. 197--224, 1988.

\bibitem{suzuki1991ahomogenization}
K.~Suzuki and N.~Kikuchi, ``A homogenization method for shape and topology
  optimization,'' \emph{Computer Methods in Applied Mechanics and Engineering},
  vol.~93, no.~3, pp. 291--318, 1991.

\bibitem{bendsoe1989optimal}
M.~P. Bends{\o}e, ``Optimal shape design as a material distribution problem,''
  \emph{Structural optimization}, vol.~1, no.~4, pp. 193--202, 1989.

\bibitem{rozvany1992generalized}
G.~I.~N. Rozvany, M.~Zhou, and T.~Birker, ``Generalized shape optimization
  without homogenization,'' \emph{Structural Optimization}, vol.~4, no. 3-4,
  pp. 250--252, 1992.

\bibitem{hassani1998areview}
B.~Hassani and E.~Hinton, ``A review of homogenization and topology
  optimization {I} - homogenization theory for media with periodic structure,''
  \emph{Computers and Structures}, vol.~69, no.~6, pp. 707--717, 1998.

\bibitem{rozvany2001aims}
G.~I.~N. Rozvany, ``Aims, scope, methods, history and unified terminology of
  computer-aided topology optimization in structural mechanics,''
  \emph{Structural and Multidisciplinary Optimization}, vol.~21, no.~2, pp.
  90--108, 2001.

\bibitem{rozvany2009acritical}
------, ``A critical review of established methods of structural topology
  optimization,'' \emph{Structural and Multidisciplinary Optimization},
  vol.~37, no.~3, pp. 217--237, 2009.

\bibitem{sethian2000structural}
J.~Sethian and A.~Wiegmann, ``Structural boundary design via level set and
  immersed interface methods,'' \emph{Journal of Computational Physics}, vol.
  163, no.~2, pp. 489--528, 2000.

\bibitem{wang2003alevel}
\BIBentryALTinterwordspacing
M.~Y. Wang, X.~Wang, and D.~Guo, ``A level set method for structural topology
  optimization,'' \emph{Computer Methods in Applied Mechanics and Engineering},
  vol. 192, no.~1, pp. 227 -- 246, 2003. [Online]. Available:
  \url{http://www.sciencedirect.com/science/article/pii/S0045782502005595}
\BIBentrySTDinterwordspacing

\bibitem{chapman1994genetic}
C.~D. Chapman, K.~Saitou, and M.~J. Jakiela, ``Genetic algorithms as an
  approach to configuration and topology design,'' \emph{Journal of Mechanical
  Design}, vol. 116, pp. 1005--1012, December 1994.

\bibitem{jakiela2000continuum}
\BIBentryALTinterwordspacing
M.~J. Jakiela, C.~Chapman, J.~Duda, A.~Adewuya, and K.~Saitou, ``Continuum
  structural topology design with genetic algorithms,'' \emph{Computer Methods
  in Applied Mechanics and Engineering}, vol. 186, no.~2, pp. 339 -- 356, 2000.
  [Online]. Available:
  \url{http://www.sciencedirect.com/science/article/pii/S0045782599003904}
\BIBentrySTDinterwordspacing

\bibitem{andreassen2011efficient}
E.~Andreassen, A.~Clausen, M.~Schevenels, B.~S. Lazarov, and O.~Sigmund,
  ``Efficient topology optimization in {MATLAB} using 88 lines of code,''
  \emph{Structural and Multidisciplinary Optimization}, vol.~43, pp. 1--16,
  2011.

\bibitem{sigmund2001a99}
O.~Sigmund, ``A 99 line topology optimization code written in {MATLAB},''
  \emph{Structural and Multidisciplinary Optimization}, vol.~21, no.~2, pp.
  120--127, 2001.

\bibitem{sirovich1987lowdimensional}
L.~Sirovich and M.~Kirby, ``Low-dimensional procedure for the characterization
  of human faces,'' \emph{Journal of Optical Society of America {A}}, vol.~4,
  no.~3, pp. 519--524, March 1987.

\bibitem{kirby1990application}
M.~Kirby and L.~Sirovich, ``Application of the {K}arhunen-{L}oeve procedure for
  the characterization of human faces,'' \emph{{IEEE} Transactions on Pattern
  Analysis and Machine Intelligence}, vol.~12, no.~1, pp. 103--108, 1990.

\bibitem{turk1991eigenfaces}
M.~Turk and A.~Pentland, ``Eigenfaces for recognition,'' \emph{Journal of
  Cognitive Neuroscience}, vol.~3, no.~1, pp. 71--86, 1991.

\bibitem{allen2003thespace}
B.~Allen, B.~Curless, and Z.~Popovi\'{c}, ``The space of human body shapes:
  Reconstruction and parameterization from range scans,'' \emph{{ACM}
  Transactions on Graphics}, vol.~22, no.~3, pp. 587--594, July 2003.

\bibitem{guest2010reducing}
J.~K. Guest and L.~C. Smith~Genut, ``Reducing dimensionality in topology
  optimization using adaptive design variable fields,'' \emph{International
  Journal for Numerical Methods in Engineering}, vol.~81, no.~8, pp.
  1019--1045, 2010.

\bibitem{jolliffe2005principal}
I.~Jolliffe, \emph{Principal Component Analysis}.\hskip 1em plus 0.5em minus
  0.4em\relax Wiley Online Library, 2005.

\bibitem{cox1994multidimensional}
T.~Cox and M.~Cox, \emph{Multidimensional Scaling}.\hskip 1em plus 0.5em minus
  0.4em\relax London: Chapman and Hall, 1994.

\bibitem{tenenbaum1998advances}
J.~B. Tenenbaum, ``Mapping a manifold of perceptual observations,'' in
  \emph{Advances in neural information processing systems}, 1998, pp. 682--688.

\bibitem{roweis2000nonlinear}
S.~T. Roweis and L.~K. Saul, ``Nonlinear dimensionality reduction by locally
  linear embedding,'' \emph{Science}, vol. 290, no. 5500, pp. 2323--2326, 2000.

\bibitem{blanz1999amorphable}
V.~Blanz and T.~Vetter, ``A morphable model for the synthesis of 3d faces,'' in
  \emph{Proceedings of the 26th annual conference on Computer graphics and
  interactive techniques}.\hskip 1em plus 0.5em minus 0.4em\relax ACM
  Press/Addison-Wesley Publishing Co., 1999, pp. 187--194.

\bibitem{bickel2010design}
\BIBentryALTinterwordspacing
B.~Bickel, M.~B\"{a}cher, M.~A. Otaduy, H.~R. Lee, H.~Pfister, M.~Gross, and
  W.~Matusik, ``Design and fabrication of materials with desired deformation
  behavior,'' \emph{ACM Trans. Graph.}, vol.~29, no.~4, pp. 63:1--63:10, Jul.
  2010. [Online]. Available: \url{http://doi.acm.org/10.1145/1778765.1778800}
\BIBentrySTDinterwordspacing

\bibitem{skouras2013computational}
\BIBentryALTinterwordspacing
M.~Skouras, B.~Thomaszewski, S.~Coros, B.~Bickel, and M.~Gross, ``Computational
  design of actuated deformable characters,'' \emph{ACM Trans. Graph.},
  vol.~32, no.~4, pp. 82:1--82:10, Jul. 2013. [Online]. Available:
  \url{http://doi.acm.org/10.1145/2461912.2461979}
\BIBentrySTDinterwordspacing

\bibitem{schumacher2015microstructures}
\BIBentryALTinterwordspacing
C.~Schumacher, B.~Bickel, J.~Rys, S.~Marschner, C.~Daraio, and M.~Gross,
  ``Microstructures to control elasticity in 3d printing,'' \emph{ACM Trans.
  Graph.}, vol.~34, no.~4, pp. 136:1--136:13, Jul. 2015. [Online]. Available:
  \url{http://doi.acm.org/10.1145/2766926}
\BIBentrySTDinterwordspacing

\bibitem{panetta2015elastic}
\BIBentryALTinterwordspacing
J.~Panetta, Q.~Zhou, L.~Malomo, N.~Pietroni, P.~Cignoni, and D.~Zorin,
  ``Elastic textures for additive fabrication,'' \emph{ACM Trans. Graph.},
  vol.~34, no.~4, pp. 135:1--135:12, Jul. 2015. [Online]. Available:
  \url{http://doi.acm.org/10.1145/2766937}
\BIBentrySTDinterwordspacing

\bibitem{ulu2018designing}
\BIBentryALTinterwordspacing
N.~G. Ulu, S.~Coros, and L.~B. Kara, ``Designing coupling behaviors using
  compliant shape optimization,'' \emph{Computer-Aided Design}, vol. 101, pp.
  57 -- 71, 2018. [Online]. Available:
  \url{https://www.sciencedirect.com/science/article/pii/S0010448518301386}
\BIBentrySTDinterwordspacing

\bibitem{prevost2013make}
\BIBentryALTinterwordspacing
R.~Pr{\'e}vost, E.~Whiting, S.~Lefebvre, and O.~Sorkine-Hornung, ``Make it
  stand: Balancing shapes for 3d fabrication,'' \emph{ACM Trans. Graph.},
  vol.~32, no.~4, pp. 81:1--81:10, Jul. 2013. [Online]. Available:
  \url{http://doi.acm.org/10.1145/2461912.2461957}
\BIBentrySTDinterwordspacing

\bibitem{bacher2014spinit}
\BIBentryALTinterwordspacing
M.~B\"{a}cher, E.~Whiting, B.~Bickel, and O.~Sorkine-Hornung, ``Spin-it:
  Optimizing moment of inertia for spinnable objects,'' \emph{ACM Trans.
  Graph.}, vol.~33, no.~4, pp. 96:1--96:10, Jul. 2014. [Online]. Available:
  \url{http://doi.acm.org/10.1145/2601097.2601157}
\BIBentrySTDinterwordspacing

\bibitem{chen2013spec2fab}
\BIBentryALTinterwordspacing
D.~Chen, D.~I.~W. Levin, P.~Didyk, P.~Sitthi-Amorn, and W.~Matusik, ``Spec2fab:
  A reducer-tuner model for translating specifications to 3d prints,''
  \emph{ACM Trans. Graph.}, vol.~32, no.~4, pp. 135:1--135:10, Jul. 2013.
  [Online]. Available: \url{http://doi.acm.org/10.1145/2461912.2461994}
\BIBentrySTDinterwordspacing

\bibitem{medeiros2015adaptive}
\BIBentryALTinterwordspacing
A.~Medeiros E~S\'{a}, V.~M. Mello, K.~Rodriguez~Echavarria, and D.~Covill,
  ``Adaptive voids,'' \emph{Vis. Comput.}, vol.~31, no. 6-8, pp. 799--808, Jun.
  2015. [Online]. Available: \url{http://dx.doi.org/10.1007/s00371-015-1109-8}
\BIBentrySTDinterwordspacing

\bibitem{zhang2015medial}
\BIBentryALTinterwordspacing
X.~Zhang, Y.~Xia, J.~Wang, Z.~Yang, C.~Tu, and W.~Wang, ``Medial axis tree-an
  internal supporting structure for 3d printing,'' \emph{Comput. Aided Geom.
  Des.}, vol.~35, no.~C, pp. 149--162, May 2015. [Online]. Available:
  \url{http://dx.doi.org/10.1016/j.cagd.2015.03.012}
\BIBentrySTDinterwordspacing

\bibitem{langlois2016stochastic}
\BIBentryALTinterwordspacing
T.~Langlois, A.~Shamir, D.~Dror, W.~Matusik, and D.~I.~W. Levin, ``Stochastic
  structural analysis for context-aware design and fabrication,'' \emph{ACM
  Trans. Graph.}, vol.~35, no.~6, pp. 226:1--226:13, Nov. 2016. [Online].
  Available: \url{http://doi.acm.org/10.1145/2980179.2982436}
\BIBentrySTDinterwordspacing

\bibitem{xu2015interactive}
H.~Xu, Y.~Li, Y.~Chen, and J.~Barbi\v{c}, ``Interactive material design using
  model reduction,'' \emph{{ACM} Trans. on Graphics}, vol.~34, no.~2, 2015.

\bibitem{smith2002creating}
\BIBentryALTinterwordspacing
J.~Smith, J.~Hodgins, I.~Oppenheim, and A.~Witkin, ``Creating models of truss
  structures with optimization,'' \emph{ACM Trans. Graph.}, vol.~21, no.~3, pp.
  295--301, Jul. 2002. [Online]. Available:
  \url{http://doi.acm.org/10.1145/566654.566580}
\BIBentrySTDinterwordspacing

\bibitem{rosen2007design}
D.~W. Rosen, ``Design for additive manufacturing: a method to explore
  unexplored regions of the design space,'' in \emph{Eighteenth Annual Solid
  Freeform Fabrication Symposium}, 2007, pp. 402--415.

\bibitem{choi2002structural}
W.-S. Choi and G.-J. Park, ``Structural optimization using equivalent static
  loads at all time intervals,'' \emph{Computer methods in applied mechanics
  and engineering}, vol. 191, no. 19-20, pp. 2105--2122, 2002.

\bibitem{campbell2011could}
T.~Campbell, C.~Williams, O.~Ivanova, and B.~Garrett, ``Could 3d printing
  change the world,'' \emph{Technologies, Potential, and Implications of
  Additive Manufacturing, Atlantic Council, Washington, DC}, 2011.

\bibitem{seepersad2014challenges}
C.~C. Seepersad, ``Challenges and opportunities in design for additive
  manufacturing,'' \emph{3D Printing and Additive Manufacturing}, vol.~1,
  no.~1, pp. 10--13, 2014.

\bibitem{lipson2013fabricated}
H.~Lipson and M.~Kurman, \emph{Fabricated: The new world of 3D printing}.\hskip
  1em plus 0.5em minus 0.4em\relax John Wiley \& Sons, 2013.

\bibitem{hasan2010physical}
M.~Ha\v{s}an, M.~Fuchs, W.~Matusik, H.~Pfister, and S.~Rusinkiewicz, ``Physical
  reproduction of materials with specified subsurface scattering,'' \emph{ACM
  Trans. Graph.}, vol.~29, no.~4, pp. 61:1--61:10, Jul. 2010.

\bibitem{dong2010fabricating}
Y.~Dong, J.~Wang, F.~Pellacini, X.~Tong, and B.~Guo, ``Fabricating
  spatially-varying subsurface scattering,'' \emph{ACM Trans. Graph.}, vol.~29,
  no.~4, pp. 62:1--62:10, Jul. 2010.

\bibitem{shigley2004mechanical}
J.~E. Shigley, R.~G. Budynas, and C.~R. Mischke, \emph{Mechanical engineering
  design}.\hskip 1em plus 0.5em minus 0.4em\relax McGraw-Hill, 2004.

\bibitem{simpson2001kriging}
T.~W. Simpson, T.~M. Mauery, J.~J. Korte, and F.~Mistree, ``Kriging models for
  global approximation in simulation-based multidisciplinary design
  optimization,'' \emph{AIAA journal}, vol.~39, no.~12, pp. 2233--2241, 2001.

\bibitem{queipo2005surrogate}
N.~V. Queipo, R.~T. Haftka, W.~Shyy, T.~Goel, R.~Vaidyanathan, and P.~K.
  Tucker, ``Surrogate-based analysis and optimization,'' \emph{Progress in
  Aerospace Sciences}, vol.~41, no.~1, pp. 1 -- 28, 2005.

\bibitem{wang2007review}
G.~G. Wang and S.~Shan, ``Review of metamodeling techniques in support of
  engineering design optimization,'' \emph{Journal of Mechanical Design}, vol.
  129, no.~4, pp. 370--380, 2007.

\bibitem{mueller2014matsumoto}
J.~Mueller, ``{MATSuMoTo}: The {MATLAB} surrogate model toolbox for
  computationally expensive black-box global optimization problems,''
  \emph{arXiv preprint arXiv:1404.4261}, 2014.

\bibitem{myers1995response}
R.~H. Myers and D.~C. Montgomery, \emph{Response Surface Methodology: Process
  and Product in Optimization Using Designed Experiments}, 1st~ed.\hskip 1em
  plus 0.5em minus 0.4em\relax New York, NY, USA: John Wiley \& Sons, Inc.,
  1995.

\bibitem{simpson2001metamodels}
T.~Simpson, J.~Poplinski, P.~N. Koch, and J.~Allen, ``Metamodels for
  computer-based engineering design: Survey and recommendations,''
  \emph{Engineering with Computers}, vol.~17, no.~2, pp. 129--150, 2001.

\bibitem{powell1990thetheory}
M.~J. Powell, \emph{The theory of radial basis function approximation in
  1990}.\hskip 1em plus 0.5em minus 0.4em\relax University of Cambridge.
  Department of Applied Mathematics and Theoretical Physics, 1990.

\bibitem{mullur2005extended}
A.~A. Mullur and A.~Messac, ``Extended radial basis functions: more flexible
  and effective metamodeling,'' \emph{AIAA journal}, vol.~43, no.~6, pp.
  1306--1315, 2005.

\bibitem{haykin1999neural}
S.~Haykin, \emph{Neural Networks: A Comprehensive Foundation}.\hskip 1em plus
  0.5em minus 0.4em\relax Prentice-Hall, 1999.

\bibitem{martin2004ontheuse}
J.~D. Martin and T.~W. Simpson, ``On the use of kriging models to approximate
  deterministic computer models,'' in \emph{International Design Engineering
  Technical Conferences and Computers and Information in Engineering
  Conference}.\hskip 1em plus 0.5em minus 0.4em\relax American Society of
  Mechanical Engineers, 2004, pp. 481--492.

\bibitem{girosi1998anequivalence}
F.~Girosi, ``An equivalence between sparse approximation and support vector
  machines,'' \emph{Neural computation}, vol.~10, no.~6, pp. 1455--1480, 1998.

\bibitem{jin2001comparative}
R.~Jin, W.~Chen, and T.~W. Simpson, ``Comparative studies of metamodelling
  techniques under multiple modelling criteria,'' \emph{Structural and
  Multidisciplinary Optimization}, vol.~23, no.~1, pp. 1--13, 2001.

\bibitem{muller2014influence}
J.~Muller and C.~A. Shoemaker, ``Influence of ensemble surrogate models and
  sampling strategy on the solution quality of algorithms for computationally
  expensive black-box global optimization problems,'' \emph{Journal of Global
  Optimization}, vol.~60, no.~2, pp. 123--144, 2014.

\bibitem{mullur2006metamodeling}
A.~A. Mullur and A.~Messac, ``Metamodeling using extended radial basis
  functions: a comparative approach,'' \emph{Engineering with Computers},
  vol.~21, no.~3, pp. 203--217, 2006.

\bibitem{schramm2006recent}
U.~Schramm and M.~Zhou, ``Recent developments in the commercial implementation
  of topology optimization,'' in \emph{{IUTAM} Symposium on Topological Design
  Optimization of Structures, Machines and Materials}, ser. Solid Mechanics and
  Its Applications, M.~P. Bendsoe, N.~Olhoff, and O.~Sigmund, Eds., vol.~{\bf
  137}.\hskip 1em plus 0.5em minus 0.4em\relax Springer {N}etherlands, 2006,
  pp. 239--248.

\bibitem{richardson2010robust}
J.~N. Richardson, R.~F. Coelho, and S.~Adriaenssens, ``Robust topology
  optimization of 2d and 3d continuum and truss structures using a spectral
  stochastic finite element method,'' in \emph{10th World Congress on
  Structural and Multidisciplinary Optimization ({WCSMO} 10), Orlando, Florida,
  USA, May 19}, vol.~24, 2013, p. 2012.

\bibitem{aage2013parallel}
N.~Aage and B.~S. Lazarov, ``Parallel framework for topology optimization using
  the method of moving asymptotes,'' \emph{Structural and Multidisciplinary
  Optimization}, vol.~47, no.~4, pp. 493--505, 2013.

\bibitem{bishop2006pattern}
C.~M. Bishop and N.~M. Nasrabadi, \emph{Pattern recognition and machine
  learning}.\hskip 1em plus 0.5em minus 0.4em\relax springer New York, 2006,
  vol.~1.

\bibitem{hibbeler2015structural}
R.~C. Hibbeler and T.~Kiang, \emph{Structural analysis}.\hskip 1em plus 0.5em
  minus 0.4em\relax Prentice Hall, 2015.

\bibitem{rockafellar2015convex}
R.~T. Rockafellar, \emph{Convex Analysis}.\hskip 1em plus 0.5em minus
  0.4em\relax Princeton University Press, 2015.

\bibitem{crane2013geodesics}
K.~Crane, C.~Weischedel, and M.~Wardetzky, ``{Geodesics in Heat: A New Approach
  to Computing Distance Based on Heat Flow},'' \emph{ACM Trans. Graph.},
  vol.~32, 2013.

\bibitem{sved1968structural}
\BIBentryALTinterwordspacing
G.~Sved and Z.~Ginos, ``Structural optimization under multiple loading,''
  \emph{International Journal of Mechanical Sciences}, vol.~10, no.~10, pp. 803
  -- 805, 1968. [Online]. Available:
  \url{http://www.sciencedirect.com/science/article/pii/0020740368900210}
\BIBentrySTDinterwordspacing

\bibitem{kirsch1990on}
U.~Kirsch, ``On singular topologies in optimum structural design,''
  \emph{Structural and Multidisciplinary Optimization}, vol.~2, no.~3, pp.
  133--142, 1990.

\bibitem{zhou2005large}
\BIBentryALTinterwordspacing
K.~Zhou, J.~Huang, J.~Snyder, X.~Liu, H.~Bao, B.~Guo, and H.-Y. Shum, ``Large
  mesh deformation using the volumetric graph laplacian,'' \emph{ACM Trans.
  Graph.}, vol.~24, no.~3, pp. 496--503, Jul. 2005. [Online]. Available:
  \url{http://doi.acm.org/10.1145/1073204.1073219}
\BIBentrySTDinterwordspacing

\bibitem{zhang2010spectral}
\BIBentryALTinterwordspacing
H.~Zhang, O.~Van~Kaick, and R.~Dyer, ``Spectral mesh processing,''
  \emph{Computer Graphics Forum}, vol.~29, no.~6, pp. 1865--1894, 2010.
  [Online]. Available: \url{http://dx.doi.org/10.1111/j.1467-8659.2010.01655.x}
\BIBentrySTDinterwordspacing

\bibitem{hsu2005interpreting}
M.-H. Hsu and Y.-L. Hsu, ``Interpreting three-dimensional structural topology
  optimization results,'' \emph{Computers \& structures}, vol.~83, no.~4, pp.
  327--337, 2005.

\bibitem{lehoucq1998arpack}
R.~B. Lehoucq, D.~C. Sorensen, and C.~Yang, \emph{ARPACK users' guide: solution
  of large-scale eigenvalue problems with implicitly restarted Arnoldi
  methods}.\hskip 1em plus 0.5em minus 0.4em\relax Siam, 1998, vol.~6.

\bibitem{nocedal2006numerical}
J.~Nocedal and S.~J. Wright, \emph{Numerical Optimization, Second
  Edition}.\hskip 1em plus 0.5em minus 0.4em\relax Springer, 2006.

\bibitem{eigenweb}
G.~Guennebaud, B.~Jacob \emph{et~al.}, ``Eigen v3,''
  http://eigen.tuxfamily.org, 2010.

\bibitem{Paris2010stress}
\BIBentryALTinterwordspacing
J.~Paris, F.~Navarrina, I.~Colominas, and M.~Casteleiro, ``Stress constraints
  sensitivity analysis in structural topology optimization,'' \emph{Computer
  Methods in Applied Mechanics and Engineering}, vol. 199, no.~33, pp. 2110 --
  2122, 2010. [Online]. Available:
  \url{http://www.sciencedirect.com/science/article/pii/S0045782510000861}
\BIBentrySTDinterwordspacing

\bibitem{polyjetDatasheet}
``{PolyJet} materials data sheet,'' 2017, http://www.stratasys.com/materials.

\end{thebibliography}
\end{document}